%% file: main.tex
\documentclass[11pt,superscriptaddress]{article}

\input{setup}

\title{\bf Quench dynamics of entanglement entropy under projective charge measurements: the free fermion case}

\author{Riccardo Travaglino$^1$, Colin Rylands$^{1,2}$ and Pasquale Calabrese$^{1}$}

\date{}

\begin{document}
\maketitle
{\small
\vspace{-5mm}  \ \\
{$^{1}$}  SISSA and INFN Sezione di Trieste, via Bonomea 265, 34136 Trieste, Italy\\[-0.1cm]
\medskip{$^{2}$} Centre for Fluid and Complex Systems, Coventry University, Coventry, CV1 2TT, United\\[-0.3cm]
\medskip Kingdom
}

\begin{abstract}
We consider the effect of projective measurements on the quench dynamics of the bipartite entanglement entropy in one dimensional free fermionic systems. In our protocol, we consider projective measurements of a $U(1)$ conserved charge, the particle number, on some large subsystem, and study the entanglement entropies between  the same subsystem and its complement. We compare the dynamics emanating from  two classes of initial states, one which is an eigenstate of the charge and another which is not. Moreover, we consider the effects of a single measurement as well as multiple  which are periodically performed. Using the quasiparticle picture, we obtain analytic expressions for the behaviour of the entanglement which admit a transparent physical interpretation. In general, we find that measurements introduce two distinct types of corrections to the entanglement, which can be interpreted separately as classical and quantum contributions. The classical contribution is independent of the measurement outcome and scales logarithmically with variance of the charge distribution. In contrast, the quantum contribution depends on the specific measurement outcome and can be significant for individual realizations; however, it becomes negligible when averaged over all possible outcomes. 
Our expressions reduce to previously known results for symmetry resolved entanglement and full counting statistics in some relevant limits, and are confirmed by an exact calculation performed on the Néel initial state.
\end{abstract}
\newpage

\tableofcontents
\section{Introduction}
In recent years, the interplay between unitary evolution and measurements in quantum mechanical systems has attracted a considerable amount of interest. This is because an understanding of the combined effect of the two is crucial for the control complex quantum systems, but also since the observation of measurement-related effects is now experimentally feasible \cite{measurement1,Measurement2,measurement3,measurement4}.
 The significant theoretical effort which ensued has unveiled an extraordinary richness of phenomena driven by the interplay between unitary dynamics and measurements including the quantum Zeno effect \cite{Li_2018,Biella2021manybodyquantumzeno} and an array of measurement induced phase transitions \cite{Li_2019,Skinner_2019,Alberton_2021}.  These effects show that measurements can significantly alter the structure of many body quantum systems \cite{plenio1998,holographic_tensor_networks,zabalo2020criticalproperties,gullans2020scalableprobes,rossinivicari2020,Fan2021selforganized,nahum2021,Ware2021,li2021conformal,Sierant2022dissipativefloquet,Zabalo2022operatorscaling}. One of the most commonly used probes of many body quantum systems, and in particular their unitary dynamics, is the bipartite entanglement entropy between a subsystem and its complement. It condenses the complex correlations of a system to a much simpler form which can display universal properties through its scaling with subsystem size~\cite{vidal_2003,Plenio2005AnIT,damico2008,horodecki2009,Pasquale_Calabrese_2004,Calabrese_2009,Calabrese_2009_1}. Out of equilibrium, its scaling with time, provides insight into the quantum relaxation of a subsystem toward local equilibrium~\cite{quench1,quench2,Deutsch1991,srednicki1,Rigol:2007juv,polkovnikov,DAlessio:2015qtq,Gogolin_2016,ares2025simpler,Collura_2014,CalabreseLN}. It has, therefore been a natural choice to help understand the impact of measurements upon a system, being employed widely in the aforementioned studies, see also \cite{chan2019,measurement_induced_criticality,choi2020,Purifications,ippoliti_2020,bao2020,turkeshi2021infzeroclicks,turkeshi20203d,Altland_2022,murciano_measurement_criticality}.  
 
Because of the difficulty, however, in treating systems which combine unitary evolution and measurements, a predominance of early studies investigated the averaged dynamics in the presence of randomness, either in the system itself or the measurement protocol. The favoured setting for these studies was brickwork quantum circuits, which are models of discrete time evolution in which a robust analytical toolkit allows one to obtain several analytical and numerical results. 
However, further studies
\cite{cao_2019,Biella2021manybodyquantumzeno,Alberton_2021,murciano_measurement_criticality,Turkeshi2022,coppola2022,entanglement_transition_quasiparticles,tirrito2022measindising,turkeshi2023,Romito1, Romito2} have also investigated the effect of local measurements on free fermionic systems, predicting deviations from the  well known universal behaviour of entanglement dynamics, i.e.  linear growth at small times and saturation to volume law at large times. 

Here we consider a different but related problem concerning the effect of measurements on the otherwise unitary quench dynamics of a many body quantum system. Unlike previous studies however, we shall be concerned with  specific types of measurements, namely those of a globally conserved quantity inside an extended subsystem. This resembles in some way the approach of \cite{rajabpour2015,Rajabpour_2016,Lin2023probingsign} (see also the recent work \cite{khanna2025}), where the measurements were performed over extended subsystems. However, the measurements in those works were still of local quantities, such as the single spin configurations, while here we consider global objects, such as the total subsystem charge. In particular, we shall study the quench dynamics of a free fermion chain from two classes of initial states, which is perturbed by a number of measurements of the particle number inside  subsystem. We consider the case of both one and multiple periodically performed charge measurements. As with previous studies, we  investigate this protocol through the  bipartite the entanglement entropy and additionally do so via analytic methods.

\begin{figure}
    \centering
    \begin{tikzpicture}[scale=0.75]
   
        \draw[black,thick,->] (-1,0) -- (13,0) node[right,scale=1.3]{$x$};
         \draw[black,thick,->] (-1,0) -- (-1,5) node [left, scale=1.3]{$t$};
        \shade [top color=Fuchsia, bottom color=cyan, opacity=0.5]
(8,0) rectangle (13,5);
\shade [top color=Fuchsia, bottom color=cyan, opacity=0.5]
(-1,0) rectangle (4,5);
\shade [top color=yellow, bottom color=red, opacity=0.8]
(4,0) rectangle (8,5);
\draw[thick,black] (4,0)-- (4,5);
\draw[thick,black] (8,0)-- (8,5);
;
\draw[black,dashed,line width=2pt] (7,0)--(8,1) ;
\draw[black,dashed,line width=2pt] (7,0)--(6,1);

\draw[black,line width=2pt,->] (8,1)--(10,3) node[right]{$+k$} ;
\draw[black,line width=2pt,->] (6,1)--(4,3) node[left]{$-k$};

\draw[black,dashed,line width=2pt] (3,0)--(4,0.5) ;
\draw[black,dashed,line width=2pt] (3,0)--(2,.5);
\draw[black,line width=2pt,->] (4,0.5)--(6,1.5)  ;
\draw[black,line width=2pt,->] (2,.5)--(0,1.5);

\node [scale=1.5] at (1.5,5.5){$\overline{A}$};
\node [scale=1.5] at (10.5,5.5){$\overline{A}$};
\node [scale=1.5] at (6,5.5){$A$};
\draw[black,->,line width=2pt,dashed] (6,0)--(7,4) ;
\draw[black,->,line width=2pt,dashed] (6,0)--(5,4);

    \end{tikzpicture}
    \caption{Quasiparticle picture of quench dynamics in free fermions. In the picture, the quench acts on the initial state by producing pairs of entangled quasiparticles, relating, therefore, entanglement propagation to transport.  At each instant of time, dashed lines correspond to pairs which are not shared between $A$ and $\overline{A}$ and therefore do not contribute to the entropy. Solid lines in contrast correspond to shared pairs, which contribute to the entanglement. }
    \label{fig:qpp}
\end{figure}
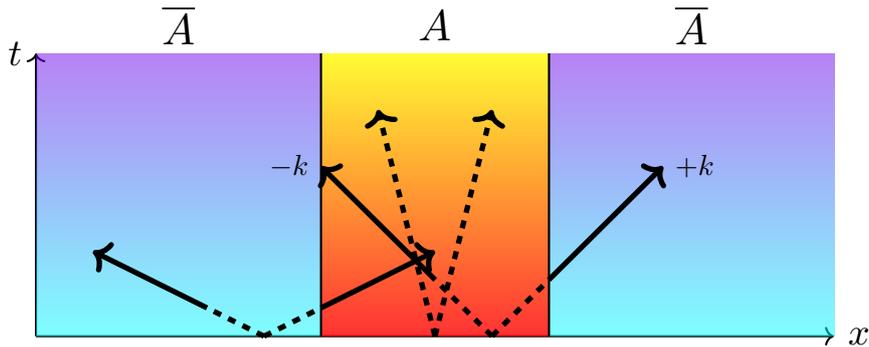

Our primary tool in this endeavour is the quasiparticle picture of entanglement dynamics and, in particular, in the recent operatorial formulation proposed in \cite{rottoli2024entanglementHamiltoniansquasiparticlepicture}. Originally developed in \cite{quench1,quench2} to describe the evolution of the Von Neumann entropy in critical systems at large space-time scales, the fundamental idea of the quasiparticle picture is that the propagation of the correlations after a quench is due to pairs of entangled quasiparticles which are created at the instant of the quench and travel ballistically across the system. From this point of view, the problem of evaluating the entanglement entropy reduces to understanding the entanglement contribution of each pair, and then to a counting problem to understand which pairs contribute to the entanglement, as depicted schematically in Figure \ref{fig:qpp}.
This picture describes the most relevant properties of entanglement evolution at the large spacetime scales in massless and free theories \cite{quench2,fagotti2008evolution,Calabrese_2016_quench}, namely its linear growth at short times and subsequent relaxation to thermodynamic values at long times. 
It was then extended to interacting integrable systems, in which the notion of stable quasiparticles can still be applied \cite{alba1,alba2} \footnote{The validity of the of the quasiparticle picture for the evolution of R\'enyi entropies in interacting systems was contested in \cite{klobas2021,PRX} on the basis of results obtained through spacetime duality \cite{PRX,bertini2023nonequilibrium,bertini_asymmetric_2024}. However, these studies confirmed the validity of the picture for the Von Neumann entropy, and in general for the evaluation of all quantities which do not require a multi-replica evaluation.} and 
 to the characterization of several entanglement related measures in free and interacting systems \cite{Bertini_2018,Entanglement_GHD_interacting,alba2019quantum,AlbaQA,parez2021exact,parez2021quasiparticle,Parez_2022,Murciano2022,ares2023entanglement,murciano2024entanglement}. Recently, the quasiparticle picture was given an operatorial formulation \cite{rottoli2024entanglementHamiltoniansquasiparticlepicture}, which allows to access operatorial characterizations of entanglement such as the entanglement and negativity Hamiltonians \cite{travaglino2024,travaglino2025}. In this work, we will show that this picture can also be used quite effectively to approach the measurement problem at the hydrodynamic scale. We note that there have already been some attempts to extend the quasiparticle picture to study open systems \cite{alba_carollo2021,Alba_2022,carollo_alba2022,alba_carollo2023}
and systems undergoing continuous measurements \cite{cao_2019,entanglement_transition_quasiparticles,carollo2022,turkeshi2023}. However, these results are still inconclusive on whether a quasiparticle picture could actually describe the continuous measuring process. Although our focus is different, our operatorial approach could shed light on the problem, as it leads to the quasiparticle picture result for the entropy in a constructive manner starting from the operatorial formulation, which is a well justified ansatz for the evolution of the density matrix itself.

The structure of the paper is the following. In Section \ref{sec:setup} we introduce the specifics of our setup, including the Hamiltonian, initial states, and measurement protocol. We also review the operatorial quasiparticle picture. In Section \ref{sec:symm_measurements} we comment on the properties of entanglement in charge symmetric states and  analyse the effect of a single measurement on the entanglement in such systems. In Section \ref {sec:multiple_symm} we extend this analysis to the case of multiple measurements on symmetric states. In Section \ref{sec:squeezed} we move on to initial states which are not symmetric and analyse the effect of both a single and multiple measurement. In Section \ref{sec:examples} we provide some examples using prototypical initial states, namely, the N\'eel, dimer and tilted ferromagnetic states. In  Section \ref{sec:concl} we summarize and conclude. Details of several calculations and some generalizations are presented in the appendices. 


\section{Setup}
\label{sec:setup}
The standard quantum quench protocol involves initializing a closed quantum system in a non-equilibrium state, $\ket{\psi}$ and allowing it to undergo unitary time evolution governed by a system Hamiltonian, $\ket{\psi(t)}=e^{-iHt}\ket{\psi}$. We study the scenario in which this protocol is modified by interspersing the unitary dynamics with  projective measurements, performed within a finite subsystem, of a globally conserved quantity. In this section we introduce the  Hamiltonian, initial states, the type of measurement  and our specific protocol for the time evolution.  
\subsection{Hamiltonian and initial states}
 Our system consists of a free fermionic chain of $L$ sites whose unitary dynamics is determined by the tight binding Hamiltonian,
\begin{equation}\label{eq:Hamiltonian}
    H=-\frac{1}{2} \sum_{x=1}^Lc^\dagger_i c_{i+1} + c^\dagger_{i+1} c_{i}=\sum_{k}\varepsilon c^\dagger_kc_k~,
\end{equation} 
where $c^\dagger_x,c_x$ are fermionic creation and annihilation operators for the site $x$ and $ c^\dagger_k,c_k$ are their Fourier space counterparts. The dispersion is given by $\varepsilon=-\cos(k)$ and we impose periodic boundary conditions. Much of what we say, however, is applicable more generally to arbitrary free fermion models, with the final results depending only on the group velocity $v_k = \frac{\partial\varepsilon}{ \partial_k}$. We shall rely on the explicit form of~\eqref{eq:Hamiltonian} and the value $v_k=\sin (k)$, only when considering some specific examples later on. The Hamiltonian conserves the particle number, $\hat{Q}$, also referred to as the charge, which is defined as
\begin{eqnarray}
    \hat{Q}=\sum_{x=1}^L c^\dagger_xc_x=\sum_{k}c^\dag_kc_k~.
\end{eqnarray}
In addition, the model has an extensive number of other conserved quantities, such as the mode occupation $c^\dag_kc_k$ making it integrable. 

We consider two different classes of initial states which have been widely utilized in studies of free fermion quenches. The first class is given by 
\begin{eqnarray}\label{eq:symmetric_state}
    \ket{\psi}=\prod_{k>0} \left(\sqrt{1-n(k)}c^\dagger_{k-\pi}+\sqrt{n(k)} c^\dagger_k \right)\ket{0},
\end{eqnarray}
with  $\ket{0}$ being the vacuum containing no fermions.  Here,  $n(k)$ is the fermion occupation function, $n(k)=\bra{\psi}c^\dag_kc_k\ket{\psi}=1-\bra{\psi}c^\dag_{k-\pi}c_{k-\pi}\ket{\psi}$, which we leave as an arbitrary function  characterizing the initial state. This state is an eigenstate of $\hat{Q}$ and has two site translational symmetry as well as particle hole symmetry. Consequently, the eigenvalue of the charge is fixed to be $q=L/2$ i.e half filling. In the remainder of the paper we refer to states within this class as symmetric states. The second class of states is given by 
\begin{equation}\label{eq:squeezed}
    \ket{\psi}= \prod_{k>0}\left( \sqrt{1-n(k)} + \sqrt{n(k)} c^\dagger_{k} c^\dagger_{-k}\right)\ket{0}~,
\end{equation} 
where, $n(k)$ is again the fermion occupation function, but which now must be an even function but is otherwise arbitrary. This state is not an eigenstate of $\hat{Q}$ but  has single site translational invariance and so the expectation value of the charge on any finite subsystem is a constant of motion, we denote the average charge density by $\bar{q}$.  We refer to states of this type state as either squeezed states or non-symmetric states. Despite their differences, both classes of states are Gaussian and so their quench dynamics can be treated exactly using free fermionic techniques. 

\subsection{Protocol}
Throughout the time evolution, we shall make a number of projective  charge measurements inside a subsystem, $A$, which we take to be formed of $\ell$ contiguous sites. The fermion number, restricted to this subsystem is denoted by $\hat{Q}_A$ and is given explicitly by 
\begin{eqnarray}
    \hat{Q}_A=\sum_{x\in A}c_x^\dagger c_x~,
\end{eqnarray}
where we have merely restricted the sum over sites to be inside $A$. This operator has integer eigenvalues, taking values in the range $q\in [0,\ell]$ which are the possible outcomes of the projective measurement. We denote the projector onto the eigenspace of $\hat{Q}_A$ with eigenvalue $q$ by $\Pi^A_q$. 

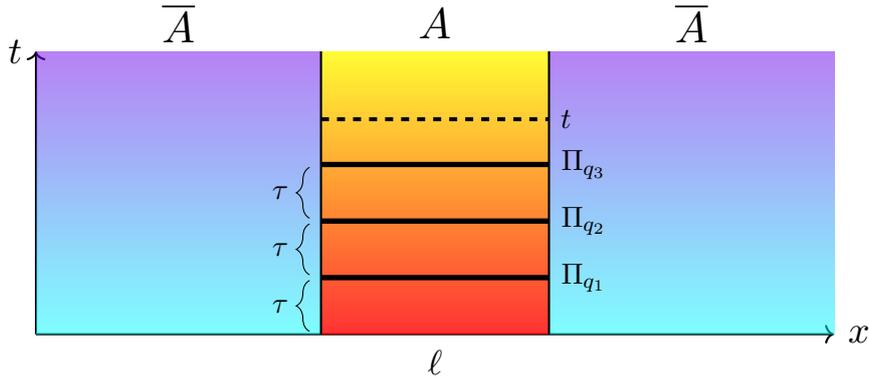
\begin{figure}
    \centering
    \begin{tikzpicture}[scale=0.75]
   
        \draw[black,thick,->] (-1,0) -- (13,0) node[right,scale=1.3]{$x$};
         \draw[black,thick,->] (-1,0) -- (-1,5) node [left, scale=1.3]{$t$};
        \shade [top color=Fuchsia, bottom color=cyan, opacity=0.5]
(8,0) rectangle (13,5);
\shade [top color=Fuchsia, bottom color=cyan, opacity=0.5]
(-1,0) rectangle (4,5);
\shade [top color=yellow, bottom color=red, opacity=0.8]
(4,0) rectangle (8,5);
\draw[thick,black] (4,0)-- (4,5);
\draw[thick,black] (8,0)-- (8,5);
;

\draw[black,line width=2pt] (4,1)--(8,1)node[right]{$\Pi_{q_1}$} ;
\draw[black,line width=2pt] (4,2)--(8,2)node[right]{$\Pi_{q_2}$} ;
\draw[black,line width=2pt] (4,3)--(8,3)node[right]{$\Pi_{q_3}$} ;
\node [scale=1.2] at (6,-0.5){$\ell$};

\draw[black,line width=1.5pt,dashed] (4,3.8)--(8,3.8)node[right]{$t$} ;

\node [scale=1.5] at (1.5,5.5){$\overline{A}$};
\node [scale=1.5] at (10.5,5.5){$\overline{A}$};
\node [scale=1.5] at (6,5.5){$A$};

\draw [decorate,decoration={brace,amplitude=5pt}]
  (3.8,0.05) -- (3.8,0.95) node[midway,xshift=-1em]{$\tau$};
  \draw [decorate,decoration={brace,amplitude=5pt}]
  (3.8,1.05) -- (3.8,1.95) node[midway,xshift=-1em]{$\tau$};
  \draw [decorate,decoration={brace,amplitude=5pt}]
  (3.8,2.05) -- (3.8,2.95) node[midway,xshift=-1em]{$\tau$};
 
    \end{tikzpicture}
    \caption{Measurement protocol \eqref{eq:full_state_multiple}. The system undergoes a non-equilibrium free evolution, from a non-equilibrium initial state. After a time step $\tau$, it is projected within the system $A$ to an eigenstate of the $U(1)$ charge. The unitary evolution-projection is then repeated an arbitrary number of times; after the $m$-th step, at some time t, the entanglement entropy of the interval $A$ is then computed.   }
    \label{fig:measurementprotocol}
\end{figure}

Our particular measurement protocol consists of periodically performed measurements of $\hat{Q}_A$ in between unitary dynamics generated by $H$. More specifically, as depicted in Figure \ref{fig:measurementprotocol}, we allow our system to undergo unitary time evolution until a time $\tau$ at which point we projectively measure the charge in $A$, obtaining a certain outcome which we denote by  $q_1$, the system then evolves unitarily until we make another measurement at time $2\tau$ which returns the value $q_2$. We repeat this until $m$ measurements are performed and after the final measurement the system undergoes a further stage of unitary evolution for a time $t-m\tau$ so that the total time for the whole protocol is $t$. The sequence of  measurement outcomes is collected in the $m$ component vector $\{q_i\}=(q_1,\dots,q_m)$. 
According to the Born rule, the final state of the full system at time $t$ is given by
\begin{align}\label{eq:full_state_multiple}
    \rho(t|\tau,\{q_i\})=\frac{e^{-iH(t-m\tau)}\Pi^A_{q_m}e^{-iH\tau}\dots \Pi^A_{q_1}e^{-iH\tau}\ket{\psi}\!\bra{\psi}e^{i\tau H}\Pi^A_{q_1}\dots e^{iH\tau}\Pi_{q_{m}}^A e^{iH(t-m\tau)}}{p(\tau,\{q_i\}\,)}
\end{align}
where $p(\tau,\{q_i\}\,)$ is the probability of that particular sequence of measurement outcomes. This is given by
\begin{eqnarray}
   p(\tau,\{q_i\})=\tr \Big[\Pi^A_{q_m}e^{-iH\tau}\dots \Pi^A_{q_1}e^{-iH\tau}\ket{\psi}\!\bra{\psi}e^{i\tau H}\Pi^A_{q_1}\dots e^{iH\tau}\Pi_{q_{m}}^A\Big]~,
\end{eqnarray}
which ensures that $\tr[\rho(t|\tau,\{q_i\})]=1$. 

Our goal in this paper is to  investigate how the measurements affect the relaxation of the system to local equilibrium. To do this we compute one of the most widely used measures of quantum relaxation in quench dynamics, the entanglement entropy between $A$ and its complement, $\bar{A}$. To this end we employ the replica trick and compute the R\`enyi entanglement entropies
\begin{eqnarray}
    S^{(\alpha)}_A(t|\tau,\{q_i\}\,)=\frac{1}{1-\alpha}\log \tr_A \Big[\{\rho_A(t|\tau,\{q_i\}\,)\}^\alpha\Big]~,~~\alpha \in \mathbb{N}
\end{eqnarray}
where $\rho_A(t|\tau,\{q_i\}\,)=\tr_{\bar{A}}[\rho_A(t|\tau,\{q_i\}\,)]$ is the reduced density matrix of the system at time $t$. Upon analytically continuing $\alpha$ to arbitrary values and taking $\alpha\to 1$ limit we obtain the von-Neumann entanglement entropy
\begin{eqnarray}
    S_A(t|\tau,\{q_i\}\,)=-\tr_A\big[\rho_A(t|\tau,\{q_i\}\,)\log \{\rho_A(t|\tau,\{q_i\}\,)\}\big]=\lim_{\alpha\to 1}S^{(\alpha)}_A(t|\tau,\{q_i\}\,)~.
\end{eqnarray}
In the absence of measurements, the quench dynamics of the entanglement entropy has been extensively studied, particularly in free fermion models. In addition, for a symmetric initial state and a single measurement made at the final time, $t=\tau$, $m=1$
 the problem is closely related to the symmetry resolution of entanglement \cite{Goldstein-sela,xavier_alcaraz_sierra}, and in particular to the characterization of the charged moments, which was studied in \cite{parez2021exact,parez2021quasiparticle}. For non-symmetric initial states, the situation more closely resembles studies of the entanglement asymmetry~\cite{ares2023entanglement} which is used to study symmetry breaking at the level of an extended subsystem. 
We shall make connections with both of these quantities to gain insight into the current scenario. To the best our knowledge the cases of $t>\tau$ and $m\geq 1$ have not been considered before. 

\subsection{Quasiparticle Picture}
To calculate $S^{(\alpha)}_A(t)$, and ultimately $S_A(t)$, we shall employ the quasiparticle picture of entanglement dynamics. This semi-classical effective theory has proven to be highly successful in not only obtaining analytic predictions but also providing  intuitive understanding. The picture emerges from exact calculations in the ballistic scaling limit of free systems where we take $L,t,\ell\to \infty$ such that $L\gg\ell,t$  and the ratio $t/\ell$ is held fixed. Alternatively, it can be used as an effective theory from the outset with final results checked numerically. This was the approach used in~\cite{rottoli2024entanglementHamiltoniansquasiparticlepicture}  where it was shown that the reduced density matrix without measurements obeys an operatorial form of the quasiparticle picture. Here we briefly review how the theory works in the absence of the measurements, i.e. under purely unitary dynamics.

The starting point of using the quasiparticle picture as an effective theory is 
 to  divide the system into mesoscopic hydrodynamic cells of size $1\ll\Delta\ll \ell$. The lattice sites are then indexed as $x={x_0+\varkappa}$ where $x_0=1,\dots,L/\Delta$ labels the fluid cell and $\varkappa=0,\dots,\Delta-1$ signifies the position within the cell. After this we define the quasiparticle creation operator as 
\begin{equation}
    b^\dag_{x_0,k} = \frac{1}{\sqrt{\Delta}}\sum_{\varkappa=0}^{\Delta-1}  e^{-ik\varkappa}c^\dagger_{x_0+\varkappa} 
    \label{eq:cellfourier}~,
\end{equation}
where we have taken the Fourier transform inside the fluid cell. We should note that the wavevector $k$ in the above expression is quantized according to the fluid cell size rather than the full system size as is the case in~\eqref{eq:Hamiltonian}. We shall, nevertheless, refer to both of these wavevectors by the same symbol as ultimately we shall take the thermodynamic limit in which case their quantization will not matter. If the correlation length of the initial state is small compared to $\Delta$, then the initial density matrix for our two classes of initial states can be expressed as product over local density matrices inside each fluid cell,
\begin{eqnarray}\label{eq:intial_rho_decomp}
\rho(0)=\prod_{x_0=1}^{L/\Delta}\prod_{k>0}\rho_{x_0,k}~,
\end{eqnarray}
where $\prod_{k>0}\rho_{x_0,k}$ is the local state within the fluid cell decomposed into its 
 different momentum sectors. 
For the symmetric initial state the local density matrix is given by~\cite{rottoli2024entanglementHamiltoniansquasiparticlepicture,travaglino2024},
\begin{eqnarray}\nonumber
\label{eq:initialdimer}
    \rho_{x_0,k} &=&n(k)\hat{n}_{x_0}(k)(1-\hat{n}_{x_0}(k-\pi)) + (1-n(k))(1-\hat{n}_{x_0}(k) )\hat{n}_{x_0}(k-\pi)\hspace{1cm}\\\label{eq:symm_presrve_pure} &&+ \sqrt{n(k)(1-n(k)) }\left(b_{x_0,k}^\dagger b_{x_0,k-\pi}+b_{x_0,k-\pi}^\dagger b_{x_0,k}\right)~,
\end{eqnarray}
with $\hat{n}_{x_0}(k)=b^\dag_{x_0,k}b_{x_0,k}$. For the squeezed state, on the other hand, we have instead 
\begin{eqnarray}\nonumber
\label{eq:initialsqueezed}
\rho_{x_0,k}&=&n(k)\hat{n}_{x_0}(k)\hat{n}_{x_0,-k}+(1-n(k))(1-\hat{n}_{x_0}(k))(1-\hat{n}_{x_0}(-k))\\
&&+\sqrt{n(k)(1-n(k))}(b_{x_0,k}^\dagger b_{x_0,-k}^\dagger + b_{x_0,-k}b_{x_0,k})~,
\end{eqnarray}
whose symmetry breaking properties are evident in the second line. The above decomposition,~\eqref{eq:intial_rho_decomp}, means that there are correlations only between quasiparticles which are from the same fluid cell and have paired momenta i.e. between the quasiparticles labelled by $x_0,k$ and $x_0,k-\pi$ for the symmetric state or $x_0,k$ and $x_0,-k$ for the squeezed state. In either case, one member of the pair has velocity $v_k$ and the other $-v_k$.

The quasiparticle picture assumes that the quasiparticles of the system propagate semi-classically and ballistically through the system. In terms of our quasiparticle creation operators, $b^\dagger_{x_0,k}$ this means that we approximate the true quantum evolution by~\cite{rottoli2024entanglementHamiltoniansquasiparticlepicture},
\begin{equation}
\label{eq:ballisticevolution}
    b^\dag_{x_0,k}{(t)} = e^{-iHt} b^\dag_{x_0,k}e^{iHt} \approx  b^\dag_{x_t(k),k}
\end{equation}
where 
\begin{equation}
x_t(k)=x_0+v_kt~,
\end{equation} and $v_k = \frac{\partial \varepsilon}{\partial_k} $.  This leads to an effective time-dependent density matrix of the form, 
\begin{equation}
\rho(t)=\prod_{k>0}\prod_{x_0}\rho_{x_0,k}(t)~,
    \label{eq_decomposition}
\end{equation}
where $\rho_{x_0,k}(t)$ is immediately obtained from the initial state which can be either of the form \eqref{eq:initialdimer} or \eqref{eq:initialsqueezed} by imposing the semi-classical evolution~\eqref{eq:ballisticevolution}. This preserves the structure of correlations which where present in the initial state, i.e. the quasiparticle which originate at $x_0$ with momentum $k$ is still only correlated with the one which also originated in $x_0$ but which has $k-\pi$ for the symmetric state or $-k$ for the squeezed state. The effect of the time evolution is that these two quasiparticles no longer lie inside the fluid cell $x_0$ but have been transported to $x_t(k)$ and $x_t(k-\pi)$ (or $x_t(-k)$ for the squeezed state).

At this point, evaluating certain quantities is fairly straightforward. For example, to compute the R\'enyi entropies  we must trace out the region $\bar A$ which can be carried out individually for each correlated quasiparticle pair as follows. For the pair indexed by $x_0,k$, suppose that $x_t(k)\in A$ while $x_t(k-\pi)\in \bar{A}$ i.e it is shared between $A$ and $\bar{A}$ then 
\begin{equation}\label{eq:local_gibbs}
    \tr_{\overline{A}} [\rho_{x_0,k}(t) ]= n(k) \hat{n}_{x_t}(k) + (1-n(k) ) (1-\hat{n}_{x_t}(k))~.
\end{equation}
If instead, $x_t(k), x_t(k-\pi)\in A$ then  $ \tr_{\overline{A}} [\rho_{x_0,k}(t)] =  \rho_{x_0,k}(t) $ whereas if $x_t(k),x_t(k-\pi)\in \bar{A}$ then $ \tr_{\overline{A}} [\rho_{x_0,k}(t) ]=  1 $. Since each pair density matrix is pure, the only contributions to the entropy arises only from the shared pairs, in which the trace of the complement induces mixing as shown in~\eqref{eq:local_gibbs}.  After summing up all pair contributions and then taking the thermodynamic limit so that $\sum_k\to \int_{-\pi}^\pi \frac{{\rm d}k}{2\pi}$, we obtain the  quasiparticle picture result, 
\begin{equation}
\label{eq:Qp_renyi}
    S^{(
    \alpha)}_A(t) = \int \frac{{\rm d}k}{2\pi} \min(2|v_k|t,\ell)s^{(\alpha)}[n(k)]~.
\end{equation}
Here,  $s^{(\alpha)}[n(k)]=\frac{1}{1-\alpha}\log\{[1-n(k)]^\alpha+[n(k)]^\alpha\}$ is the pair contribution to the R\'enyi entanglement entropy and the function $\min(2|v_k|t,\ell)$ counts how many pairs are shared between $A$ and $\bar A$ at time $t$. In the replica limit we find 
\begin{equation}
\label{eq:qppentropy}
    S_A(t) = \int \frac{{\rm d}k}{2\pi} \min(2|v_k|t,\ell)s[n(k)]~.
\end{equation}
where now
 $s[n(k)]=-n(k) \log n(k)-(1-n(k))\log(1-n(k))$ which is equivalent to the thermodynamic entropy of~\eqref{eq:local_gibbs}.  
In the next sections we will describe how to incorporate the effect of the measurements into the quasiparticle picture and how to compute the resulting entropy.

\section{A single measurement on symmetric states}
\label{sec:symm_measurements}
We begin our analysis by considering the dynamics from symmetric initial states~\eqref{eq:symmetric_state} and concentrating, in this section, on just a single measurement. Before proceeding to the quasiparticle picture calculations we make some preliminary remarks concerning the entanglement entropy in symmetric states. 

\subsection{Remarks on entanglement in symmetric states}
\label{sec:remarks}
Prior to performing the calculation of the $S_A(t|\tau,q)$ in our system, it is instructive to recap some basic properties of the entanglement entropy of symmetric states. In particular, if we quench a system from a symmetric initial state, such as~\eqref{eq:symmetric_state}, and allow it to evolve in time according to a symmetric Hamiltonian, for the moment considering the case without any measurement, the state will remain symmetric at all times. Moreover, the full state will remain an eigenstate of $\hat{Q}$ with eigenvalue $q=L/2$. Furthermore, since the charge can be split into a part acting on $A$ and a part acting on $\bar A$, the reduced state of the subsystem $A$ will also remain symmetric, 
\begin{eqnarray}
    [\rho_A(t),\hat{Q}_A]=0,~~\forall t~.
\end{eqnarray}
The state $\rho_A(t)$ is a mixed state and due to its symmetry admits a decomposition into the different charge sectors of $\hat{Q}_A$\begin{eqnarray}\label{eq:rho_decomposition}
    \rho_A(t)=\bigotimes_{\mathfrak{q}=0}^\ell p(t,\mathfrak{q})  \rho_{A,\mathfrak{q}}(t)
\end{eqnarray}
 where $\rho_{A,\mathfrak{q}}(t)$ is the normalized density matrix obtained by projecting $\rho_{A}(t)$ into the eigenspace of $\hat{Q}_A$ with eigenvalue $\mathfrak{q}$ and $p(t,\mathfrak{q})$ is its normalization,
\begin{eqnarray}\label{eq:ent_symm_1}
     \rho_{A,\mathfrak{q}}(t)&=&\frac{\Pi^A_\mathfrak{q}\rho_{A}(t)\Pi^A_\mathfrak{q}}{p(t,\mathfrak{q})}~~\\\label{eq:ent_symm_2}
     p(t,\mathfrak{q})&=&\tr_A\Big[\Pi^A_\mathfrak{q}\rho_{A}(t)\Big].
\end{eqnarray}
With this, we have $\tr_A[ \rho_{A,\mathfrak{q}}(t)]=1$. This decomposition has served as the basis for numerous studies on the symmetry resolution of entanglement~\cite{Goldstein-sela,xavier_alcaraz_sierra,castro2024symmetry}. These studies seek to understand how the entanglement entropy between $A$ and $\bar{A}$ decomposes into the different charge sectors. One of the main results of these studies is that the entanglement entropy can be expressed as a sum of two distinct terms 
\begin{eqnarray}\label{eq:symme_decomp_ent}
    S_A(t)=S_{A,{\rm num}}(t)+S_{A,{\rm conf}}(t)~.
\end{eqnarray}
The first term, $S_{A,{\rm num}}(t)$,  is called the number entropy, 
\begin{align}\label{eq:ent_num}
    S_{A,{\rm num}}(t)=-\sum_{\mathfrak{q}=0}^\ell p(t,\mathfrak{q})\log \{p(t,\mathfrak{q})\}~.
\end{align}
It arises from the fluctuations of charge between different charge sectors and coincides with the classical Shannon entropy of the charge probability distribution of the state. The second term, $S_{A,{\rm conf}}(t)$, is the configuration entropy
\begin{eqnarray}\label{eq:ent_conf_1}
     S_{A,{\rm conf}}(t)&=&\sum_{\mathfrak{q}=0}^\ell p(t,\mathfrak{q})S_{A,\mathfrak{q}}(t)~,\\\label{eq:ent_conf_2}S_{A,\mathfrak{q}}(t)&=&-\tr_{A}[\rho_{A,\mathfrak{q}}(t)\log \{\rho_{A,\mathfrak{q}}(t)\}]~.
\end{eqnarray}
This, instead, arises from considering the entanglement contained within the charge sectors themselves. The contribution of each charge sector is given by   $S_{A,\mathfrak{q}}(t)$, also known as the symmetry resolved entropy, and the total configuration entropy is the sum of these, weighted by their associated probability distributions.

Several interesting results have been derived concerning the symmetry resolution of entanglement. First, the symmetry resolved entropy experiences a time delay, namely there exists a time $t_{{\rm d},\mathfrak{q}}$ prior to which there is no contribution to entanglement for that particular charge sector,
i.e. 
\begin{eqnarray}
    S_{A,\mathfrak{q}}(t<t_{{\rm d},\mathfrak{q}})=0,~~p(t<t_{{\rm d},\mathfrak{q}},\mathfrak{q})=0~.
\end{eqnarray}
The reason for this is that when quenching from an eigenstate of the charge, there is a minimum amount of time which one must wait for the subsystem to accrue  a charge $\mathfrak{q}$. The further $\mathfrak{q}$ is from the initial value of the subsystem charge, $\ell/2$, the longer the time delay. For $\mathfrak{q}=\ell/2$ there is no time delay, $t_{{\rm d},\ell/2}=0$. Second, in the predominance of cases studied so far, an equipartition of entanglement between charge sectors has been observed. This means that, at times beyond the delay time and for charges which are close to the initial expectation value, the symmetry resolved entropy does not depend on $\mathfrak{q}$ to leading order. That is,
\begin{eqnarray}
     S_{A,\mathfrak{q}}(t> t_\mathfrak{q})=  S_{A,\ell/2}(t)+\mathcal{O}(\Delta \mathfrak{q}^2)
\end{eqnarray}
where $\Delta \mathfrak{q}=\ell/2-\mathfrak{q}$. Finally, at long enough times the charge probability distribution is approximately normally distributed with a variance, $\sigma_t^2 $, which grows linearly in time. As a result we have that the number entropy, being the Shannon entropy of the probability distribution, has a logarithmic growth in time,
\begin{eqnarray}
    S_{A,{\rm num}}(t)=\frac{1}{2}\log ( 2\pi e \sigma_t^2)\simeq\frac{1}{2}\log ( t)~,
\end{eqnarray}
where the second equality is valid in the regime $t<\ell/2$.

The aforementioned results have been derived in the case when no measurement is made on the system. However, since we perform a measurement of the charge within the subsystem, we have that the full state remains an eigenstate of $\hat{Q}$ regardless of the outcome of the measurement and similarly,
\begin{eqnarray}
    [\rho_A(t|\tau, q),\hat{Q}_A]=0,~~\forall t, \tau,q~.
\end{eqnarray}
Thus all the results quoted above can be applied to the measured system also. This allows us to make some immediate predictions for our protocol. The most straightforward is that,
\begin{eqnarray}
    p(\tau<t_{{\rm d},q},q)=0~,
\end{eqnarray}
or in words, it is not possible to measure the value $q$ inside the subsystem prior to its delay time. Second, since upon measuring the subsystem, the state is projected into a single charge sector, there will be a drop in entropy which can be written as a sum of number and configuration pieces. In particular, if $\tau <\ell/2$  it is expected that the entropy takes the form 
\begin{equation}
\label{eq:expectation}
    S_A(t|\tau,q) \approx S_{A,\rm conf} - \frac{1}{2} \log \tau + \dots 
\end{equation}
where the second term represents precisely the drop in the number entropy which takes place as the state is projected over a charge sector.
Note that this form of the correction is valid only at small times after the measurements.  As time increases, additional fluctuations determined by the evolution start playing a role and this result is expected to change. 

\subsection{Quasiparticle approach}
We now turn to the analysis of the entanglement dynamics emerging from a symmetric state with a single measurement, $m=1$, $t\geq\tau$ using the quasiparticle picture. We begin by bringing the expression for $\rho(t|\tau,q)$ to a form which is more amenable to calculation by the quasiparticle picture. To do this we use the Fourier decomposition of the projector $\Pi^A_q$
\begin{eqnarray}   \label{eq:projector_fourier} \Pi^A_q=\int_{-\pi}^\pi\frac{{\rm d}\lambda}{2\pi}e^{-i\lambda(q-\hat{Q}_A)}~.
\end{eqnarray}
In terms of this we can then write 
\begin{eqnarray}\label{eq:charge_moments}
   \tr_A \Big[\{\rho_A(t|\tau,q)\}^\alpha\Big]&=&\frac{ \int_{-\pi}^\pi \Big[\prod_{j=1}^\alpha\frac{{\rm d}\lambda_j{\rm d}\lambda_j'}{4\pi^2}e^{-i q(\lambda_j+\lambda_j')}\Big] e^{ \mathcal{F}_{\alpha}(t,\{\lambda_j,\lambda_j'\}\,)}}{\Big[\int_{-\pi}^\pi \frac{{\rm d}\lambda {\rm d}\lambda'}{4\pi^2}e^{-i q(\lambda+\lambda')} e^{\mathcal{F}_{1}(t,{\lambda},{\lambda}')}\Big]^\alpha}
\end{eqnarray}
where 
\begin{eqnarray}\label{eq:mathcalF_1}
   \mathcal F_{\alpha}(t,\{\lambda_j,\lambda_j'\}\,)=\log \Bigg\{\tr_{A}\Big[\prod_{j=1}^\alpha\Big(\tr_{\bar{A}}\big[e^{-iH(t-\tau)}e^{i\lambda_j\hat{Q}_A}\rho(\tau)e^{i\lambda'_j\hat{Q}_A}e^{iH(t-\tau)}\big]\Big)\Big]\Bigg\}
\end{eqnarray}
with $\rho(\tau)$ being the unmeasured density matrix of the full system at time $\tau$. Therefore, our task boils down to calculating $\mathcal{F}_\alpha(t,\{\lambda_j,\lambda_j'\})$  using the quasiparticle picture. This can be simplified further by noting that 
 the subsystem charge operator can be decomposed into quasiparticle operators as 
\begin{eqnarray}
    \hat{Q}_A = \sum_{x_0\in A}\sum_k \hat{n}_{x_0,k}~.
\end{eqnarray}
Combining these expressions with the decomposition of the density matrix~\eqref{eq:intial_rho_decomp} we see that the evaluation of $\mathcal{F}_\alpha(t,\{\lambda_j,\lambda_j'\})$ requires us to first understand how the operator $e^{i\lambda \hat{n}_{x_0,k}}$ modifies the quasiparticle 
 decomposition of $\rho(\tau)$. Given this we can then implement the semi-classical time evolution for a period $t-\tau$ and then perform the trace over $\bar{A}$. 

There are several different configurations to consider depending on whether one, both, no quasiparticles are inside $A$ at time $\tau$.  First, consider the case where both quasiparticles in a pair are inside $A$ at time $\tau$ and remain there at time $t$ i.e. $x_\tau(k),x_\tau(k-\pi)\in A$  while also $x_t(k),x_t(k-\pi)\in A$. In this case we find that
\begin{align}\label{eq:unshared_1}
 e^{-iH(t-\tau)}\tr_{\bar A}[e^{i\lambda \hat{n}_{x_0,k}}\rho_{x_0,k}(\tau)e^{i\lambda'\hat{n}_{x_0,k}}]e^{iH(t-\tau)}=e^{i(\lambda+\lambda')}\rho_{x_0,k}(t)~.
\end{align}
Thus the pair remain in a pure state inside the subsystem, but acquire a phase. If on the other hand one of the quasiparticles is no longer inside $A$ at time $t$ i.e. $x_\tau(k),x_\tau(k-\pi)\in A$ and  $x_t(k),\in A$ but $x_t(k-\pi),\in \bar{A}$ we have 
\begin{align}\nonumber
 e^{-iH(t-\tau)}\tr_{\bar A}[e^{i\lambda \hat{n}_{x_0,k}}\rho_{x_0,k}(\tau)e^{i\lambda'\hat{n}_{x_0,k}}]e^{iH(t-\tau)}&\\\label{eq:unshared_2}&\hspace{-2cm}
 = e^{i(\lambda+\lambda')}\big[n(k)\hat{n}_{x_t}(k)+ (1-n(k))(1-\hat{n}_{x_t}(k) )\big]
\end{align}
with similar expression for the complementary case when $x_t(k)\in \bar A$ but $,x_\tau(k-\pi)\in {A}$. Up to the overall phase in front this is the same expression as in the case without measurements~\eqref{eq:local_gibbs}. Lastly, if both quasiparticles exit the subsystem by the time $t$ then we simply have 
\begin{align}\label{eq:unshared_3}
 e^{-iH(t-\tau)}\tr_{\bar A}[e^{i\lambda \hat{n}_{x_0,k}}\rho_{x_0,k}(\tau)e^{i\lambda'\hat{n}_{x_0,k}}]e^{iH(t-\tau)}=e^{i(\lambda+\lambda')}~.
\end{align}
Thus when both particles are inside the subsystem at the time of the measurement the modification amounts to the inclusion of an overall phase factor $e^{i(\lambda+\lambda')}$. 

More substantial modifications occur, however, when only a single quasiparticle is inside $A$ at the time of the measurement. Consider first the case that this quasiparticle remains inside $A$ at time $t$, let us take for example $x_\tau(k), x_t(k)\in A$ while $x_\tau(k-\pi), x_t(k-\pi)\in \bar A$  then we find that
\begin{align}\nonumber
 e^{-iH(t-\tau)}\tr_{\bar A}[e^{i\lambda \hat{n}_{x_0,k}}\rho_{x_0,k}(\tau)e^{i\lambda'\hat{n}_{x_0,k}}]e^{iH(t-\tau)}&\\\label{eq:share_1}&\hspace{-2cm}
 =  e^{i(\lambda+\lambda')}n(k)\hat{n}_{x_t}(k)+ (1-n(k))(1-\hat{n}_{x_t}(k) )
\end{align}
In the complementary case where $x_\tau(k), x_t(k)\in \bar A$ while $x_\tau(k-\pi), x_t(k-\pi)\in A$ an analogous expression is found. On the other hand if the quasiparticle subsequently exits $A$ we have 
\begin{align}\label{eq:share_2}
 e^{-iH(t-\tau)}\tr_{\bar A}[e^{i\lambda \hat{n}_{x_0,k}}\rho_{x_0,k}(\tau)e^{i\lambda'\hat{n}_{x_0,k}}]e^{iH(t-\tau)}
 =  e^{i(\lambda+\lambda')}n(k)+ 1-n(k)~.
\end{align}
 One could also consider cases were neither member of the pair is inside $A$ at the time of the measurement. Their contribution will the same as in the unmeasured case. 

We can now combine all these different quasiparticle configurations to find the expression for $\mathcal{F}_\alpha(t,\{\lambda_j,\lambda_j'\})$. Upon taking the thermodynamic limit  we find 
\begin{eqnarray}\nonumber
\mathcal{F}_\alpha(t,\{\lambda_j,\lambda_j'\})=\int_{-\pi}^\pi\frac{{\rm d}k}{2\pi} \Big( \chi_{A\bar A}^{(1)}(t,k)f_{
\alpha}(\sum_{j=1}^\alpha[\lambda_j+\lambda_j'],k)+\chi^{(1)}_{\bar A\bar{A}}(t,k)\sum_{j=1}^\alpha  f_1(\lambda_j+\lambda_j',k)\\\nonumber
+\Big[
 \chi_{A\bar A}^{(0)}(t,k) +\chi_{A\bar A}^{(2)}(t,k)\Big] f_\alpha(0,k)
\hspace{3.59cm}\\ + \Big[\chi^{(2)}_{AA}(t,k)+\chi^{(2)}_{A\bar A}(t,k)+\chi^{(2)}_{\bar A\bar A}(t,k)\Big]\sum_{j=1}^\alpha i[\lambda_j+\lambda_j'] \Big)\quad \label{eq:Fsymm_1}
\end{eqnarray}
where we have introduced the counting functions $\chi_{BB'}^{(\rm x)}(t,k)$, here $B,B'=A~{\rm  or }~\bar A~ {\rm and}~{\rm x}=0,1,2$, which count the number of quasiparticles in a certain configuration at time $t$ along with
\beqa
f_\alpha(z,k)=\log\{[1-n(k)]^\alpha+n(k)^\alpha e^{i z}\}~,
\eeqa
which accounts for their contribution.  For the counting functions, the subscript denotes the configuration of the particles at the time $t$ i.e. $AA$ means both members of a pair are inside $A$ whereas $A\bar A$ means one is in $A$ and the other in $\bar A$ etc. The superscript, instead, denotes how many quasiparticles in the pair where inside $A$ at the time of the measurement, either $0,1,$ or $2$. For example,  $\chi_{A\overline{A}}^{(1)}(t,k)$ counts the number of quasiparticles which are inside $A$ at the time $\tau$ and $t$ and whose partner was in $\bar A$ at both $\tau$ and $t$. 
It is given by
\begin{eqnarray}
    \chi_{A\overline{A}}^{(1)}(t,k) =  \min(2|v_k|t,\ell)-\min(2|v_k|(t-\tau),\ell) ~.
    \label{eq:chiaabar}
\end{eqnarray}
The first term above is the standard expression giving the number of quasiparticles which belong to pairs which are shared between $A$ and $\bar A$. The second term is the number which became shared pairs between the times $\tau$ and $t$. We give explicit expressions for all these counting functions in Appendix~\ref{appA}. The first line in \eqref{eq:Fsymm_1} contains the contributions of pairs for which only one member was inside $A$ at the time of the measurement. The first term in the second line accounts for pairs for which both members were inside $\bar A$ at the time of the measurement and so are unaffected by it. Finally, the second term of the second line and the last line are the total contributions for pairs which were both inside $A$ at the time of the measurement.

Plugging~\eqref{eq:Fsymm_1} back into~\eqref{eq:charge_moments} one can perform the inverse Fourier transform and determine the R\'enyi entropy. In practice, it is rarely possible to perform these integrals exactly (see Sec. \ref{sec:Néel} for an exception) and so to obtain analytic predictions we can use a saddle point approximation to arrive at the final result. 

\subsection{Saddle point approximation}
In the ballistic, hydrodynamic regime which we are considering, all relevant quantities, such as $\tau, t,\ell$, and $q$ are taken to be very large allowing us to perform a saddle point approximation in the evaluation of the $\lambda_j,\lambda_j'$ integrals. In this subsection we present the results of such a calculation, leaving the details of the evaluation for appendix \ref{appA}.

The first point to note is that, the saddle point equations,
\beqa\label{eq:saddle_1}
&q=i\partial_{\lambda_j}\mathcal{F}_\alpha (t, \{\lambda_j,\lambda'_j\}),\\\label{eq:saddle_2}
&q=i\partial_{\lambda'_j}\mathcal{F}_\alpha (t, \{\lambda_j,\lambda'_j\}),
\eeqa
admit a solution only for specific values of measured charge $q$ for a given $
\tau$. In particular, a saddle point solution  exists only if
\begin{equation}
\label{eq:conditionontau}
    |\Delta q| \leq \frac{2 \tau}{\pi}, \hspace{0.5cm} \Delta q=q-\ell/2~.
\end{equation}
This condition encapsulates the time delay which was anticipated in  \ref{sec:remarks}, and was first obtained in the context of symmetry resolved entanglement in \cite{parez2021quasiparticle}. To reiterate, it states that because of the finite velocity of the propagation of charge, it is not possible to measure values of the charge which deviate too much from the initial value, $\ell/2$, if the time at which the measurement is performed is not large enough. Given the existence of the solution, we also note that the equations \eqref{eq:saddle_1}, \eqref{eq:saddle_2} suggest that all saddle points coincide i.e. $\lambda_j=\lambda_j':=\lambda_\alpha(q)/2$ and moreover they depend on the value of $q$ as well as the R\'enyi index. The fact that the saddle point solution depends on the replica index is quite unusual for free systems and complicates the analysis. It arises because the measured value of the charge needs to be the same on each replica. Nevertheless, in the replica limit, considerable simplification occurs. In particular, for any physically measurable value of the charge, i.e. one for which \eqref{eq:conditionontau} is satisfied, we find that  
\beqa
S_A(t|\tau,q) \approx S_A(t) + \int \dk \chi_{A\overline{A}}^{(1)}(t,k) \left(s[n_{\Delta q}(k)]-s[n(k)]\right) + \log \mathcal{N}(t)
\label{eq:entropydimergeneral2}.
\eeqa
   This expression is the first main result of this paper.  It states that the entanglement entropy after a single measurement can be expressed as the unmeasured value, $S_A(t)$, given in~\eqref{eq:qppentropy}, perturbed by corrections of two types, only one of which depends on the measurement outcome as we now explain.

The first correction depends on the difference of entropies evaluated using two occupation functions: the normal one, $n(k)$ and a modified one given by
\begin{equation}\label{eq:modified_occupation}
     n_{\Delta q}(k) = \frac{n(k)e^{\lambda(q)}}{n(k)e^{\lambda(q)} + 1-n(k)}, 
\end{equation}
where $-i\lambda(q)$ is the solution of the saddle point equations, in the replica limit. This modified occupation function is the one which can be associated to the symmetry resolved density matrix, $\rho_{A,q}$. Assuming that the measured value is close to $\ell/2$ this can be approximated as
\beqa\label{eq:saddle_point_approx}
\lambda(q) \approx \frac{\Delta q}{\sigma_\tau^2}~.
\eeqa
 This  correction depends on both the measurement outcome, through $n_{\Delta q}(k)$ and the time at which we calculate the entanglement, through $\chi^{(1)}_{A\bar A}(t)$. We can note, however that if the measured value is exactly the initial value then this term vanishes i.e. for $\Delta q=0$  we have $n_{0}(k)=n(k)$.  On the other hand if $\lambda(q)\to\pm \infty$ meaning that $|\Delta q|$ is large, then $n_{\Delta q}(k)\to 0$ and the change in the entanglement is most significant, however it should be noted that the saddle point approximation will break down for large enough $\lambda(q)$, corresponding to a significant deviation from the average of the measured charge. In addition, if $t<\ell/2$ then $\chi_{A \bar A}^{(1)}(t,k)=2|v_k|\tau$ and the $t$ dependence is absent. In light of $\eqref{eq:expectation}$, we can understand this term as being due to the change in configuration contribution to the entanglement entropy. 
 

 The second correction term is instead related to the number entropy of the subsystem at the time of the measurement, and arises from the contribution of the Hessian matrix in the saddle point evolution. In fact, as  shown in the appendix, for $t= \tau$
 \begin{equation}
     \log \mathcal{N}(\tau) \approx  -S_{A,{\rm  num}}(\tau) + O(1) 
     \label{eq:logNsymmetric}
 \end{equation}
 where explicitly,
\begin{eqnarray}
    S_{A,{\rm  num}}(\tau)&=&\frac{1}{2}\log [2\pi e \sigma_\tau^2],~\\\label{eq:Charge_variance}\sigma_\tau^2&=&\int \dk \,{\rm min}(2|v_k|\tau,\ell) n(k)[1-n(k)]~.
\end{eqnarray}
This drop in the entanglement entropy, was anticipated earlier~\eqref{eq:expectation} and arises from the destruction of charge fluctuations in the measured state. In the two regimes regime $\tau <\ell/2$ and $\tau \gg \ell$, it contributes factor $-\frac{1}{2}\log \tau$ and  $-\frac{1}{2}\log \ell$ respectively. 
Compared to the standard evolution of symmetry resolved entanglement, however, our protocol has additional structure, as we consider an extra time evolution, that is $t\neq \tau$. 
In general, it is not obvious that $\log \mathcal{N}(t)$ follows \eqref{eq:logNsymmetric} in general, however, as also shown in the appendix, in the regime if $\tau < \ell/2$ it is always true that
\begin{equation}
    \log \mathcal{N}(t) \approx -\frac{1}{2}\log \sigma_\tau^2 + O(1) \hspace{0.5cm} \forall t\in[\tau,\ell/2],
    \label{eq:logtermsymm}
\end{equation}
where $\sigma_\tau^2 \propto \tau$ in this regime.
On the other hand, for larger values of $t$, the time evolution leads to a decreasing of the value of $\log \mathcal{N}(t)$ as a consequence of the resurgence of fluctuations as induced by the time evolution.   Eventually, for $t \to \infty$,  the correction disappears reflecting that the effect of the measurement is washed away at long times if the total system is much larger than $A$. In particular, for $t>\ell/2>\tau$ it can be expressed as
\begin{equation}
    \log \mathcal{N}(t) \approx - \frac{1}{2}\log \frac{\sigma_\tau^2} {\sigma_\tau^2 + \sigma_{t-\tau}^2 - \sigma_t^2}+ ...
\end{equation}
which disappears at long times by observing that $\sigma_t^2\sim \sigma_{t-\tau}^2$ for $t\to \infty$. Note that, in all the above regimes, this correction is independent of the measurement outcome which is in contrast to the first correction. 

To summarize, we find that a single measurement of the charge inside the subsystem leads to two corrections which we attribute to the change in number and configurational entropies. At short times and to leading order the number entropy correction is the classical entropy of the subsystem charge distribution, we thus refer to this as the classical contribution. In contrast, the other term which strongly depends on the measurement outcome is is referred to as the quantum contribution.   In section \ref{sec:examples} and in Appendix \ref{appB}, we will show that in quenches from the Néel state it is possible to solve the $\lambda,\lambda'$ integrals exactly, without making use of the saddle point approximation. This exact solution reproduces correctly the saddle point expression \eqref{eq:entropydimergeneral2} in the limit of large $\tau, \ell$ and $t$, thus confirming our general result. 

 \subsection{Average Entanglement Entropy}
The change in entropy can differ drastically depending on the outcome of the measurement. While the classical contribution is independent of the outcome, the quantum part is most prominent if the outcome significantly deviates from $\ell/2$. One can expect such outcomes, however, to be much more unlikely or, in some cases, forbidden due to the time delay. To investigate this further, in this subsection we calculate the entanglement entropy averaged over all possible outcomes. Denoting this by $\braket{S_A(t|\tau,q)}$, it is given by the expression
\begin{align}\label{eq:average}
\braket{S_A(t|\tau,q)}=S_A(t)+\log \mathcal{N}(t)+\int {\rm d}q\,p(\tau,q)\int \dk \chi_{A\overline{A}}^{(1)}(t,k) \left(s[n_q(k)]-s[n(k)]\right) .
\end{align}
To evaluate this we need the probability distribution $p(\tau,q)$.  Referring again to  Appendix~\ref{appA}, we find that, for sufficiently large $\tau$ such that the time delay does not play a role, 
\begin{equation}
    p(\tau,q)\approx \frac{1}{\sqrt{2\pi \sigma_\tau^2}} \exp{-\frac{\Delta q^2}{2\sigma_\tau^2}}~.
    \end{equation}
    Namely, the measurement outcomes are approximately normally distributed about   $\ell/2$ with variance $\sigma_\tau^2$. 
To evaluate~\eqref{eq:average} we expand $s[n_q(k)]$ for small $\Delta q/\tau $ and then perform the Gaussian integral. As might be expected from the equipartition of entanglement, the leading order term is $\Delta q^2$. Then 
using $\braket{\Delta q^2} = \sigma_\tau^2$ we find 
\begin{multline}
    \braket{ S_A(t|\tau,q)}\!=\!S_A(t)+\log \mathcal{N}(t) - \frac{\sigma_t^2-\sigma_{t-\tau}^2}{2\sigma_\tau^2}\\
   +  \frac{1}{2\sigma_\tau^2} \int \dk \chi_{A\overline{A}}^{(1)}(t,k) n(k)[1-n(k)][1-2n(k)]\log\Big(\frac{1-n(k)}{n(k)}\Big) \label{eq:versionwithvariances},
\end{multline}
where $\sigma_x^2,$ are the charge variances given by~\eqref{eq:Charge_variance}, evaluated at times $x=t,
\tau,t-\tau$.  Naturally, the first two terms in \eqref{eq:versionwithvariances} are unaffected by the average, while to understand the second two, we consider as a starting point the regime $\tau,t<\ell/2$. Therein, the counting function simplifies to the value $2|v_k|\tau$ while, additionally,  the variance is linear in $\tau$, $\sigma_\tau^2=2\tau  D$. Here $D$ is known as the Drude self weight which is related to the two point function of the charge current~\cite{doyon2020lecture}.  Using this we find that the measurement time drops out of the last two terms 
 \begin{align}
\braket{S_A(t|\tau,q)}=S_A(t)+\log \mathcal{N}(t)-\frac{1}{2} + \frac{1}{2D}\int \dk |v_k| n(1-n) (1-2n)\log\Big(\frac{1-n}{n}\Big)  
 \label{eq:correctionisonlyo(1)}~,
 \end{align}
 which implies that the average correction due to the configuration entropy term is of $O(1)$ in time. We will evaluate such terms for specific states in section \ref{sec:examples}.
 
 We therefore come to the conclusion that the average change in entropy is dominated by contribution from the change in number entropy, i.e. the classical contribution. 
 This is ultimately a consequence of the fact that the standard deviation of the charge distribution is very small compared to the measurement time, as $ \sigma_\tau \propto \sqrt{\tau}$ in the short time regime. 
 In principle, projecting over values of $\Delta q$ comparable to $\tau$ would lead to much greater values of $s[n_{\Delta q}(k)]-s[n(k)]$, and possibly the dominant contribution coming from the configuration term. Hence this feature is just a consequence of the decreased likelihood of measuring charges which deviate significantly from the expected value in the type of quenches and initial states we consider.

\section{Multiple measurements on symmetric states}
\label{sec:multiple_symm}
Having seen the effect of a single measurement on the entanglement entropy in symmetric states, we now extend our analysis to the case of multiple measurements i.e. $m>1$, $t>m\tau$. Note that, despite the more complicated protocol the state remains symmetric at all times and in particular always admits a decomposition  similar to~\eqref{eq:rho_decomposition}. This will allow us to interpret the resulting expression in terms of contributions from the number and configurational entropies, as was the case for a single measurement.

\subsection{Quasiparticle approach}
In the presence of several projective measurements, the essence of the reasoning remains unaltered, but the technical structure is significantly more complicated.  For $m$ measurements performed at times $t=l\tau, l=1,\dots, m$ with outcomes $q_l,$ we have that
\begin{equation}\label{eq:charge_moments_2}
     \tr_A \Big[\{\rho_A(t|\tau,\{q_i\})\}^\alpha\Big] =\frac{ \int_{-\pi}^\pi \Big[\prod_{l=1}^m\prod_{j=1}^\alpha\frac{{\rm d}\lambda_{l,j} {\rm d}\lambda_{l,j}'}{4\pi^2}e^{i q_l(\lambda_{l,j}+\lambda_{l,j}')}\Big] e^{ \mathcal{F}_{\alpha}(t,\{\lambda_{l,j},\lambda_{l,j}'\})}}{\Big[\int_{-\pi}^\pi \prod_{l=1}^m\frac{{\rm d}\lambda_l {\rm d}\lambda_l'}{4\pi^2}e^{i q_l(\lambda_l+\lambda_l')} e^{\mathcal{F}_{1}(t,{\lambda}_l,{\lambda}_l')}\Big]^\alpha}, 
\end{equation}
where the function $\mathcal{F}_{\alpha}(t,\{\lambda_{l,j},\lambda_{l,j}'\})$ is a generalization of \eqref{eq:mathcalF_1} to the multiple measurement case and has a complicated, yet straightforward expression in terms of the quasiparticle picture generalizing the expression of the single measurement~\eqref{eq:Fsymm_1}. We again proceed via a saddle point approximation in order to evaluate the $\lambda_{l,j},\lambda_{l,j}$ integrals. We present the full expression for $\mathcal{F}_{\alpha}(t,\{\lambda_{l,j},\lambda_{l,j}'\})$ as well as details of the saddle point calculation in Appendix~\ref{appA2} and discuss here only the results.  

 The solution shares some fundamental similarities with the case of a single measurement. In particular, not only is there a time delay for the first measurement, but also for all subsequent measurements.  Each delay time depends on the outcome of the previous measurement. In particular we have that the saddle point exists only if, 
\begin{equation}
    |\Delta q_i| = |q_i - q_{i-1}|\leq \frac{2\tau}{\pi} ,\hspace{0.5cm} q_0=\ell/2~.
\end{equation}
Such a restriction is natural, since after each measurement the state is projected back to a single charge sector and one must allow a sufficient amount of time to between measurements for the charge to build up inside the subsystem in order to measure a certain value.  Furthermore, the saddle points are still dependent on the R\'enyi index and also on the measurement outcomes. However, as in the  $m=1$ case, they simplify considerably in the replica limit, $\alpha\to 1$, and admit a compact expression upon expanding about $|\Delta q_i|/\tau$ being small, 
\begin{equation}
\label{eq:saddlepointnmeas}
\sum_{l=k}^m \lambda_l 
\approx \frac{\Delta q_k}{\sigma_{k\tau}^2-\sigma_{(k-1)\tau}^2}\ ,
\end{equation}
where the dependence of the saddle points on the charges is left implicit to lighten the notation.
The denominator is the difference in charge variance evaluated at times $t=(k-1)\tau $ and $t=k\tau$.  In the regime $m\tau <\ell/2 $, this simplifies, and $\sigma^2_{k\tau}-\sigma^2_{(k-1)\tau}=2\tau D\hspace{0.2cm} \forall k$.

Proceeding as outlined in Appendix~\ref{appA2}, and assuming that $\tau$ is large enough to avoid issues of the time delay, i.e. $\{q_i\}$ corresponds to a physically allowable sequence of measurement outcomes, we arrive at the following expression for the entanglement entropy,
\begin{equation}
\label{eq:finalmultiplesymmetric}
    S_A(t|\tau,\{q_l\}) = S_A(t) +\sum_{l=1}^m \Big\{\log \mathcal{N}_l(t)+\int \dk \chi_{A\overline{A}}^{(1,l)} (t,k) (s[n_{\Delta q_l}(k)]-s[n(k)])\Big\}.
\end{equation}
This is the second main result of this paper. It states that,  as in the single measurement case the entanglement entropy can be expressed as the unmeasured value perturbed by series of corrections for each measurement performed and which take two different forms. We explain these fully below. 

The first correction is due to the drop in number entropy experienced at each measurement.
This can be easily understood by analogy with the single measurement case. Each time a measurement is made, the state is projected onto a single charge sector, thereby destroying charge fluctuations in the subsystem and reducing the entanglement by the number entropy term. After the measurement, the ballistic propagation of charge still takes place within the system, allowing charge fluctuations to build up within $A$ and specifically, a linear in time growth of the subsystem charge variance. When the next measurement is made, these newly acquired charge fluctuations are destroyed, once again leading to a reduction of the entanglement through the number entropy. These contributions all sum up to give a correction  $\sim-(m/2)\log(D \tau)$ in the regime $t <\ell/2$.

The second type of correction depends on both the outcomes of the measurements $\{q_l\}$ as well as the time at which the entropy is computed. It consists of entropy differences  $s[n_{\Delta q_l}(k)]-s[n(k)]$, i.e. entropy difference between the state and the modified state, defined by 
 \begin{equation}
     n_{\Delta q_l}(k) = \frac{n(k)e^{\sum_{s=l}^m \lambda_s}}{n(k)e^{\sum_{s=l}^m \lambda_s} + 1-n(k)}  = \frac{n(k)e^{ \frac{\Delta q_l}{\sigma_{l\tau}^2-\sigma_{(l-1)\tau}^2}}}{n(k)e^{ \frac{\Delta q_l}{\sigma_{l\tau}^2-\sigma_{(l-1)\tau}^2}} + 1-n(k)}~.
 \end{equation}
This takes the same form as for the single measurement case~\eqref{eq:modified_occupation}. The differences are weighted by the counting functions $\chi_{A\bar A}^{(1,l)}(t,k)$ which count the number of quasiparticles which were part of shared pairs between $A$ and $\bar A$ at the time of the $l^{\rm th}$ measurement and which remain shared at the time $t$. By comparing to the single measurement case, we are naturally lead to interpret this term as coming from the change in configuration entropy of the state after each measurement. We note that if $q_l=q_{l-1},~\forall l$ then this term vanishes and the correction is solely due to the number entropy term.  For large deviations between the measured values, however, the contribution significant and will dominate over the number entropy term.

\subsection{Average entanglement entropy}
 We can also calculate the  change in the entanglement upon averaging over all measurement outcomes. For this we need the probability 
 distribution which generalizes straightforwardly from the $m=1$ case.  In particular, within the saddle point approximation it simply reduces to the product of Gaussians, such that each is centred in the outcome of the previous measurement,
 \begin{equation}
    p(\tau,\{q_l\}) \propto \exp\left\{-\sum_{l=1}^m\frac{\Delta q_l^2}{2(\sigma_{l\tau}^2-\sigma_{(l-1)\tau}^2)}\right\},
\end{equation}
In the regime $m\tau < \ell/2$ reduces to a suggestive form  given by a product of Gaussians of same variance, dependent only on $\tau$ and on the Drude weight,
 \begin{equation}
    p(\tau,\{q_l\}) \propto \exp\left\{-\frac{\sum_{l=1} \Delta q_l^2}{4D\tau}\right\}.
\end{equation}
Such a product structure of the probability implies that taking averages follows directly from the $n=1$ case. In fact, this implies immediately
\begin{eqnarray}
    \int \Big(\prod_{s=1}^m{\rm d}q_{s}\Big)\, p( \tau,\{q_l\}
    ) \sum_{l=1}^m\int \dk \chi_{A\overline{A}}^{(1,l)} (t,k) (s[n_{\Delta q_l}(k)]-s[n(k)])  \\
    = -\frac{m}{2} + \frac{m}{2D}\int \dk |v_k| n(k)[1-n(k)] [1-2n(k)]\log\Big(\frac{1-n(k)}{n(k)}\Big)  \label{eq:averagemultiple}
\end{eqnarray}
which is the sum of single measurement terms. Thus, the average correction to the entropy is dominated by the number entropy term and the whole correction in the regime $m\tau<\ell/2$ is simply a sum of single measurement corrections.

\section{Measurements on symmetry breaking states}
\label{sec:squeezed}
Up till now we have investigated the dynamics from states which were eigenstates of the total charge. Now we turn our attention to consider the other class of states which are not charge symmetric, the squeezed states~\eqref{eq:initialsqueezed}. We proceed by first making some general remarks on entanglement in non-symmetric states and then consider the single measurement case. After that we look at the multiple measurement case.

\subsection{Remarks on entanglement in symmetry breaking states}
For symmetric states we were able to decompose the reduced density matrix into the different eigenspaces of $\hat{Q}_A$. In turn, this lead to a decomposition of the entanglement entropy into a number and configuration part. When the state is not symmetric, this decomposition of the reduced density matrix is no longer possible. Nevertheless, a useful decomposition of the entanglement entropy exists. Specifically, we can write the entanglement entropy for any state, symmetric or not, as 
\begin{eqnarray}\label{eq:symm_decomp_breaking}
    S_A(t)=S_{A,{\rm num}}(t)+S_{A,{\rm conf}}(t)-\Delta S_A(t)~.
\end{eqnarray}
The first two terms are the number and configuration contributions which are defined in the same way as previously, through~\eqref{eq:ent_num}-\eqref{eq:ent_conf_2} using \eqref{eq:ent_symm_1} and \eqref{eq:ent_symm_2}. We should note that although the first two terms are defined as in the symmetric case, their properties can be different. In particular, for symmetry breaking states there is, in general, no delay time. The initial state already contains charge fluctuations and so any value of the charge can be measured. For typical symmetry breaking states we have that $S_{A,\rm num}(0)\sim \frac{1}{2}\log (\ell)$ while $S_{A,\rm num}(t\to\infty)\sim \frac{1}{2}\log (\ell/2)$ \cite{horvath2024full}.  The last term in \eqref{eq:symm_decomp_breaking} is new.  It is called the entanglement asymmetry and is defined as the relative entropy between $\rho_A(t)$ and an auxiliary state, $\rho_{A,Q}(t)$, which is constructed from $\rho_A(t)$ by symmetrization.  Explicitly,
\begin{eqnarray}\label{eq:ent_asymm}
    \Delta S_A(t)=S(\rho_{A,Q}(t)||\rho_A(t)) ,~~~
    \rho_{A,Q}(t)=\sum_{\mathfrak{q}} p(t,\mathfrak{q})\rho_{A,\mathfrak{q}}(t)~,
    \end{eqnarray}
where $\rho_{A,\mathfrak{q}}(t)$  and $ p(t,\mathfrak{q})$ are given in \eqref{eq:ent_symm_1} and \eqref{eq:ent_symm_2} respectively.  This quantity was introduced in \cite{ares2023entanglement} (see also \cite{Gour:2009abc,Casini:2019kex,Casini:2020rgj,Marvian:2014awa}) as a probe of symmetry breaking in extended subsystems of many body quantum systems. In the present context, however, it measures the reduction in  entanglement of the state due to its symmetry breaking. If the state is symmetric, $\rho_{A,Q}(t)=\rho_A(t)$ meaning that $\Delta S_A(t)=0$ and the previous expression is recovered~\eqref{eq:symme_decomp_ent}. 

The entanglement asymmetry has since been investigated in a wide range of scenarios, both in and out of equilibrium~\cite{ares2023entanglement,Capizzi:2023xaf,Ferro:2023sbn,Chen:2023gql,Fossati:2024xtn,Benini:2024xjv,Fossati:2024ekt,Kusuki:2024gss,Ares:2023ggj,Russotto:2024pqg,ares2023lack,bertini_asymmetric_2024,rylands2024microscopic,shion2,murciano2024entanglement,Summer:2025wsa} with many of its properties now well understood. Notably, being a relative entropy it is non-negative, $\Delta S_A(t)\geq 0$ and typically for symmetry breaking states $\Delta S_A(0)\sim \frac{1}{2}\log \ell$ while $\Delta S_A(t\to\infty)\to 0$.  Moreover, for an initial product state we have that, 
\begin{eqnarray}
    S_{A,\rm num}(0)=\Delta S_A(0)
\end{eqnarray}
although their finite time dynamics differs, as could be guessed from their aforementioned long time behaviour

From the decomposition~\eqref{eq:symm_decomp_breaking} and the general properties of the associated quantities we can make some predictions for the behaviour of the entropy in the case of a single measurement. First, we can expect that due to the absence of a time delay, all possible eigenvalues of $\hat{Q}_A$ can be measured at any time, independently of $\tau$. Second, after the measurement is made, the subsystem no longer contains charge fluctuations and so we can expect a drop in the entropy coming from the absence of the number entropy. However, this can be compensated for by the presence of $\Delta S_A(t)$ especially for times which are small compared to $\ell$. At long times however, the asymmetry vanishes while the number entropy remains finite. Thus, for $\tau$ large we can expect a drop in the entropy due to destruction of the number fluctuations. Lastly, after the first measurement, the subsystem has no charge fluctuations and so we can expect that a delay time is present for any subsequent measurements.

\subsection{Single measurement}
We begin our analysis by considering the case of a single measurement. The philosophy of the calculation is the same as in the case of the symmetric states, but with some modifications coming from the structure of the state~\eqref{eq:squeezed}. The most notable of these arises when both of the quasiparticles in a pair are inside $A$ at the time of the measurement. As an example, consider the case where both quasiparticles in a pair are inside $A$ at time $\tau$ but only one remains there at time $t$ i.e. $x_\tau(k),x_\tau(k-\pi)\in A$  while also $x_t(k)\in A,~x_t(k-\pi)\in\bar{A}$. In this case we find that
\begin{eqnarray}\nonumber
 e^{-iH(t-\tau)}\tr_{\bar A}[e^{i\lambda \hat{n}_{x_0,k}}\rho_{x_0,k}(\tau)e^{i\lambda'\hat{n}_{x_0,k}}]e^{iH(t-\tau)}\\\label{eq:unshared_1_breaking}
 &\hspace{-3cm}
 = e^{2i(\lambda+\lambda')}n(k)\hat{n}_{x_t}(k)+ (1-n(k))(1-\hat{n}_{x_t}(k) )~.
\end{eqnarray}
This can be contrasted with~\eqref{eq:unshared_2}, the analogous expression in the symmetric case which just contributes a phase. After reconsidering all such possible configurations we obtain an expression akin to~\eqref{eq:charge_moments}, where now 
\beqa\nonumber
\mathcal{F}_\alpha(t,\{\lambda_j,\lambda_j'\})=\int_{-\pi}^\pi\frac{{\rm d}k}{2\pi} \Big( \chi_{A\bar A}^{(1)}(t,k)f_{
\alpha}(\sum_j^\alpha[\lambda_j+\lambda_j'],k)+\chi^{(1)}_{\bar A\bar{A}}(t,k)\sum_{j=1}^\alpha  f_1(\lambda_j+\lambda_j',k)\\\nonumber
+
 \chi_{A\bar A}^{(0)}(t,k)  f_\alpha(0,k)+\chi_{A\bar A}^{(2)}(t,k)f_{
\alpha}(2\sum_j^\alpha[\lambda_j+\lambda_j'],k)
\\ + \chi^{(2)}_{AA}(t,k)\sum_{j=1}^\alpha f_1(2[\lambda_j+\lambda_{j+1}'],k)+\chi^{(2)}_{\bar A\bar A}(t,k)\sum_{j=1}^\alpha f_1(2[\lambda_j+\lambda_{j}'],k)\Big).\quad \label{eq:squeezedtaugreatzero}
\eeqa
In this expression, the first line and the first term of the second line are the same as for the symmetric state~\eqref{eq:Fsymm_1}. This means that, when either zero or one members of a quasiparticle pair are inside $A$ at the time of the measurement, there is no difference between symmetric and non-symmetric cases.  In contrast, when both members of the pair are present in $A$ at the time of the measurement there is a significant difference, which is seen in the remainder of the terms. For example, the configuration considered in~\eqref{eq:unshared_1_breaking} gives rise to the second term in the second line of~\eqref{eq:squeezedtaugreatzero}.

Inserting this back into~\eqref{eq:charge_moments} we can then proceed, as before, via a saddle point approximation. In this case, the saddle point equations take the same form as before, ~\eqref{eq:saddle_1}, \eqref{eq:saddle_2}, however their solutions are different owing to the structure of $\mathcal{F}_\alpha(t,\{\lambda_j,\lambda_j'\})$. Similar to the symmetric case, one can verify that as before all the saddle points coincide $\lambda_j=\lambda'_k,~\forall j,k$ and moreover that they depend on the replica index. In the non-symmetric case however, one can always find a solution to the equations meaning that there is no time delay. In the replica limit, the expression simplifies considerably. For small $\Delta q=q-\bar q$, with $\bar q$ being the average charge in the subsystem, we find
\begin{equation}
    \lambda(q) \approx\frac{\Delta q}{\sigma_\tau^2},
\end{equation}
Schematically, this has the same form as~\eqref{eq:saddle_point_approx}, although importantly, the expression for the variance is now given by,
\begin{equation}
    \sigma_\tau^2 = \int \dk\left[2\ell-\min(2|v_k|\tau,\ell) \right] n(k) [1-n(k)].
\end{equation}
From this we see that the variance is finite and extensive in the initial state saturating to half its value in the limit $t\to\infty$. Thus the saddle point solution does not diverge as $t\to\infty$ for finite values of $\Delta q$. 

Providing details in the Appendix~\ref{sec:appA3} we state the  result for the Von Neumann entropy for non-symmetric states, 
\beqa\nonumber
    S_A(t|\tau,q) 
    &=& S_A(t)+\int \dk \chi^{(1)}_{A\bar A}(t,k) (s[n_{\Delta q}(k)]-s[n(k)])\\
    && +\int \dk \chi^{(2)}_{A\bar A}(t,k) (s[n_{2\Delta q}(k)]-s[n(k)]) + \log \mathcal{N}(t).
\label{eq:single_squeezed}
\eeqa
Here we see that there are three terms contributing to the correction. The first of these is the same as before coming from a single member of a pair being present at the time of the measurement. The second term is new to the non-symmetric case and arises from the fact that the measurement acts non-trivially when both members of a pair are inside the interval. The last term,  $\log \mathcal{N}(t)$, represents the logarithmic corrections arising from the Hessian matrix in the saddle point evaluation. As proven in appendix \ref{sec:appA3}, at time $t=\tau$ it can be expressed in terms of the entanglement asymmetry and the number entropy as 
\begin{equation}
    \log \mathcal{N}(\tau)= \Delta S_A(\tau) -S_{A,{\rm num} }(\tau) 
\end{equation}
as was expected from general considerations,
where the entanglement asymmetry can be approximated at short times as
\beqa 
\label{eq:asymmetry}
\Delta S_A(\tau)&=&\frac{1}{2}\log[2\pi e \mathcal{X}_
\tau],\\
\mathcal{X}_\tau &=&2\int \dk\left[\ell-\min(2|v_k|\tau,\ell) \right] n(k) [1-n(k)].
\eeqa
For $\tau=0$, it coincides with $S_{A,\rm num}(0)$, thereby cancelling the number contribution exactly, and also suppressing it for $\tau\ll\ell$. For $t>\tau$, $\log \mathcal{N}(t)$ will acquire a time dependence which can be  controlled in the various relevant limits, which are investigated in \ref{sec:appA3}. As mentioned above, the presence of this non-trivial time dependence is expected, since in our protocol we are adding additional time evolution compared to the standard situations in which entanglement asymmetry and symmetry resolution are considered; and this time evolution is at the origin of the revival of fluctuations which are reduced by the measurement. 

\subsection{Multiple measurements}
Although symmetry breaking states are structurally different from symmetric ones, when performing a measurement they are projected over a symmetry sector. Therefore, if we consider a protocol in which measurements are repeated periodically with period $\tau$, the evolution of the system after the first measurement should share some features with that analysed in section \ref{sec:multiple_symm}. 
This expectation is confirmed by the saddle point calculation as detailed in Appendix~\ref{sec:appA3}.  From there we see that, while the first measurement carries no restriction, from the second measurement on a time delay appears,
\begin{equation}
    |\Delta q_i| = |q_{i}-q_{i-1}| \leq \frac{2\tau}{\pi}  \hspace{0.3cm}\forall i>1.
\end{equation}
This is the fundamental signature of the state being projected to a sector of definite charge over $A$. However, this does not imply that the result for the entropies become equivalent to that of the symmetric case, since squeezed states are collections of 
pairs of particles, while symmetric states are collections of particle-hole excitations. 
As a consequence, the solution will contain several terms depending on whether pairs or single particles were contained in $A$ at the time of each measurements. 
In the case of two measurements, the solution is therefore 
\begin{multline}
   S_A(t|\tau,\{q_i\})  =S_A(t) +  \int \dk \chi_{A\overline{A}}^{(2,2)}(s[n_{2\lambda_1 + 2\lambda_2}(k)]-s[n(k)])+ \int \dk \chi_{A\overline{A}}^{(2,1)}(s[n_{2\lambda_1 + \lambda_2}(k)]-s[n(k)])\\
    + \int \dk \chi_{A\overline{A}}^{(1,1)}(s[n_{\lambda_1+\lambda_2}(k)]-s[n(k)])+\int \dk \chi_{A\overline{A}}^{(0,1)}( s[n_{\lambda_2}(k)]-s[n(k)])   + \log \mathcal{N},
    \label{eq:2meassqueezed}
\end{multline}
where the values of the saddle points satisfy more complicated expressions compared to the symmetric case,
\begin{align}
         q_1 - \overline{q} = \lambda_1 \sigma_\tau^2 + \lambda_2 \sigma_{2\tau}^2, \\
         q_2-q_1 = \lambda_1(2\sigma_\tau^2-\sigma_\infty^2).
\end{align}
This increase in the difficulty is again a consequence of the presence of full pairs which are affected  non-trivially by  the measurements.

In general, the final solution for arbitrary $t$ and arbitrary number of measurements will contain all possibility of combinations, depending on how many particles of each pair were inside the interval at the time of each measurement. Compactly, this can be expressed as 
\begin{equation}
    S_A(t|\tau,\{q_i\}) = S_A(t) + \sum_{\vec l} \int \dk \chi_{A\overline{A}}^{({\vec l})} \left(s[n_{\vec{l} \cdot\vec{\lambda} }(k)] -s[n(k)]\right) + \log \mathcal{N}(t)
\end{equation}
where $\vec l$ are the strings which can be constructed of 0,1,2
 which represent the number of particles which were inside the interval at the time of each measurement, on the same line of \eqref{eq:2meassqueezed}.

\section{Examples}
\label{sec:examples}
Having laid out the general theory of our time evolution protocol, in this section we present some explicit examples. For this we choose two prototypical symmetric initial states, the N\'eel and dimer states, and a non-symmetric state, the tilted ferromagnet. 

\subsection{N\'eel State}
\label{sec:Néel}
Perhaps the simplest initial state that can be considered is the N\'eel state. In real space this takes the form,
\begin{eqnarray}
    \ket{\psi}&=&\prod_{x=1}^{L/2} c^\dag_{2x}\ket{0}\\
    &=&\prod_{k>0} \frac{c^\dagger_{k-\pi}+ c^\dagger_k }{\sqrt{2}}\ket{0} \label{eq:neelstate}~.
\end{eqnarray}
In the second line we have switched to Fourier space and from which we see that the mode occupation function is particularly simple $n(k)=1/2$, meaning that 
\beqa 
f_\alpha (z,k)=\log[1+e^{iz}]-\alpha\log(2)~.
\eeqa
This lack of $k$ dependence allows one to carry out both the $k$ integral in $\mathcal{F}_\alpha(t,\{\lambda_j,\lambda_j'\})$ and the $\lambda$ integrals without the  need for the saddle point approximation. Details of the exact calculation are given in Appendix~\ref{appB}. We shall use this to compare to the saddle point approximation but now proceed with the saddle point calculation as outlined in previous sections. 

Considering a single measurement as a starting point, we have from \eqref{eq:saddle_1} that the saddle point equations are
\begin{equation}
    \Delta q = \int \dk  \Big[\chi^{(2)}_{AA}(t,k)+\chi^{(2)}_{A\bar A}(t,k)+\chi^{(2)}_{\bar A\bar A}(t,k)\Big]+\chi_{A\overline{A}}^{(1)}(t,k) \frac{e^{\alpha\lambda_\alpha(q)}}{e^{\alpha\lambda_\alpha(q)}+1}~.
\end{equation}
where we have used the fact that all the saddle points should be the same and denoting this by $\lambda_\alpha(q)$ to signify that it depends on both the replica index and the measurement outcome. These can be easily solved for $t <\ell/2$,
\beqa
    \Delta q &=&  \frac{e^{\alpha\lambda_\alpha(q)}}{e^{\alpha\lambda(q)}+1}\int \dk 2|v_k|\tau - \frac{2\tau}{\pi} \\
    &=& \frac{2\tau}{\pi} \operatorname{tanh}(\alpha\lambda_\alpha(q)/2)
\eeqa
from which we can obtain the exact saddle point as a function of the charge variation,
\begin{equation}
\label{eq:alphaNéel}
    \lambda_\alpha(q) = \frac{2}{\alpha}\operatorname{arctanh}\left(\frac{\pi \Delta q}{2\tau}\right) \Rightarrow  \lambda_\alpha(q)  = \frac{1}{\alpha}\lambda(q).
\end{equation}
where $\lambda(q)$ is the saddle point in the replica limit. Since both $s[n(k)]$ and $s[n_{\Delta q}(k)]$ are $k$-independent, 
\begin{equation}
\label{eq:neelcorrection}
    S_A(t|\tau,q) =S_A(t) + (s[n_{\Delta q}]-\log2) \int \dk \chi_{A\overline{A}}^{(1)} (t,k) +\log \mathcal{N}(t)
\end{equation}
where the $\log \mathcal{N}$ term will be evaluated shortly.
\begin{figure}[h]
    \centering
\includegraphics[width=.8\linewidth]{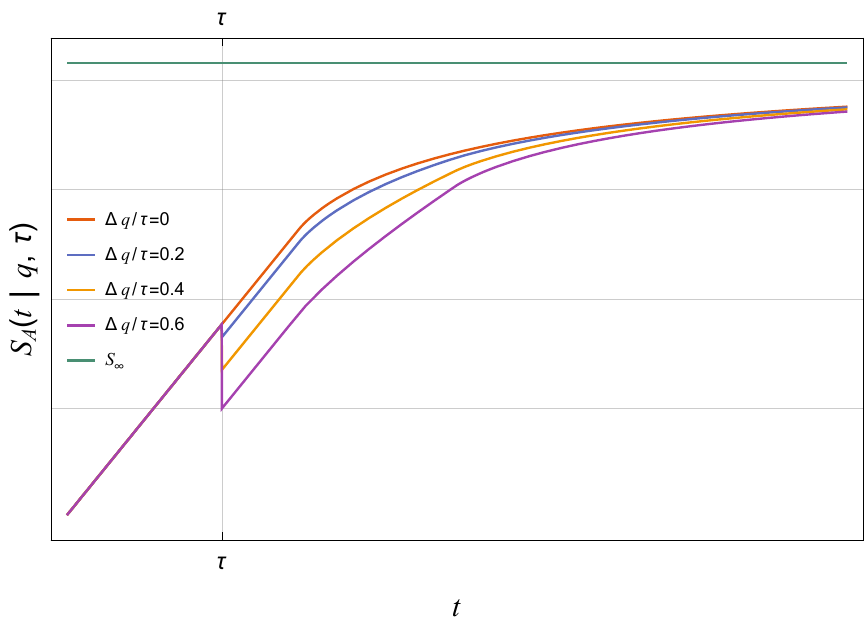}
    \caption{$S_A(t|\tau,q)$ in the Néel state for several values of $\Delta q/\tau$, at fixed $\tau=\ell/3$, and varying time. Immediately after the measurement, the entropy is decreased and the difference with respect to $S_A(t)$ remains constant up to $t=\ell/2$. Successively, all curves tend to the same stationary value. Since in \eqref{eq:neelcorrection} the entropic contributions are momentum-independent, the approach to the stationary value is dictated for any $q$ by the integral of the counting function, implying that the ratios of the distances between the curves remains constant in time.  }
    \label{fig:corr_neel}
\end{figure}

The entropic contribution $s[n_{\Delta q}] - \log 2$ can be evaluated exactly from \eqref{eq:alphaNéel} after some algebra,
\begin{equation}
    s[n_{\Delta q}] - \log 2 =\frac{\pi \Delta q}{4\tau} \log\left(1-\frac{2\pi\Delta q}{\pi \Delta q + 2\tau}\right) -\frac{1}{2 }\log\left(1-\left(\frac{\pi \Delta q}{2\tau}\right)^2\right)
    \label{eq:sqminuslog2}
\end{equation}
where we have made use of \eqref{eq:alphaNéel}. In the limit of small $\alpha$, this reduces to 
\begin{equation}
    s[n_{\Delta q}] - \log 2 \approx - \frac{\pi^2 \Delta q^2} {8\tau^2} + O(\Delta q/\tau)^4
\end{equation}
from which we obtain as expected equipartition at first order,
\begin{equation}
    S_A(t|\tau,q)  \approx  S_A(t) - \frac{\pi \Delta q^2}{2\tau} +\log \mathcal{N}(t).
\end{equation}
The logarithmic correction is evaluated in this time regime by simply substituting the specifics of the state in the general result \eqref{eq:logtermsymm}, giving 
\begin{equation}
    \log \mathcal{N}(t) = - \frac{1}{2}\log 2\tau
\end{equation}
result which is also confirmed by the exact evaluation.

\begin{figure}
    \centering
    \includegraphics[width=\linewidth]{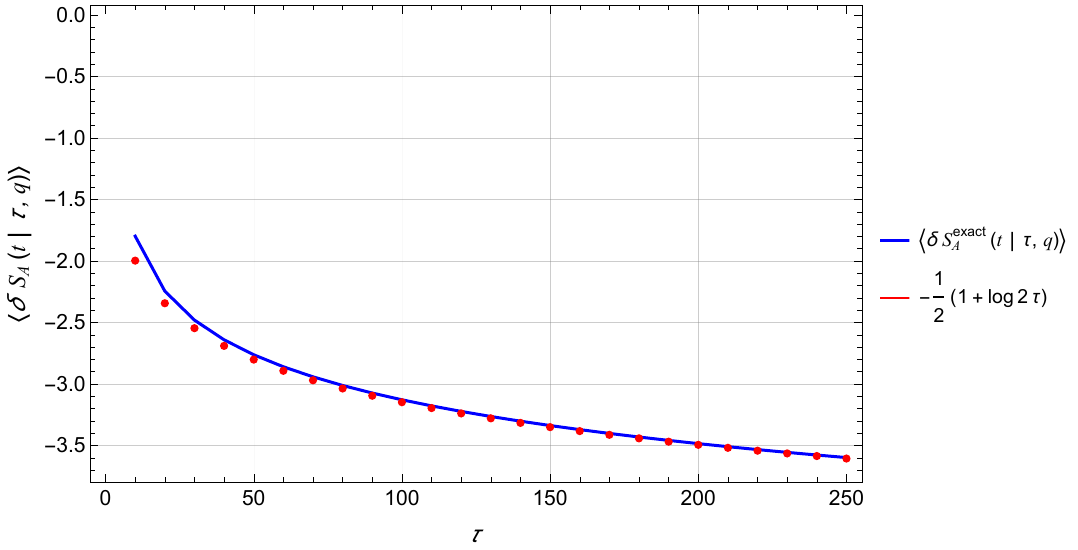}
    \caption{Comparison between the exact solution for the average correction to the entanglement entropy, $\langle\delta S_A(t|\tau,q)\rangle =\langle S_A(t|\tau,q)\rangle-S_A(t)$ of the Néel state, obtained from the exact solution discussed in appendix \ref{appB}, and the average of the quasiparticle/saddle point prediction \eqref{eq:correctionneel}. The saddle point result is accurate, exhibiting a logarithmic correction in the time of the measurement.  }
    \label{fig:neel_exact}
\end{figure}

In this situation, it is also possible to obtain the probability distribution, owing to the specific form of the saddle point \eqref{eq:alphaNéel}, which leads to 
\begin{equation}
    p(\tau,q) \propto \exp\left\{\left(\Delta q - \frac{2\tau}{\pi} \right) \log \left(1-\frac{\pi \Delta q}{ 2\tau}\right)-\left(\Delta q + \frac{2\tau}{\pi} \right) \log \left(1+\frac{\pi \Delta q}{ 2\tau}\right)\right\}.
\end{equation}
Interestingly, this expression for the probability clearly exhibits the reason for the time delay which was obtained above on general grounds: the values for which $\tau<\frac{\pi|\Delta q|}{2}$ lead to ill-defined logarithms. Hence the time delay corresponds to the impossibility to measure certain values of charge, as claimed heuristically above.
At the Gaussian level simply reduces to 
\begin{equation}
    p(\tau,q) \approx \frac{1}{\sqrt{2\tau}} \exp\left\{-\frac{\pi \Delta q^2}{2\tau}\right\}
\end{equation}
which shows that $\sigma_\tau^2 = \tau/\pi$ in this regime. Performing the average of the entropy, using that $\braket{\Delta q^2}=\tau/\pi$, and exploiting the fact that in this case \eqref{eq:versionwithvariances} simplifies, since $(1-2n)=0$, the general correction can be expressed as
\beqa
    \braket{  S_A(t|\tau,q)} &=& S_A(t)-
     \frac{\sigma_t^2-\sigma_{t-\tau}^2}{2\sigma_\tau^2}- \frac{1}{2}\log 2\tau \\
     &=&S_A(t)-1/2 - \frac{1}{2}\log(2\tau)+ O(1/\tau) \label{eq:correctionneel}
\eeqa
which shows clearly that the classical contribution is the dominant correction. Figure \ref{fig:neel_exact} shows that this value reproduces very accurately the exact result even for relatively small values of $\tau$. Since all time dependence in this situation is in $S_A(t)$, the deviation $\delta S_A(t|\tau,q) = S_A(t|\tau,q) -S_A(t) $ has in fact no time dependence.

In the case of several measurements, we would have the same structure (see also appendix \ref{appB} for the exact solution in the case of several measurements),
\begin{equation}
    \braket{ S_A(t|\tau,\{q_i\})}\approx S_A(t) -m\left(\frac{1}{2}+\frac{1}{2}\log2\tau\right).
\end{equation}

Note that in this particularly simple initial case, it is also possible to evaluate the R\'enyi entropies with relative ease,  since the various saddle points are all related in a rather simple way, $\lambda_1 =\alpha\lambda_{\alpha}$. In particular, 
\beqa
\label{eq:RenyiNeel}
    S^{(\alpha)}_A(t|\tau,\{q_i\}) =  S_A(t|\tau,\{q_i\})
\eeqa
which shows that the independence of the R\'enyi entropies from the R\'enyi index, which is a feature of the unitary quench evolution of the Néel state, is also preserved by the presence of the measurements. 

In appendix \ref{appB} we  prove that, owing to the momentum independence of the occupation functions of this state, it is possible to obtain an exact result for the R\'enyi entropies in $t< \ell/2$ regime, which takes the form 
\begin{equation}
     S^{(\alpha)}_A(t|q) =S_{A} (t) - \log \left( 2^{\frac{4\tau}{\pi}} \frac{\Gamma(\frac{2\tau}{\pi}+\Delta q +1)\Gamma(\frac{2\tau}{\pi}-\Delta q +1)}{\Gamma(\frac{4\tau}{\pi}+1)}\right),
\end{equation}
from which it is clear that even in the presence of the measurement there is no dependence on the R\'enyi index. This result can be expanded using Stirling formula for large $\tau$; postponing the details to the appendix, this gives
\beqa
 S^{(\alpha)}_A(t|q) &\approx&S_{A}^{(\alpha)} (t) - (\frac{2\tau}{\pi}+\Delta q) \log\left(1+\frac{\pi\Delta q}{2\tau}\right) -(\frac{2\tau}{\pi}-\Delta q) \log\left(1-\frac{\pi\Delta q}{2\tau}\right) \nonumber\\
 &-& \log \left[\sqrt{\frac{\pi^2}{2\tau}\left(\frac{4\tau^2}{\pi^2} -\Delta q^2\right)}\right]. \label{eq:logcorr}
\eeqa 
which is exactly identical to \eqref{eq:neelcorrection}, and reproduces the result of \cite{parez2021quasiparticle}. 
Moreover, it confirms that the claimed value of the classical correction, since
\begin{equation}
    \log\mathcal{N}(t)=-\log \left[\sqrt{\frac{\pi^2}{2\tau}\left(\frac{4\tau^2}{\pi^2} -\Delta q^2\right)}\right] \approx-\frac{1}{2} \log 2\tau + O((\Delta q/\tau)^2).
    \end{equation}
Therefore the saddle point prediction is confirmed to be the large $\tau$ expansion of the exact solution. A similar solution can also be obtained in the case of several measurements, confirming the solution also in that situation.

\subsection{Dimer State}
\begin{figure}[h]
    \centering
    \includegraphics[width=.8\linewidth]{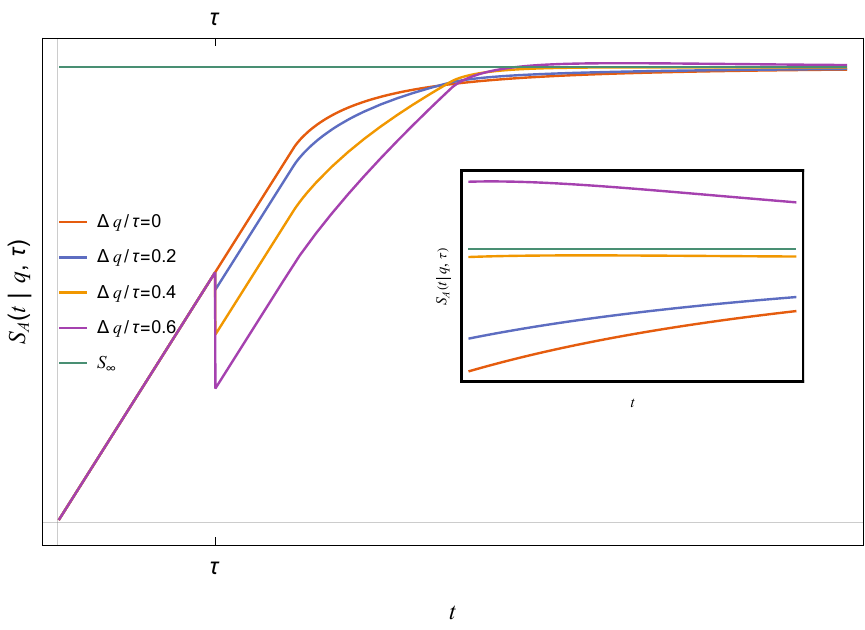}
    \caption{Behaviour of $S_A(t|\tau,q)$ for several values of $\Delta q/\tau$, at fixed $\tau=\ell/4$, and varying time. We see that immediately after the measurement the entropy is decreased and the difference with respect to $S_A(t)$ remains constant up to $t=\ell/2$. At this point the structure of the counting function becomes more complex, and this leads to a faster increase in the entropy of the measured state. In particular, this surpasses both the unitary value and the asymptotic $t\to \infty$ value, which is then approached from above, as shown in the inset. Note that the deviations from the unitary value appear quite significant for $t<\ell/2$; it is however necessary to consider that the variance of the probability distribution is only $\sigma^2 =\frac{2\tau}{3\pi}$, therefore all values shown in the plot are highly unlikely. In particular, the logarithmic correction is truly negligible for such values.     }
    \label{fig:dimer_correction}
\end{figure}

The second explicit state we consider is the dimer state, which belongs to the class of symmetric states of the form \eqref{eq:dimer}, with occupation functions $n(k) = \frac{1-\cos k}{2}$. It takes therefore the form 
\begin{equation}\label{eq:dimer}
    \ket{\psi}=\prod_{x=1}^{L/2}\left(\frac{c^\dag_{2x-1}-c^\dag_{2x}}{\sqrt{2}}\right)\ket{0}=\prod_{k>0} \left(\sqrt{\frac{1+\cos k}{2}}c^\dagger_{k-\pi}+\sqrt{\frac{1-\cos k}{2}} c^\dagger_k \right)\ket{0} 
\end{equation}
In this context, the main advantage of this state which allows to obtain analytical solutions is the feature ${\rm d}n = -\frac{v_k}{2} {\rm d}k$, which in principle can be used to simplify integrals in the regime $\tau<\ell/2$ in which the effective velocity appears explicitly.
In fact, focusing on the regime $t<\ell/2$, the saddle point equation in the replica limit is 
\beqa
   \Delta q = \int \dk  2|v_k|\tau\left(\frac{n e^{\lambda(q)}}{ne^{\lambda(q)} +(1-n)}-\frac{1}{2}\right),
\eeqa
which can be solved quite easily by change of variables $k\to n$,
\begin{equation}
    \Delta q = \frac{2\tau}{\pi} \frac{\sinh \lambda(q) - \lambda(q)}{\cosh \lambda(q) -1} \approx \frac{2\tau}{3\pi} \lambda(q) + O(\lambda(q)^3) \Rightarrow \lambda(q) \approx \frac{3\pi}{2\tau} \Delta q,
\end{equation}
which implies $\sigma_\tau^2 =\frac{2\tau}{3\pi}$. The entropy correction can be expressed in a simple form for $\tau<\ell/2$ by changing variables,
\begin{equation}
\label{eq:entropymultiplemeasurement}
    S_A(t|\tau,q) = S_A(t) + \frac{4\tau}{\pi} \int_0^1 {\rm d}n (s[n_{\Delta q}] - s[n])+\log \mathcal{N}(t)
\end{equation}
It is not difficult to convince oneself that the correction is always negative. Therefore the effect of the measurements at short times will be that of a sudden decrease of the entropy, as would be naturally expected. 
At longer times, however, this is not obvious. In fact, there is a subtle interplay between the sign of $s[n_{\Delta q}(k)]- s[n(k)]$, which is not always positive, and the structure of the counting function, which tends to truncate fast modes (which correspond to measured particles which exit the interval the quickest, and which therefore cease to contribute to the entropy). The interplay of these  two is shown in figure \ref{fig:dimer_correction}: while the entropy is initially decreased by the measurement, it then outgrows the unitary value, and tends to the stationary value from above.

Using the value of the variance in the probability distribution it is then possible to extract the average behaviour of the entropy correction, which is expressed at leading order,
\begin{equation}
    \braket{ S_A(t|\tau,q)} =  S_A(t)+\tau \int \dk |v_k|\frac{(n-1)n(1+2(2n-1)\operatorname{arctanh}(1-2n))}{\sigma_\tau^2} + \log \mathcal{N}(t).
\end{equation}
which shows that the $\tau$ dependence disappears completely; solving the integral leads to 
\begin{equation}
   \braket{S_A(t|\tau,q)} = S_A(t) - \frac{1}{4} + \log \mathcal{N}(t)
\end{equation}
which is immediately generalized to the case of several measurements,
\begin{equation}
   \braket{S_A(t|\tau,\{q_i\})} = S_A(t) - \frac{m}{4} +m \log \mathcal{N}(t).
\end{equation}
As above $\log \mathcal{N}$ has an easy expression for small times, and is dominated by a factor $-\frac{1}{2}\log \tau$. Therefore also in this case the logarithmic correction is the dominant one on average.

\subsection{ Tilted ferromagnet}
An interesting state belonging to the second class of states is the tilted ferromagnet, which has been studied in several instances as it provides the simplest realization of the quantum Mpemba effect\cite{ares2023entanglement,ares2023lack}. In fermionic language, it can be expressed as 
\begin{equation}
    \ket{TF_\theta} = \prod_{x=1}^L (\sin\frac{\theta}{2} + \cos \frac{\theta}{2} c_x^\dagger)\ket{0} .
\end{equation}
From this, one can construct the ``cat" version
\begin{equation}
    \ket{TF} = \frac{\ket{TF_\theta}-\ket{TF_{-\theta}}}{\sqrt{2}}
\end{equation}
which can be shown to be the ground state of a free Hamiltonian, and therefore is Gaussian (see for instance \cite{tiltedferro1,tiltedferro2} for the discussion on the spin formulation of the state, which is shown to be the ground state of a specific XY Hamiltonian). In particular, it is a squeezed state, as it breaks particle number symmetry. The occupation functions are given by \cite{ares2023lack}
\begin{equation}
    n(k) = \frac{1-\cos \Theta_k}{2},\hspace{0.5cm} \cos \Theta_k = \frac{2\cos\theta -(1+(\cos\theta)^2)\cos k}{1-2\cos\theta \cos k+(\cos\theta)^2}.
\end{equation}
For brevity, we focus on the maximally tilted state, for which $\theta=\pi/2$. For this value of the parameter, the occupation functions simplify significantly, as $\cos \Theta_k = \cos k$ and therefore $n(k) = \frac{1-\cos k}{2}$. Note that the occupation functions are the same that appear in the dimer state, but the structure of the state is different, as it is of the form \eqref{eq:squeezed}. 

Specifying the general expression obtained in the previous sections to such occupation functions, we see that the saddle point equation for $\alpha=1$ reduces to 
\begin{equation}
\label{eq:saddlepointtilted}
    \lambda(q) = \frac{\Delta q}{ \sigma_\tau^2}= \begin{cases}
         \frac{q-\ell/2}{\ell/4-\frac{2\tau}{3\pi}} \hspace{0.3cm} \tau <\ell/2,\\
           \frac{q-\ell/2}{\ell/8} \hspace{0.5cm} \tau \gg\ell.
    \end{cases}
\end{equation}
Note that for $\tau <\ell/2$ the integrals appearing in the saddle point equation can be solved exactly, giving 
\begin{equation}
    q = \ell \frac{e^{\lambda(q)}}{e^\lambda(q)+1} + 2\tau \frac{1-\lambda(q) \coth \lambda(q)}{\pi \sinh \lambda(q)} 
\end{equation}
which reduces to \eqref{eq:saddlepointtilted} expanding for small $\lambda(q)$. Focusing on a single measurement, in the regime $t < \ell/2$, we also have that the two relevant counting functions are $\chi_{A\overline{A}}^{(2)} = |v_k|(t-\tau)$ and  $\chi_{A\overline{A}}^{(1)} = 2|v_k|\tau$.  The correction to the entropy is therefore given by \eqref{eq:single_squeezed}: 
\beqa\nonumber
      S_A(t|\tau,q)
    &=& S_A(t)+ 2\tau\int \dk |v_k|(s[n_{\Delta q}(k)]-s[n(k)])\\ &&+(t-\tau)\int \dk |v_k|(s[n_{2\Delta q}(k)]-s[n(k)]) +\log \mathcal{N} .
\eeqa
Substituting the saddle point value and performing the integral leads to the compact solution 
\begin{equation}
    S_A(t|\tau,q)  \approx S_A(t)-\frac{\Delta q^2}{6\pi \left(\ell/4 - \frac{2\tau}{3\pi}\right)^2} (2t-\tau)+ \log \mathcal{N},
\end{equation}
which has a more interesting time dependence compared to the symmetric case. In particular, there is an explicit dependence on the time $t$, arising from the pairs which were both contained in $A$ at the time of the measurement and get progressively shared with the complement.
However, on average, the effect is still small, since the standard deviation is also in this case of order the square root of the expectation value of the measurement. In fact taking the average simply gives
\begin{equation}
    \braket{S_A(t|\tau,q)} \approx S_A(t) -\frac{2t-\tau}{6\pi \left(\ell/4 - \frac{2\tau}{3\pi}\right)} +\log \mathcal{N}
\end{equation}
where the first term is $O(1)$ for $t<\ell/2$. In this case, the evaluation of the logarithmic correction is more involved but can still be performed in various relevant limits, as discussed in the appendix.

\section{Conclusions}\label{sec:concl}
In this work we have investigated the effect of measurements of a $U(1)$ conserved charge, performed in an extended subsystem, on a quenched free fermion chain. In particular, we have computed analytic expressions for the  bipartite entanglement entropy between the measured subsystem and its complement. This was done for two different classes of initial states, one which is symmetric with respect to the measured charge and another which is not. We studied the correction to the growth of entanglement for arbitrary measurements outcomes as well as the averaged entanglement. These results were then extended to the case of multiple periodically performed measurements, and confirmed by exact analytical calculations on the Néel state. 

For initial states which are eigenstates of the charge and a single measurement, we find that the entanglement entropy can be written as the standard expression, perturbed by two distinct corrections.  
One of these can be attributed to the destruction of charge fluctuations within the subsystem. It depends on the time at which the measurement was performed but is independent of the outcome of the measurement. At short times it scales logarithmically with the measurement time. The other term can be attributed to the change in configurational entropy of the state, this term scales linearly with the measurement time and depends both on the measurement outcome as well as the time at which the entropy is computed. For typical outcomes, the first correction dominates but for outcomes which deviate sufficiently from the average charge in the subsystem the latter term is more significant. If the measured value is exactly the average value then only the first correction is present.  

For non-symmetric initial states we also find that the entropy can be written in terms of a number of perturbations of the unitary value. In this case the configuration entropy term is also present but at short times the term arising from the destruction of charge fluctuations is compensated by the entanglement asymmetry of the state. This means that the drop in entropy coming from the destruction of charge fluctuations is compensated for the increase in entropy coming from the destruction of coherences between the different charge sectors. Upon averaging over all measurement outcomes at short times, we find that the entropy receives a negligible correction.  At long times, the unitary evolution causes the state to become symmetric, as a result, the entanglement asymmetry vanishes and the number entropy term is present. 
For both choices of initial states, it appears that the post-selection of specific measurement results is the only way to obtain significant deviations from the unperturbed results: taking averages over measurement outcomes generically  leads to negligible corrections. As noted above, this ultimately results from the strongly peaked charge distributions  which characterizes the states under consideration. 

To obtain these results we have performed analytic calculations using a hydrodynamic framework. In particular, we employed the operator form of the quasiparticle picture which has been shown to accurately describe the reduced density matrix in the ballistic scaling limit. As a special case of our results, we can recover easily previously known results on full counting statistics and symmetry resolution of entanglement, which provides a strong confirmation to our results.

Further lines of research include the extension of the discussion to the interacting regime, which can be performed by merging insight coming from the free case analysed in this work and from spacetime duality \cite{bertini_asymmetric_2024,bertini2023nonequilibrium}. Moreover, the techniques used in this work appear promising as a way to systematically approach several different problems involving quench evolution in free models, considering for example dissipative systems and systems undergoing different measurement protocols.

\noindent {\bf Acknowledgments:} 
We thank Filiberto Ares, Lorenzo Piroli, Michele Mazzoni and Ali Rajabpour for useful discussions. 
RT thanks the organizers of the "Student Workshop on Integrability" (June 2025) in Budapest, where a part of the project was carried out. 
PC and CR acknowledge support from European Union-NextGenerationEU, in the framework of the PRIN 2022 Project HIGHEST no. 2022SJCKAH\_002.

\appendix
\section{Evaluating the saddle point}
\label{appA}
\subsection{Symmetric states - Single measurement}
\label{appA1}
In this appendix we perform in detail the evaluation of the integrals in \eqref{eq:charge_moments} by saddle point. Since the various saddle point required for the different kind of measurements considered in the various sections are all rather similar, we perform in detail only the one for a single measurement on symmetric states. 

In this case the exponent of the integrand takes the form \eqref{eq:Fsymm_1},
\begin{multline}
\mathcal{F}_\alpha(t,\{\lambda_j,\lambda_j'\})=\int_{-\pi}^\pi\frac{{\rm d}k}{2\pi} \Big( \chi_{A\bar A}^{(1)}(t,k)f_{
\alpha}(\sum_j^\alpha[\lambda_j+\lambda_j'],k)+\chi^{(1)}_{\bar A\bar{A}}(t,k)\sum_{j=1}^\alpha  f_1(\lambda_j+\lambda_j',k)\\
+\Big[
 \chi_{A\bar A}^{(0)}(t,k) +\chi_{A\bar A}^{(2)}(t,k)\Big] f_\alpha(0,k)
\\ + \Big[\chi^{(2)}_{AA}(t,k)+\chi^{(2)}_{A\bar A}(t,k)+\chi^{(2)}_{\bar A\bar A}(t,k)\Big]\sum_{j=1}^\alpha i[\lambda_j+\lambda_j'] \Big).
\end{multline}
The counting functions can be obtained by simple physical arguments. In fact, they are related by the simple expressions:
\begin{equation}
    \begin{cases}
        \chi_{A\bar A}^{(1)}(t,k) + \chi^{(1)}_{\bar A\bar{A}}(t,k) = \min(2|v_k|\tau,\ell)\\
        \chi_{A\bar A}^{(1)}(t,k) +  \chi_{A\bar A}^{(0)}(t,k) +\chi_{A\bar A}^{(2)}(t,k)  = \min(2|v_k|t,\ell)\\
        \chi^{(2)}_{AA}(t,k)+\chi^{(2)}_{A\bar A}(t,k)+\chi^{(2)}_{\bar A\bar A}(t,k) = \frac{1}{2}\left[\ell-\min(2|v_k|\tau,\ell)\right]
    \end{cases}
\end{equation}
where the properties in the first and second lines arise because the sum of the counting functions represent the total number of pairs which are shared between $A$ and $\overline{A}$ at times $\tau$ and $t$ respectively, while the last line counts the total number of full pairs contained in $A$ at time $\tau$. Also, it is easy to convince oneself that the counting functions relating to the part which is not influenced by the measurement are obtained by counting the number of shared pairs, but starting from time $\tau$, and therefore are given by a simple modification of the usual expression,
\begin{equation}
    \chi_{A\bar A}^{(0)}(t,k) +\chi_{A\bar A}^{(2)}(t,k) = \min(2|v_k|(t-\tau),\ell)
\end{equation}
from which for instance the counting function discussed in the main text is obtained \eqref{eq:chiaabar},
\begin{eqnarray}
    \chi_{A\overline{A}}^{(1)}(t,k) =  \min(2|v_k|t,\ell)-\min(2|v_k|(t-\tau),\ell) ~,
\end{eqnarray}
and all others can be found subsequently.

The value of saddle point is obtained by differentiating the argument of the exponential with respect to each $\lambda_j$ and $\lambda_j'$. Since the two clearly give the same saddle point results, we can directly merge the two variables as $\lambda^*_j = \lambda^*_j + (\lambda'_j)^*$, where the star refers to the solution to the saddle point equations, which are 
\beqa \nonumber
     q &=& \int \dk  \chi_{A\overline{A}}^{(1)} (t,k) \partial_{\lambda_i }f_{\alpha}( \sum_j\lambda_j,k)  + \int \dk\chi^{(1)}_{\bar A\bar{A}}(t,k)\partial_{\lambda_j} f_1(\lambda_j)
     \\ &+&\frac{1}{2}\int \dk (\ell -\min(2|v_k|\tau,\ell)).
\eeqa
From this it is easy to convince oneself that the saddle point is dependent on the R\'enyi index, as 
\beqa \nonumber
       q &=& \int \dk  \chi_{A\overline{A}}^{(1)} (t,k) \frac{e^{i\sum_k \lambda^*_k} n^\alpha }{e^{i\sum_k \lambda^*_k} n^\alpha  + (1-n)^\alpha}  + \int \dk \chi^{(1)}_{\bar A\bar{A}}(t,k)\frac{e^{i \lambda^*_j} n}{e^{i \lambda^*_j} n + (1-n)}
     \\ &+&\frac{1}{2}\int \dk (\ell -\min(2|v_k|\tau,\ell)). 
\eeqa
Although the solution depends on the R\'enyi index, at fixed $\alpha$ the equation above leads to the same saddle point for each $\lambda_j$, which we write as $i\lambda_j^*:=\lambda_\alpha(q)$. Hence the saddle point solution reduces to 
\beqa \nonumber
       q &=& \int \dk  \chi_{A\overline{A}}^{(1)} (t,k) \frac{e^{\alpha \lambda_\alpha(q)} n^\alpha }{e^{\alpha \lambda_\alpha(q)} n^\alpha  + (1-n)^\alpha}  + \int \dk \chi^{(1)}_{\bar A\bar{A}}(t,k)\frac{e^{ \lambda_\alpha(q)} n}{e^{\lambda_\alpha(q)} n + (1-n)}
     \\ &+&\frac{1}{2}\int \dk (\ell -\min(2|v_k|\tau,\ell)). 
\eeqa
Introducing $\Delta q = q-\ell/2$, where $\ell/2$ is the charge of the initial state, this simplifies in the case for $\alpha=1$, which is needed in order to evaluate the Von Neumann entropy,
\beqa
\label{eq:saddlepointqsymmetric}
   \Delta q 
   &=& \int \dk  \min(2|v_k|\tau,\ell)  \left(\frac{n e^{\lambda(q)}}{ne^{\lambda(q)} +(1-n)}-\frac{1}{2}\right).
\eeqa
where we have used $\chi^{(1)}_{\bar A\bar{A}}(t,k) + \chi_{A\overline{A}}^{(1)} (t,k) = \min(2|v_k|\tau,\ell)$, and defined $\lambda(q) := \lambda_1(q)$. At this point, it is important to realize that this saddle point equation does not in general have a solution for arbitrary $\Delta q$ and $\tau$. In fact, restricting to the case $t <\ell/2$, the equation reduces to
\begin{equation}
\label{eq:deltaqdimercondition}
   \Delta q = 2\tau \int \dk |v_k|  \frac{n }{n +(1-n)e^{-\lambda(q)}}-\frac{2\tau}{\pi}.
\end{equation}
Considering for clarity the dimer state, in which $n = \frac{1+\cos k}{2}$, and noting that ${\rm d}n=-\frac{v_k}{2}dk$, a change of variables gives
\begin{equation}
  \Delta  q = \frac{4\tau }{\pi} \int_0^1 {\rm d}n \frac{n}{n +(1-n)e^{-\lambda(q)}}- \frac{2\tau}{\pi} = \frac{4\tau}{\pi} I(\lambda(q))-\frac{2\tau}{\pi} = \frac{2\tau}{\pi}(2I(\lambda(q))-1).
\end{equation}
Since $I(\lambda(q))$ is a monotonically increasing function of $\lambda(q)$, ranging from 0 to 1, the saddle point will exist only if
\begin{equation}
    \tau \geq \frac{\pi |\Delta q|}{2},
\end{equation}
as was already predicted for the symmetry resolution of entanglement in \cite{parez2021quasiparticle}.  This result has a very simple interpretation: since the initial state has a definite charge $\ell/2$, in the ballistic regime we are considering, the probability to measure a value of charge that deviates significantly from this is exponentially suppressed if the time of the measurement is not large enough. Hence this is simply a signature of the presence of a finite maximal velocity of propagation of the quasiparticles. Although we have considered the dimer to simplify the integral, the same conclusion \eqref{eq:conditionontau} applies to any state, as can be seen by considering the limits $\alpha \to \pm \infty $ of the integral in \eqref{eq:deltaqdimercondition} and its monotonicity. 

Once this issue is settled, the saddle point equation can be solved approximately by expanding for small $\lambda(q)$ ,
\beqa
   \Delta q &\approx&  \int \dk  \min(2|v_k|\tau,\ell) (n(k)-\frac{1}{2})  + \lambda(q) \int \dk  \min(2|v_k|\tau,\ell) n(k)(1-n(k)) .\hspace{1cm}
   \label{eq:expansionofdeltaq}
\eeqa
The first integral of the second line is zero: in fact, exploiting the property $n(k-\pi) =1 -n(k)$ which characterizes such states (as obvious from \eqref{eq:symm_presrve_pure}) one obtains
\beqa
    \int \dk   \min(2|v_k|\tau,\ell) n(k) &=&\int \dk \min(2|v_k|\tau,\ell) (1-n(k)) 
\eeqa
from which immediately
\begin{equation}
   \int \dk   \min(2|v_k|\tau,\ell) n(k) = \frac{1}{2}  \int \dk   \min(2|v_k|\tau,\ell).
\end{equation}
The second integral on the other hand simply represents the standard deviation of the charge distribution, 
which, for $\tau<\ell/2$, reduces to the Drude self weight \cite{drude_weights,doyon2020lecture}, since in this regime
\begin{equation}
\label{eq:drude}
   2\tau \int \dk  |v_k| n(k)(1-n(k)) = 2\tau D. 
\end{equation}
Therefore, the saddle point can be expressed in an particularly neat form as
\begin{equation}
\label{eq:saddle_generic}
    \lambda(q) \approx \frac{\Delta q}{\sigma_\tau^2}.
\end{equation}
Notably, this result is constant in time, and only depends on the time of the measurement. 

The saddle point solution can then be carrried on by evaluating the argument of the integrals in \eqref{eq:charge_moments} with the saddle point \eqref{eq:saddle_generic}. In particular 
\begin{eqnarray}
\label{eq:saddle_solution_renyi}
 \log  \tr_A \Big[\{\rho_A(t|\tau,q)\}^\alpha\Big]&=& \log\left\{\frac{  e^{ -i q \alpha \lambda_\alpha +\mathcal{F}_{\alpha}(t,\lambda_\alpha)}}{e^{-i q \alpha \lambda_1+\alpha \mathcal{F}_{1}(t,\lambda_1)}}\right\} + \log\sqrt{\frac{\left(2\pi |\mathcal{F}''_{1}(t,\lambda_1)|\right)^\alpha}{2\pi |\mathcal{F}''_{\alpha}(t,\lambda_\alpha)|}} \hspace{0.4cm}
\end{eqnarray}
where we have expressed compactly as $|\mathcal{F}''_{\alpha}(t,\lambda_\alpha)|$ the determinant of the hessian of the exponent of the integral, which commonly appear in saddle point evaluations. In general, in most studies involving the quasiparticle picture, this term contributes subleading terms, and is therefore fully neglected. Here, on the other hand, this term is not completely negligible, and we will keep it until the end referring to it as $\mathcal{N}_\alpha := \sqrt{\frac{\left(2\pi |\mathcal{F}''_{1}(t,\lambda_1)|\right)^\alpha}{2\pi |\mathcal{F}''_{\alpha}(t,\lambda_\alpha)|}}$. 

In general, in the case of arbitrary R\'enyi index \eqref{eq:saddle_solution_renyi} will depend on two separate saddle points $\lambda_\alpha(q)$ and $\lambda_1(q)$. This would lead in general to a complicated solution, in which the quasiparticle structure would not be manifest.
In the case $\alpha\to1$, however, the problem simplifies. In fact, the Von Neumann entropy can be obtained via the De l'Hopital's rule,
\begin{eqnarray}
 S_A(t|\tau,q) = \lim_{\alpha \to 1} \frac{1}{1-\alpha}\log  \tr_A \Big[\{\rho_A(t|\tau,q)\}^\alpha\Big] = - \partial_\alpha\log  \tr_A \Big[\{\rho_A(t|\tau,q)\}^\alpha\Big]\Big|_{\alpha=1}.
\end{eqnarray}
Performing the derivative explicitly and evaluating in $\alpha=1$ leads to the simplification of several terms, leading to 
\begin{multline}
S_A(t|\tau,q) = - \int \dk \chi_{A\overline{A}}^{(1)}(t,k) \left( \log(1-n+ne^{\lambda(q)}) - \frac{(1-n)\log(1-n)+ne^{\lambda(q)}\log(ne^{\lambda(q)})}{1-n+ne^{\lambda(q)}}\right) \\
- \int \dk \min(2|v_k|(t-\tau,\ell) ((1-n)\log(1-n)+n\log n) +\log \mathcal{N},
\end{multline}
where we recall that $\lambda(q) = \lambda_1(q)$, and we have defined $\log \mathcal{N}:= \lim_{\alpha \to 1} \frac{1}{1-\alpha}\log \mathcal{N}_\alpha$. At this point, we note that the contribution in the second line is simply the standard pair contribution to the Von Neumann entropy in the absence of measurement, 
\begin{equation}
\label{eq:pairentropy}
    s[n(k)]=-n(k)\log n(k) -(1-n(k))\log (1-n(k)),
\end{equation}
while the first line can be written as a similar contribution, namely a pair entropy of the form \eqref{eq:pairentropy} but evaluated on a modified state 
\begin{equation}
    n_{\Delta q}(k) = \frac{n(k) e^{\lambda(q)}}{n(k) e^{\lambda(q)}+(1-n)},
\end{equation}
which depends on the outcome of the measurement through the saddle point \eqref{eq:saddle_generic}.
Hence the entropy can be finally expressed as
\beqa
S_A(t|\tau,q) &=&  \int \dk \chi_{A\overline{A}}^{(1)}(t,k) s[ n_{\Delta q}(k)] +\int \dk \min(2|v_k|(t-\tau),\ell)s[n(k)]+ \log \mathcal{N}\nonumber \\
&=& S_A(t) + \int \dk \chi_{A\overline{A}}^{(1)}(t,k) \left[s[ n_{\Delta q}(k)] -s[ n(k)]\right] + \log \mathcal{N}
\eeqa
where in the last line we have used $ \chi_{A\overline{A}}^{(1)}(t,k) + \min(2|v_k|(t-\tau),\ell) = \min(2|v_k|t,\ell)$, which gives rise to the standard quasiparticle entropy  $S_A(t) $. This is the result which was claimed in the main text, in which the normalization factor coming from the saddle point plays the role of the number entropy.
In order to understand the average effect of the measurement on the entropy, it is also necessary to evaluate the probability of each outcome in the saddle point approximation. This is given by
\begin{equation}
\label{eq:probability_symmetric_state}
     p(t,q) = \Tr\left[e^{-i H(t-\tau)}\Pi_q \rho(\tau) \Pi_q e^{i H(t-\tau)}\right] = \Tr\left[\Pi_q \rho(\tau) \Pi_q \right] =  p(\tau,q),
\end{equation}
which is simply given by the $\alpha\to 1$ limit of expression for the charged moments. 
Evaluating by saddle point analogously as above, and using the same saddle point value $\lambda(q)$, this reduces to 
\beqa
   p(\tau,q), \approx \frac{1}{\sqrt{2\pi\sigma_\tau^2}} \exp\left\{-\frac{\Delta q^2}{2\sigma_\tau^2}\right\}.
\label{eq:probabilityGaussiansymmetric}
\eeqa
Hence the probability is a Gaussian centered about the value of the charge in the initial state, $q_0=\ell/2$, and its width is precisely the variance of the charge distribution. For $\tau<\ell/2$, as shown in \eqref{eq:drude}, $\sigma_\tau^2\propto \tau$, and therefore the variance of the distribution is precisely the measurement time, as would be naturally expected.

The final aspect which needs to be investigated is the form of the correction arising from the term $\log\mathcal{N}_\alpha$. We distinguish two regimes, the one in which $t<\ell/2$ and the one in which $t>\ell/2$. In the first regime, since $\chi^{(1)}_{\bar A\bar{A}}(t,k)$ is zero, the integral effectively only depends on $\sum_j\lambda_j$, and hence can be mapped to a one-dimensional integral. Hence the hessian reduces to the absolute value of the second derivative of the exponent,
\begin{equation}
    \lim_{\alpha\to 1}\frac{1}{1-\alpha}\log \sqrt{\frac{(2\pi |\mathcal{F}''_1(\lambda_1)|)^\alpha}{2\pi |\mathcal{F}''_\alpha(\lambda_\alpha)|}} = \frac{1}{2}\left(\partial_ \alpha\log{|\mathcal{F}_\alpha''(\lambda_\alpha)|} - \log{|\mathcal{F}''_1(\lambda_1)|}\right)_{\alpha=1} - \log \sqrt{2\pi}.
\end{equation}
From the explicit form of $\mathcal{F}''$, it is simple to convince oneself that the first term in the right hand side is $O(1)$, while the second one is $-\frac{1}{2}\log2\tau + O(1)$ for any symmetric initial state. Hence we see that this correction corresponds precisely to the number entropy which was expected to arise in \eqref{eq:expectation}.

In the regime $\tau<\ell/2$ and $t> \ell/2$, the contributions relative to $\chi^{(1)}_{\bar A\bar{A}}(t,k)$ are no longer zero, and therefore the integral does not simplify. The full hessian matrix is
\begin{multline}
\label{eq:hessian}
    \partial_i \partial_j \mathcal{F}(t,\{\lambda_k\}) = -\int \dk \chi_{A\overline{A}}^{(1)}(t,k)\frac{n^{\alpha} (1-n)^{\alpha} e^{i\sum_k\lambda_k}}{((1-n)^{\alpha}+e^{i\sum_k\lambda_k}  n^{\alpha})^2} \\- \delta_{ij}\int \dk \chi^{(1)}_{\bar A\bar{A}}(t,k)\frac{n (1-n) e^{\lambda_j}}{((1-n)+e^{\lambda_j} n)^2},
\end{multline}
which can be seen as the sum of a matrix with constant entries and a diagonal matrix, which are the first and second terms of the equation above. The determinant of the sum of a diagonal matrix $\operatorname{diag}(D_{ii})$ and a rank 1 matrix S of constant entries $s$ can be evaluated exactly as
\begin{equation}
\label{eq:determinantofsum}
    \det(D+S) = \det(D) \left(1+ \sum_i \frac{s}{D_{ii}}\right) = \left(\prod_i D_{ii}\right)\left(1+ \sum_i \frac{s}{D_{ii}}\right).
\end{equation}
Using this property in \eqref{eq:hessian} and using the saddle point values $i\lambda^*_j = \lambda_\alpha$, the hessian at the saddle point is expressed as
\begin{equation}
  |\det(\partial_i\partial_j F)_{i\lambda_j=\alpha_\alpha}| = A_\alpha^\alpha + \alpha A_\alpha^{\alpha-1}B_\alpha \overset{\alpha=1}{=} A_1+B_1
\end{equation}
where 
\begin{equation}
\label{eq:AandB}
    \begin{cases}
        B_\alpha= \int \dk \chi_{A\overline{A}}^{(1)}(t,k)\frac{n^\alpha  (1-n)^\alpha e^{\alpha\lambda_\alpha}}{((1-n)^\alpha+e^{\alpha\lambda_\alpha}  n^\alpha )^2} \vspace{0.3cm}\\
        A_\alpha =\int \dk \chi^{(1)}_{\bar A\bar{A}}(t,k)\frac{n (1-n) e^{\lambda_\alpha}}{((1-n)+e^{\lambda_\alpha} n)^2}.
    \end{cases}
\end{equation}
 In this case the form of the corrective term would be
\begin{equation}
\label{eq:logN_general}
  \frac{1}{1-\alpha}\log\mathcal{N}_\alpha= \frac{1}{1-\alpha}  \log  \sqrt{(2\pi)^{\alpha-1}\frac{(A_1+B_1)^\alpha}{A_\alpha^\alpha+\alpha A_\alpha^{\alpha-1}B_\alpha}}.
\end{equation}
Again, since we are intereted in the Von Neumann limit we can take the limit $\alpha \to 1$ to obtain
\begin{multline}
\label{eq:logN_one}
    \lim_{ \alpha\to 1} \frac{1}{1-\alpha}  \log  \sqrt{\frac{(A_1+B_1)^\alpha}{A_\alpha^\alpha+\alpha A_\alpha^{\alpha-1}B_\alpha}}=\\ =\frac{1}{2}\left(\log\left[\frac{A_1}{A_1+B_1}\right]+\frac{B_1}{A_1+B_1}  +\frac{\partial_\alpha (A_\alpha+B_\alpha)|_{\alpha=1}}{(A_1+B_1)} \right). \end{multline}
Compared to the result for $t<\ell/2$, here the correction exhibits a stronger suppression in all terms, and in particular all relevant terms vanish in the long time limit, since when $t\gg \ell$, $B_\alpha\to 0$, and therefore the correction tends simply to a $O(1)$. In particular, expanding \eqref{eq:AandB} at zero order in the saddle point and substituting the expressions for the counting functions, the correction becomes
\begin{equation}
    \log \mathcal{N}(t) \approx - \frac{1}{2}\log \frac{\sigma_\tau^2} {\sigma_\tau^2 + \sigma_{t-\tau}^2 - \sigma_t^2}+ ...
\end{equation}
which disappears in the long time limit since
\begin{equation}
    \sigma_t^2\approx \sigma_{t-\tau}^2 \text{ for } t\to \infty. 
\end{equation}

From the discussion just performed it is easy to convince oneself that the crucial aspect in the evaluation depends on whether the term $\chi_{\bar{A}\bar{A}}^{(1)}$ is zero or not. In the regime $t<\ell/2$, this is always zero, leading to the easy solution discussed above. On the other hand, another simple regime is that for which $t=\tau$, for all values of the ratio $t/\ell$. In this regime the term $\chi_{\bar{A}\bar{A}}^{(1)}$ is always absent, and therefore the discussion above can be repeated leading to 
\begin{equation}
    \log \mathcal{N}(t=\tau) \approx - \frac{1}{2}\log \sigma_\tau^2 + O(1)
\end{equation}
confirming the identification with the number entropy. Interestingly, as long as $t<\ell/2$, this form is preserved by the time evolution. However, the structure changes as t increases due to the activation of the $\chi_{\bar{A}\bar{A}}^{(1)}$ terms.

\subsection{Symmetric states - Multiple measurements}\label{appA2}
Since the main features of the generalization to several measurements are already clear considering two measurements, we begin by analysing this situation. If the outcomes of the two measurements are $q_1$ and $q_2$ respectively, expressing the projectors as
\begin{eqnarray}    \Pi^A_{q_n}=\int_{-\pi}^\pi\frac{{\rm d}\lambda_n}{2\pi}e^{i\lambda_n(q_n-\hat{Q}_A)}~,
\end{eqnarray}
which leads to
\begin{equation}
     \tr_A \Big[\{\rho_A(t|\tau,q_1,q_2)\}^\alpha\Big] =\frac{ \int_{-\pi}^\pi \Big[\prod_{n=1}^2\prod_{j=1}^\alpha\frac{{\rm d}\lambda_{n,j} {\rm d}\lambda_{n,j}'}{4\pi^2}e^{i q_n(\lambda_{n,j}+\lambda_{n,j}')}\Big] e^{ \mathcal{F}_{\alpha}(t,\{\lambda_{n,j},\lambda_{n,j}'\}\,)}}{\Big[\int_{-\pi}^\pi \prod_{n=1}^2\frac{{\rm d}\lambda_n {\rm d}\lambda_n'}{4\pi^2}e^{i q_n(\lambda_n+\lambda_n')} e^{\mathcal{F}_{1}(t,{\lambda}_n,{\lambda}_n')}\Big]^\alpha}.
\end{equation}
Clearly, $\mathcal{F}_{\alpha}(t,\{\lambda_{n,j},\lambda_{n,j}'\}\,)$ will contain more different classes of contributions compared to the case of a single measurement. In particular, one has to distinguish between the pairs in which both measurement have an effect, and those in which only one (either the first or the second) acts on the state. 
As for the case considered above, we note that it is actually possible in general to change coordinates to $\lambda_j+\lambda_j'$, thus reducing the number of degrees of freedom. With this simplification  $\mathcal{F}_{\alpha}$ can be expressed in terms of a single set of variables $\{\lambda_{n,j}\}_{\{n=1,2;\hspace{0.05cm}j=1...\alpha\}}$, as
\beqa
  \mathcal{F}_{\alpha}(t,\{\lambda_{n,j}\}\,) &=& \int \dk \left[\chi_{A\overline{A}}^{(1,1)} f_\alpha(\sum_j [\lambda_{1,j} + \lambda_{2,j}],k) + \chi_{A\overline{A}}^{(0,1)} f_\alpha(\sum_j \lambda_{2,j},k) +\left(\chi_{A\overline{A}}^{(0)} + \chi_{A\overline{A}}^{(2)} \right)f_\alpha(0)  \right.  \nonumber\\&+&   \chi^{(1,0)}_{\bar A\bar{A}} \sum_j f_1(\lambda_{1,j},k) +\chi^{(1,1)}_{\bar A\bar{A}}\sum_j f_1(\lambda_{1,j}+\lambda_{2,j},k)+\chi^{(0,1)}_{\bar A\bar{A}}  \sum_j f_1(\lambda_{2,j},k) \nonumber \\ &+& 
     \left.  \frac{i}{2} \left[\ell - \min(2|v_k|\tau,\ell)\right] \sum_j \lambda_{1,j}+ \frac{i}{2} \left[\ell - \min(4|v_k|\tau,\ell)\right] \sum_j \lambda_{2,j}  \right] .
\eeqa
The counting functions are the natural generalizations of those which where defined for the case of a single measurement. In particular, two indices represent the number of particles of each pair which were affected by the first and second measurement respectively. They are related by similar expressions, such as
\begin{equation}
\label{eq:relationsmultiple}
    \begin{cases}
      \chi_{A\overline{A}}^{(1,1)}  +\chi_{A\overline{A}}^{(0,1)} + \chi_{A\overline{A}}^{(0)} + \chi_{A\overline{A}}^{(2)}  =\min(2|v_k|t,\ell) \\
      \chi_{A\overline{A}}^{(1,1)} + \chi_{A\overline{A}}^{(0,1)} + \chi^{(1,1)}_{\bar A\bar{A}} + \chi^{(0,1)}_{\bar A\bar{A}} = \min(4|v_k|\tau,\ell) \\
      \chi_{A\overline{A}}^{(1,1)}  + \chi^{(1,1)}_{\bar A\bar{A}} + \chi^{(1,0)}_{\bar A\bar{A}} = \min(2|v_k|\tau,\ell)
    \end{cases}
\end{equation}
which as above allow to fix their precise form.

Regarding the saddle point evaluation, restricting to the case $2\tau<\ell/2$, the time-delay exhibited by the single measurement solution appears in a similar fashion. 
In fact, it is immediate to see that the saddle points are again independent on $j$, in the sense that $i\lambda_{1,j}^* = \lambda_1$ and $i\lambda_{2,j}^* = \lambda_2$, where we now leave the dependence of the saddle point values on $q_1$ and $q_2$ implicit. Then the saddle point equations are  
\beqa
\label{eq:deltacharges}
        q_1 - \frac{\ell}{2} &=& \int \dk 2|v_k|\tau\frac{e^{\lambda_1+\lambda_2} n}{e^{\lambda_1+\lambda_2} n +(1-n)} - \frac{2\tau}{\pi}\vspace{0.3cm} \\
        q_2 - \frac{\ell}{2}  &=& \int \dk 2|v_k|\tau\left(  \frac{e^{\lambda_1+\lambda_2} n}{e^{\lambda_1+\lambda_2} n +(1-n)}+\frac{e^{\lambda_2} n}{e^{\lambda_2} n +(1-n)}\right ) -  \frac{4\tau}{\pi}.
   \eeqa
    The first line immediately implies $\tau\geq \frac{\pi |\Delta q_1|}{2}$, where $\Delta q_1 = q_1 -\ell/2$. But subtracting the first equation from the second one we also obtain that:
\begin{equation}
    \Delta q_2  = q_2 - q_1 =2\tau \int \dk |v_k|\frac{e^{\lambda_2} n}{e^{\lambda_2} n +(1-n)}- \frac{2\tau}{\pi}
\end{equation}
which implies that also $\tau \geq \frac{\pi|\Delta q_2|}{2} $. Hence at each measurement, the saddle point condition has a solution only if the variation from the charge measured in the previous measurement satisfies the same fixed condition. This clearly generalizes to an arbitrary number of measurements, giving the general relation 
\begin{equation}
   |\Delta q_i|= |q_{i+1}-q_i| \leq \frac{2\tau}{\pi}.
\end{equation}
The solution of the saddle point equations can then be performed by expanding the integrals for small $\lambda_1$ and $\lambda_2$, giving for arbitrary $\tau,t$
\begin{equation}
\label{eq:saddlepoint2meas}
    \begin{cases}
        \lambda_1+\lambda_2 = \frac{\Delta q_1}{\sigma_\tau^2}  \\
        \lambda_2 = \frac{\Delta q_2}{\sigma_{2\tau}^2 - \sigma_\tau^2}
    \end{cases}
\end{equation}
The general solution is obtained by noting that when solving for the Von Neumann entropy by saddle point only terms containing $f_\alpha$ do not cancel with the normalization. The final result is then of the form 
\begin{multline}
S_{A}(t|q_1,q_2) = \int \dk\chi_{A\overline{A}}^{(1,1)}s[n_{\lambda_1+\lambda_2}(k)] +  \int \dk\chi_{A\overline{A}}^{(0,1)} s[n_{\lambda_2}(k)]\\ +  \int \dk\left(\chi_{A\overline{A}}^{(0,0)}+\chi_{A\overline{A}}^{(2,2)}\right) s[n(k)] + \log \mathcal{N},
\end{multline}
where, because of the explicit values of the saddle points \eqref{eq:saddlepoint2meas}, the occupation functions are
\begin{equation}
\begin{cases}
    n_{\lambda_1+\lambda2}(k) = \frac{n e^{\lambda_1+\lambda_2}}{n e^{\lambda_1+\lambda_2} + (1-n)}=\frac{n e^{\Delta q_1/\sigma_\tau^2}}{n e^{\Delta q_1/\sigma_\tau^2} + (1-n)} := n_{\Delta q_1}(k) \\
    n_{\lambda_2}(k) = \frac{n e^{\lambda_2}}{n e^{\lambda_2} + (1-n)}=\frac{n e^{\Delta q_2/(\sigma_{2\tau}^2 - \sigma_\tau^2)}}{n e^{\Delta q_2/(\sigma_{2\tau}^2 - \sigma_\tau^2)} + (1-n)} :=  n_{\Delta q_2}(k)
\end{cases}
\end{equation}
Using the fact that by \eqref{eq:relationsmultiple} the sum of the counting functions appearing is simply $\min(2|v_k|t,\ell)$, since they count the total number of pairs which are shared at time t, the result can again be expressed as
\begin{multline}
    S_{A}(t|q_1,q_2) = S_A(t) + \int \dk\chi_{A\overline{A}}^{(1,1)}[   s[n_{\Delta q_1}(k)] -s(k)]\\ +  \int \dk\chi_{A\overline{A}}^{(0,1)}[  s[n_{\Delta q_2}(k)]-s(k)] +\log \mathcal{N}.
\end{multline}

At this point, the generalization to $n$ measurements is straightforwardly obtained. The saddle points are the natural generalizations of \eqref{eq:saddlepoint2meas}, in terms of the $\{\lambda_i\}_{i=1,...n}$ which label the $n$ different integrals,
\begin{equation}
\label{eq:saddlepointnmeas}
    \begin{cases}
        \lambda_1+\lambda_2+ ... \lambda_ = \frac{\Delta q_1}{\sigma_\tau^2}  \\
\lambda_2+\lambda_3+ ... \lambda_{n} = \frac{\Delta q_2}{\sigma_{2\tau}^2 - \sigma_\tau^2}  \\
 &\vdots\\
        \lambda_n = \frac{\Delta q_n}{\sigma_{m\tau}^2 - \sigma_{(n-1)\tau}^2}
    \end{cases}
\end{equation}
and the final result for the entropy 
\beqa
 S_{A}(t|\{q_i\}) &=& S_A(t) + \sum_{l=1}^m\int \dk\chi_{A\overline{A}}^{(\sum_{i=l}^m)}(t,k)\left\{ s[n_{\sum_{i=l}^m \lambda_i}(k)] -s[n(k)]\right\}+\log \mathcal{N} \nonumber
 \eeqa
 where $ s[n_{\sum_{i=k}^m \lambda_i}]$ is evaluated over the state
 \begin{equation}
     n_{\sum_{i=l}^m \lambda_i}= \frac{ne^{\sum_{i=l}^m \lambda_i}}{ne^{\sum_{i=l}^m \lambda_i} + (1-n)} = \frac{ne^{\frac{\Delta q_l}{2D\tau}}}{ne^{\frac{\Delta q_l}{2D\tau}} + (1-n)}  := n_{\Delta q_l}
 \end{equation}
 where we have restricted to the case $m\tau<\ell/2$, for which $\sigma_{k\tau}^2 - \sigma_{(k-1)\tau}^2=2D\tau$, just for simplicity of notation. 
Hence we reduce to the solution of the main text,
\beqa
 S_{A}(t|\{q_i\}) &=& S_A(t) + \sum_{l=1}^m\int \dk\chi_{A\overline{A}}^{(1,l)}(t,k)\left\{ s[ n_{\Delta q_l}(k)] -s[ n(k)]\right\}+ \dots
 \eeqa
 where the counting functions $\chi_{A\overline{A}}^{(1,\ell)}(t,k)$ count the number of pairs which underwent $l$ measurements. In the small time limit $t<\ell/2$, they are all identical, and equal to $2|v_k|\tau$.
 
 It is also rather immediate from our discussion to obtain the probability of measuring a string of values $\{q_i\}_{i=1...m}$. In the Gaussian limit, this reduces to a product of Gaussians, each of which is centered in the outcome of the previous measurement,
 \begin{equation}
    p(\tau,\{q_i\}) \propto \exp\left\{-\sum_i\frac{(\Delta q_i)^2}{2D\tau}\right\},\hspace{0.3cm} \Delta q_i=q_i - q_{i-1},
\end{equation}
where again we have expressed the solution in the regime  $m\tau<\ell/2$. This allows to take averages very easily, since the solution for the entropy only depends on $\Delta q_i$. Also in this situation, the logarithmic correction can be found by generalizing straightforwardly the single measurement result. As we will show in the next section in the exact solution for the Néel state, this correction is simply given by $n$ times the correction of the single measurements where $n$ is the number of measurements.

\subsection{Symmetry breaking states}
\label{sec:appA3}
Here we briefly discuss the saddle point evaluation in the symmetry breaking case. Since the discussion is not too different from the two situations considered above, we will only sketch the main ideas.
The main difficulty arising in the symmetry breaking case is the presence of significantly different contributions arising depending on wether only one or two particles of each pair are in $A$ at the time of the measurement. The complete expression is given by \eqref{eq:squeezedtaugreatzero},
\beqa\nonumber
\mathcal{F}_\alpha(t,\{\lambda_j,\lambda_j'\})=\int_{-\pi}^\pi\frac{{\rm d}k}{2\pi} \Big( \chi_{A\bar A}^{(1)}(t,k)f_{
\alpha}(\sum_j^\alpha[\lambda_j+\lambda_j'],k)+\chi^{(1)}_{\bar A\bar{A}}(t,k)\sum_{j=1}^\alpha  f_1(\lambda_j+\lambda_j',k)\\\nonumber
+
 \chi_{A\bar A}^{(0)}(t,k)  f_\alpha(0,k)+\chi_{A\bar A}^{(2)}(t,k)f_{
\alpha}(2\sum_j^\alpha[\lambda_j+\lambda_j'],k)
\\ + \chi^{(2)}_{AA}(t,k)\sum_{j=1}^\alpha f_1(2[\lambda_j+\lambda_{j+1}'],k)+\chi^{(2)}_{\bar A\bar A}(t,k)\sum_{j=1}^\alpha f_1(2[\lambda_j+\lambda_{j}'],k)\Big)\quad
\eeqa
where as in the above the various counting functions can be related by simple expressions deriving from simple physical motivations, such as
\begin{equation}
\label{eq:relationsmultiplebreak}
    \begin{cases}
       \chi_{A\bar A}^{(1)}  + \chi^{(1)}_{\bar A\bar{A}}= \min(2|v_k|\tau,\ell) \\
        \chi_{A\bar A}^{(0)}  +  \chi_{A\bar A}^{(1)}+  \chi_{A\bar A}^{(1)} +\chi_{A\bar A}^{(2)}  = \min(2|v_k|t,\ell) \\ \chi^{(2)}_{AA}+\chi^{(2)}_{\bar A\bar A}+ \chi^{(2)}_{ A\bar A} = \frac{1}{2}\left[\ell- \min(2|v_k|\tau,\ell)\right],
    \end{cases}
\end{equation}
which as above allow to fix the specific values. Again, the saddle point equations turn out to be independent on $j$ and also to be equivalent between the $\lambda$ and $\lambda'$ coordinates. With such simplifications the equations reduce to 
\beqa
    q= \int  \frac{dk}{2\pi}\left\{ \left(\chi^{(2)}_{AA}(t,k) +\chi^{(2)}_{\bar A\bar A}(t,k)\right) \frac{2 n e^{2i\lambda_\alpha(q)}}{ne^{2i\lambda_\alpha(q)} +(1-n)} + \chi_{A\overline{A}}^{(2)} \frac{2 n^\alpha  e^{2i \alpha \lambda_\alpha(q)}}{n^\alpha e^{2i\alpha\lambda_\alpha(q)} +(1-n)^p}\right.\nonumber \\ \left.
  +\chi^{(1)}_{\bar A\bar A}(t,k)\frac{ n e^{i\lambda_\alpha(q)}}{ne^{i\lambda_\alpha(q)} +(1-n)}  +\chi_{A\overline{A}}^{(1)} \frac{n^\alpha  e^{i \alpha \lambda_\alpha(q)}}{n^\alpha e^{i\alpha \lambda_\alpha(q)} +(1-n)^p}
    \right\}.
\eeqa
As for the symmetric states, considering arbitrary Rényi index this cannot be further simplified. However, in the limit $\alpha\to 1 $, expanding for small $\lambda :=\lambda_1$ and using \eqref{eq:relationsmultiplebreak} gives
\begin{equation}
    q = \overline{q} + \lambda(q)\sigma_\tau^2 \Rightarrow \lambda(q) = \frac{\Delta q}{\sigma_\tau^2},
\end{equation}
 where the average charge and variance were defined in section \ref{sec:squeezed}. Substituting the saddle point value in the exponential, and evaluating the Rényi entropies, leads to an expression which is similar to the symmetric case. The main difference arises from the presence in  \eqref{eq:squeezedtaugreatzero} of both terms of the form $f_\alpha(\lambda)$ and $f_\alpha(2\lambda)$. Therefore the final result will contain both, and in particular 
 \beqa\nonumber
     S_A(t|\tau,q) 
    &=& S_A(t) + \int \dk \chi_{A\overline{A}}^{(1)}(t,k) \left[s[ n_{\Delta q}(k)] -s[ n(k)]\right] \\ &+& \int \dk \chi_{A\overline{A}}^{(2)}(t,k) \left[s[ n_{2\Delta q}(k)] -s[ n(k)]\right]+\log \mathcal{N}
 \eeqa
which reproduces \eqref{eq:single_squeezed}.
To study the evaluation of the logarithmic term, we focus for simplicity on the case $\tau=0$, namely the case in which the measurement is performed on the initial state. In this regime the measurement acts only on full pairs, leading to a simple structure in the $t<\ell/2$ regime,
\beqa\nonumber
\mathcal{F}_\alpha(t,\{\lambda_j,\lambda_j'\})=\int_{-\pi}^\pi\frac{{\rm d}k}{2\pi} \Big( 
 \chi_{A\bar A}^{(0)}(t,k)  f_\alpha(0,k)+\chi_{A\bar A}^{(2)}(t,k)f_{
\alpha}(2\sum_j^\alpha[\lambda_j+\lambda_j'],k)
\\ + \chi^{(2)}_{AA}(t,k)\sum_{j=1}^\alpha f_1(2[\lambda_j+\lambda_{j+1}'],k)\Big).\quad
\eeqa
Upon performing a coordinate change $\lambda_j + \lambda_{j+1}' \to \lambda_j$, this can then be studied with the same approach of the symmetric case. In particular, \eqref{eq:logN_general} and \eqref{eq:logN_one} are still valid, with the only difference being the counting functions,
\begin{equation}
    \begin{cases}
        B^{\rm asym}_\alpha= \int \dk 4\chi_{A\overline{A}}^{(2)}(t,k)\frac{n^\alpha  (1-n)^\alpha e^{2\alpha\lambda_\alpha(q)}}{((1-n)^\alpha+e^{2\alpha\lambda_\alpha(q)}  n^\alpha )^2} \vspace{0.3cm}\\
        A^{\rm asym}_\alpha =\int \dk  4\chi^{(2)}_{AA}(t,k)\frac{n (1-n) e^{2\lambda_\alpha(q)}}{((1-n)+e^{2\lambda_\alpha(q)} n)^2}.
    \end{cases}
\end{equation}
In particular, this implies that at t=0 the logarithmic part of \eqref{eq:logN_one} disappears completely, as was heuristically motivated in the main text, since $\chi_{A\overline{A}}^{(2)}(t,k)$ vanishes. 

In the opposite limit $t\gg \ell$, the functions are instead simply 
\beqa\nonumber
\mathcal{F}_\alpha(t,\{\lambda_j,\lambda_j'\})=\int_{-\pi}^\pi\frac{{\rm d}k}{2\pi} \Big(
 \chi_{A\bar A}^{(0)}(t,k)  f_\alpha(0,k)+\chi^{(2)}_{\bar A\bar A}(t,k)\sum_{j=1}^\alpha f_1(2[\lambda_j+\lambda_{j}'],k)\Big)\quad
\eeqa
which is even simpler than above, as the Hessian matrix is simply diagonal. Hence the determinant is simply 
\begin{equation}
    |\det(\partial_i\partial_j F)_{i\lambda_j=\alpha_\alpha}| = \left( C^{\rm asym}_\alpha\right)^\alpha, \hspace{0.5cm}  C^{\rm asym}_\alpha=\int \dk  2\ell\frac{n (1-n) e^{2\lambda_\alpha}}{((1-n)+e^{2\lambda_\alpha} n)^2}
\end{equation}
where we have used that in this regime $\chi^{(2)}_{\bar A\bar A}(t,k) = \ell/2$,
and hence
\begin{equation}
  \frac{1}{1-\alpha}\log\mathcal{N}_\alpha= \frac{1}{1-\alpha}  \log  \sqrt{\frac{(C_1^{\rm asym})^\alpha}{(C_\alpha^{\rm asym})^\alpha}} - \log \sqrt{2\pi}.
\end{equation}
This simplifies in the Von Neumann limit,
\begin{equation}
    \lim_{\alpha \to 1} \frac{1}{1-\alpha}\log\mathcal{N}_\alpha = \frac{1}{2} \frac{\partial_\alpha C_\alpha^{\rm asym}|_{\alpha=1}}{C_1^{\rm asym}}
\end{equation}
which contributes just a (suppressed) $O(1)$ term, since all $O(\ell)$ terms simplify. This is a further signal that the effect of the measurement is washed away in the long time limit.

A further regime of interest in this evaluation is that in which the time of the measurement is much larger than the size of the system, namely $\tau \gg \ell$. In this regime, no full pairs remain in the system at the time of the measurement, and therefore
\begin{multline}
\mathcal{F}_\alpha(t,\{\lambda_j,\lambda_j'\})=\int_{-\pi}^\pi\frac{{\rm d}k}{2\pi} \Big( \chi_{A\bar A}^{(1)}(t,k)f_{
\alpha}(\sum_j^\alpha[\lambda_j +\lambda_j'],k)  \\ \quad+\chi^{(1)}_{\bar A\bar{A}}(t,k)\sum_{j=1}^\alpha  f_1(\lambda_j+\lambda_j',k)+
 \chi_{A\bar A}^{(0)}(t,k)  f_\alpha(0,k)\Big)\quad
\end{multline}
where $\chi_{A\bar A}^{(1)}(t,k) + \chi^{(1)}_{\bar A\bar{A}}(t,k) = \ell$. For $t=\tau$, repeating the above reasoning leads to a correction of order $\log \ell$, as would be expected considering that the charge distribution at the time of the measurement has saturated to variance $\ell$. For increasing t, this correction is further decreased, until it vanishes, as in all the situations considered above.

Finally, we consider the case $t=\tau$. In this regime, the two non-zero contributions in the Hessian are simply the one containing $\chi_{A\bar{A}}^{(1)}$ and $\chi_{AA}^{(2)}$. The discussion of section \eqref{appA1} generalizes straightforwardly leading to a logarithmic term
\begin{equation}
    \log \mathcal{N}(t) \approx -\frac{1}{2}\log \frac{\int\dk \left(\chi_{A\bar{A}}^{(1)} + 4\chi_{AA}^{(2)}\right) n(1-n)}{4\int\dk \chi_{AA}^{(2)}n(1-n)}
\end{equation}
where we have expanded at zero-th order the argument of the integrals, which are of the form \eqref{eq:AandB} (with different counting functions). Using the explicit forms of $\chi_{A\bar{A}}^{(1)}$ and $\chi_{AA}^{(2)}$, we see that this becomes
\begin{equation}
   \log \mathcal{N}(t) \approx -\frac{1}{2}\left(\log \sigma_\tau^2 - \log\mathcal{X}_\tau\right) = \Delta S_A(\tau) - S_{\rm num}(\tau) 
\end{equation}
where we have used the notation of \eqref{eq:asymmetry}. This confirms that for $t=\tau$ the entanglement asymmetry appears explicitly in the evaluation. For $t>\tau$, as discussed above, it is rather tifficult to obtain explicit results, except for some specific regimes.

Finally, we can consider the extension to several measurements. As mentioned in the main text, since the measurement projects the state effectively on a symmetry sector, it is expected that the effect of all measurements from the second on will have significantly different effects compared to the first measurement. This is already captured by considering just two measurements. Focusing for simplicity on the regime $2\tau\leq t<\ell/2$, we have (performing implicitly the variable change $\lambda_{n,j} + \lambda_{n,j+1} \to \lambda_{n,j}$)
\begin{multline}
   \mathcal{F}_\alpha(t,\{\lambda_{1,j},\lambda_{2,j}\}) = \int\frac{dk}{2\pi}\left(\chi^{(2,2)}_{AA} \sum_{j}  f_1(2\lambda_{1,j} +2\lambda_{2,j},k)+ \chi^{(2,2)}_{A\overline{A}}   f_\alpha(\sum_{j}[2\lambda_{1,j}+2\lambda_{2,j}],k)\right. \\ +\chi_{A\overline{A}}^{(2,1)} f_\alpha(\sum_j[2\lambda_{1,j}+\lambda_{2,j}],k) +  \left.\chi_{A\overline{A}}^{(1,1)} f_\alpha(\sum_j[\lambda_{1,j} + \lambda_{2,j}],k) +  \chi_{A\overline{A}}^{(0,1)} f_\alpha(\sum_j \lambda_{2,j},k)  \right) . \nonumber
\end{multline}
where the notation for the counting functions is the same which was used above, with the two indices referring to the two measurements. The saddle point equations for $p\to 1$ take the form
\begin{equation}
\label{eq:conditionssqueezed}
\begin{cases}
q_1 = \int \dk \left(2(\chi^{(2,2)}_{AA} +\chi^{(2,2)}_{A\overline{A}} )n_{2\lambda_{1}+2\lambda_{2}} + 2\chi_{A\overline{A}}^{(2,1)} n_{2\lambda_{1}+\lambda_{2}} + \chi_{A\overline{A}}^{(1,1)} n_{\lambda_{1}+\lambda_{2}}\right)\\
q_2 = \int \dk \left(2(\chi^{(2,2)}_{AA} +\chi^{(2,2)}_{A\overline{A}} ) n_{2\lambda_{1}+2\lambda_{2}} + \chi_{A\overline{A}}^{(2,1)} n_{2\lambda_{1}+\lambda_{2}} + \chi_{A\overline{A}}^{(1,1)} n_{\lambda_1 + \lambda_2} + \chi_{A\overline{A}}^{(0,1)} n_{\lambda_2}    \right)
\end{cases}
\end{equation}
where $\lambda_1$ and $\lambda_2$ are the ($j$-independent) saddle points for the two variables, and we have expressed compactly the various measurement-altered occupation functions as
\begin{equation}
    n_\xi = n_\xi(k) = \frac{ne^{\xi}}{ne^\xi+1-n} .
\end{equation}
While the possible values of $q_1$ are not constrained as in the symmetric case, subtracting the first equation from the second leads to 
\begin{equation}\label{eq:deltaq2squeezed}
    \Delta q_2 = q_2 - q_1 = \int \dk  \left(\chi_{A\overline{A}}^{(0,1)} n_{\lambda_2} - \chi_{A\overline{A}}^{(2,1)} n_{2\lambda_{1}+\lambda_{2}} \right) = \tau\int \dk |v_k| \left( n_{\lambda_2} - n_{2\lambda_{1}+\lambda_{2}} \right)  .
\end{equation}
where in the second equality we have used the expression of the counting functions in the small time regime. Noting that the argument of the bracket is bounded as
$
    -1\leq \left( n_{\lambda_2} - n_{2\lambda_{1}+\lambda_{2}} \right) \leq 1
$, this can be used to bound the variation of the charge
\begin{equation}
    -\frac{2\tau}{\pi} \leq \Delta q_2 \leq \frac{2\tau}{\pi}
\end{equation}
which is precisely the same time delay which was predicted for the symmetric case, thus confirming the natural intuition that after the first measurement the system gains some features of symmetric states.  
However, the final result is not of the form \eqref{eq:finalmultiplesymmetric}, since the state is still a collection of pairs of quasiparticles and not a collection of particle-hole excitations as in the symmetric states. 
This is already apparent from the evaluation of the saddle points, which as above can be performed by linearizing \eqref{eq:conditionssqueezed}, leading to
\begin{equation}
    \begin{cases}
        \Delta q_1 = q_1 - \overline{q} = \lambda_1 \sigma_\tau^2 + \lambda_2 \sigma_{2\tau}^2, \\
        \Delta q_2 = q_2-q_1 = \lambda_1(2\sigma_\tau^2-\sigma_\infty^2).
    \end{cases}
\end{equation}
This solution makes apparent that the main difference with respect to the symmetric case arises from the presence of the full pairs. In fact, in the limit $\tau \to \infty$ full pairs are completely absent at the time of the measurements, and the saddle point equations reduce to 
\begin{equation}
    \begin{cases}
       \lambda_1+ \lambda_2 = \frac{\Delta q_1}{\sigma_\infty^2}, \\
       \lambda_1=\frac{\Delta q_2}{\sigma_\infty^2}.
    \end{cases}
\end{equation}
which is identical to the symmetric solution.

Also the final solution will depend on several different contributions depending on the number of quasiparticles which experienced each measurements. For instance, in the case of $2$ measurements  
\beqa
   S_A(t|\tau,q_1,q_2) &=& S_A(t)+ \int \dk \chi_{A\overline{A}}^{(2,2)}(s[n_{2\lambda_1 + 2\lambda_2}(k)]-s[n(k)]) 
   \nonumber\\ &+&
   \int \dk \chi_{A\overline{A}}^{(2,1)}(s[n_{2\lambda_1 + \lambda_2}(k)]-s[n(k)]) \nonumber\\ &+&  \int \dk \chi_{A\overline{A}}^{(1,1)}(s[n_{\lambda_1+\lambda_2}(k)]-s[n(k)]) 
    \nonumber\\ &+&
    \int \dk \chi_{A\overline{A}}^{(0,1)}( s[n_{\lambda_2}(k)]-s[n(k)]) + \log \mathcal{N}
\eeqa
In general, the final solution for arbitrary $t$ and arbitrary number of measurements will contain all possibility of combinations, 
\begin{equation}
    S_A(t|\tau,\{q_i\}) = S_A(t) + \sum_{\vec l} \int \dk \chi_{A\overline{A}}^{({\vec l})} \left(s[n_{\vec{l} \cdot\vec{\lambda} }(k)] -s[n(k)]\right) + \log \mathcal{N}(t)
\end{equation}
where $\vec l$ are the strings which can be contructed of 0,1,2
 which represent how many particles of each pair were inside the interval at the time of each measurement.

\section{Exact solution of the Néel state}
\label{appB}
In this appendix we provide an exact solution for the integral \eqref{eq:charge_moments} in the case of the Néel state \eqref{eq:neelstate}. We will then show that this exact result confirms the saddle point prediction in the hydrodynamic regime, thus confirming our results. Consider the case of a single measurement with $\tau <\ell/2$, in which all counting functions simplify significantly. Exploiting the momentum-independence of the occupation functions, $n=1/2$, which implies that all the functions $f_{\alpha}(\alpha)$ are themselves independent on momentum, and the property $\int \dk \chi_{A\overline{A}}^{(1)}=\frac{4\tau}{\pi}$ in this regime, the full moments can be expressed as
\begin{multline}
\label{eq:momentsforexactsolution}
    \tr_A \Big[\{\rho_A(t|\tau,q)\}^\alpha\Big]= \frac{1}{p(\tau,q\,)^\alpha} \int \frac{{\rm d}\vec{\lambda}}{(2\pi)^{\alpha}}
    \exp\left\{-i\Delta q\sum_j \lambda_j 
    \right.  \\ \left. 
+ \frac{4\tau}{\pi}  \left(\log\left(e^{i\sum_j\lambda_j}+1 \right)-\alpha\log 2 \right)
 + 4\hspace{0.1cm}\frac{t-\tau}{\pi} (1-\alpha )\log2 -i\frac{2\tau}{\pi}\sum_j\lambda_j \right\},
\end{multline}
where we have expressed compactly ${\rm d}\vec\lambda = \prod_j {\rm d}\lambda_j$ .
From this rewriting we can already see some structure of the solution arising, since $4t/\pi (1-\alpha)\log 2= (1-\alpha)S^{(\alpha)}(t)$ at short times. This leads to the rewriting
\begin{equation}
   S^{(\alpha}_A(t|q)=  S_A^{(\alpha)}(t)  +\frac{1}{1-\alpha}\log\left( \frac{ \int \frac{{\rm d}\vec{\lambda}}{(2\pi)^{\alpha}} \exp{\left(-i \Delta q\sum_j \lambda_j\right) }\left[ \cosh\left(i\sum_j \frac{\lambda_j}{2}\right)\right]^{\frac{4\tau}{\pi}} } 
 { \left(\int \frac{{\rm d}\alpha}{2\pi} \exp{\left(-i \Delta q\alpha \right) \left[\cosh(i\alpha/2)\right]^{\frac{4\tau}{\pi}}}\right)^{\alpha}} \right). \nonumber
\end{equation}
Note that, since the hyperbolic cosine arises from the exponential of a logarithm of an exponential, its argument is defined modulo $2\pi$, with the fundamental range of definition between $-\pi$ and $\pi$. 
 As a check, it is easy to convince oneself that substituting the saddle point \eqref{eq:alphaNéel} and expanding for small $\Delta q/\tau$ leads again to \eqref{eq:neelcorrection}. This can be further simplified to 
 \begin{equation}
 \label{eq:obvious}
     S^{(\alpha}_A(t|q) =   S_A^{(\alpha)}(t)  +\frac{1}{1-\alpha}\log\left( \frac{ \int \frac{{\rm d} \vec{\lambda}}{(2\pi)^{\alpha}} \exp{\left(-i \Delta q\sum_j \lambda_j\right) }\left[ \cos\left(\sum_j \frac{\lambda_j}{2}\right)\right]^{\frac{4\tau}{\pi}} } 
 { \left(\int \frac{d\alpha}{2\pi} \exp{\left(-i \Delta q\alpha \right) \left[\cos(\alpha/2)\right]^{\frac{4\tau}{\pi}}}\right)^{\alpha}} \right) 
 \end{equation}
noting that the integral in the line above depends only on the sum of all $\alpha$ we can change variables, 
\begin{equation}
     S^{(\alpha}_A(t|q) =   S_A^{(\alpha)}(t)  +\frac{1}{1-\alpha}\log\left( \frac{ \int_{-\alpha\pi}^{\alpha\pi} \frac{{\rm d}\lambda}{2\pi} \exp{\left(-i \Delta q\lambda \right) }\left[ \cos\left(\lambda/2\right)\right]^{\frac{4\tau}{\pi}}V_{\alpha}(\lambda) } 
 { \left(\int_{-\pi}^{\pi} \frac{{\rm d}\lambda}{2\pi} \exp{\left(-i \Delta q\lambda \right) \left[\cos(\lambda/2)\right]^{\frac{4\tau}{\pi}}}\right)^\alpha} \right) 
\end{equation}
where $V_{\alpha}(\lambda)$ is a volume factor which enters integrating out all other variables. This is the volume of the surface $\sum_j\lambda_j=\lambda$, which can be expressed in terms of the Irwin-Hall distribution:
\begin{equation}
V_{\alpha}(\lambda)=\rho_{IH}^{\alpha}\left(\frac{\lambda + \alpha\pi}{2\pi}\right),\hspace{0.5cm}  \rho_{IH}^{\alpha}(\lambda) = \frac{1}{(\alpha-1)!} \sum_{k=0}^{\lfloor \lambda\rfloor}(-1)^k \begin{pmatrix}
    \alpha \\ k
\end{pmatrix} (\lambda-k)^{\alpha-1}.
\end{equation}
From the structure of the distribution one can convince oneself that it satisfies the fundamental property
\begin{equation}
\begin{cases}
    V_\alpha(\lambda) + V_\alpha(2\pi-\lambda) +  V_\alpha(2\pi+\lambda)  + \dots + V_\alpha(\alpha\pi-\lambda)=1 \hspace{0.5cm} \alpha \text{ even}\vspace{0.3cm} \\
     V_\alpha(\lambda) + V_\alpha(2\pi-\lambda) +  V_\alpha(2\pi+\lambda) + \dots V_\alpha((\alpha-1)\pi + \lambda) =1 \hspace{0.5 cm} \alpha \text{ odd }.
 \end{cases}
\end{equation}
This allows by a simple coordinate change to obtain (note that this relation would be naturally expected from the periodicity of the integrand already in \eqref{eq:obvious})
\begin{equation}
   \int_{-\alpha\pi}^{\alpha\pi} \frac{{\rm d}\lambda}{2\pi} \exp{\left(-i \Delta q\lambda \right) }\left[ \cos\left(\lambda/2\right)\right]^{\frac{4\tau}{\pi}}V_\alpha(\lambda) = \int_{-\pi}^{\pi} \frac{{\rm d}\lambda}{2\pi} \exp{\left(-i \Delta q\lambda \right) \left[\cos(\lambda/2)\right]^{\frac{4\tau}{\pi}}}.
\end{equation}
This allows for a complete simplification of the R\'enyi index, 
\begin{equation}
    S^{(\alpha}_A(t|q) =S_A^{(\alpha)}(t) + \log \left(\int_{-\pi}^{\pi} \frac{{\rm d}\lambda}{2\pi} \exp{\left(-i \Delta q\lambda \right) \left[\cos(\lambda/2)\right]^{\frac{4\tau}{\pi}}}\right).
\end{equation}
The integral can be solved by making use of the Euler Beta function \cite{abramowitz1948handbook,Andrews_Askey_Roy_1999}, as was already noted in  a similar setting in \cite{parez2021quasiparticle},
\begin{equation}
    \int_{-\pi}^{\pi} \frac{{\rm d}\lambda}{2\pi} \exp{\left(-i \Delta q\lambda \right) \left[\cos(\lambda/2)\right]^{\frac{4\tau}{\pi}}} = \frac{2}{2^x x B(\frac{x+y+1}{2},\frac{x-y+1}{2})}
\end{equation}
where $x=\frac{4\tau}{\pi}+1$,  $y=2\Delta q$, and $B$ is the beta function, defined in terms of the standard gamma function as
\begin{equation}
    B(x,y) = \frac{\Gamma(x) \Gamma(y)}{\Gamma(x+y)}.
\end{equation}
This leads to the exact solution for all R\'enyi entropies, which takes the rather simple form
\begin{equation}
    S^{(\alpha)}_A(t|\tau,q) =S_{A}^{(\alpha)} (t) - \log \left( 2^{\frac{4\tau}{\pi}} \left(\frac{4\tau}{\pi}+1\right) B\left(\frac{2\tau}{\pi} + \Delta q +1 ,\frac{2\tau}{\pi} - \Delta q +1\right)\right),
\end{equation}
which can be further simplified exploiting the definition in terms of Gamma functions and their properties, 
\begin{equation}
     S^{(\alpha)}_A(t|\tau,q) =S_{A}^{(\alpha)} (t) - \log \left( 2^{\frac{4\tau}{\pi}} \frac{\Gamma(\frac{2\tau}{\pi}+\Delta q +1)\Gamma(\frac{2\tau}{\pi}-\Delta q +1)}{\Gamma(\frac{4\tau}{\pi}+1)}\right).
\end{equation}
In the ballistic regime, for which all quantities appearing are assumed to be large, this can be simplified using Stirling formula,
\begin{equation}
    \Gamma(z+1) \approx \sqrt{2\pi z}\hspace{0.2cm}(z/e)^z \hspace{0.5cm} z\gg1,
\end{equation}
which allows to directly re-obtain and confirm the saddle point solution:
\beqa
 S^{(\alpha)}_A(t|\tau,q)&\approx&S_{A}^{(\alpha)} (t) - (\frac{2\tau}{\pi}+\Delta q) \log\left(1+\frac{\pi\Delta q}{2\tau}\right) -(\frac{2\tau}{\pi}-\Delta q) \log\left(1-\frac{\pi\Delta q}{2\tau}\right) \nonumber\\
 &-& \log \left[\sqrt{\frac{\pi^2}{2\tau}\left(\frac{4\tau^2}{\pi^2} -\Delta q^2\right)}\right]. \label{eq:logcorr}
\eeqa 
In the first line we recognize precisely the term $ s[ n_{\Delta q}(k)]-\log 2$ which appears in \eqref{eq:sqminuslog2}, while the second term is a logarithmic correction,
\begin{equation}
    \delta S^{(\alpha)}_A(t|q)   =  ( s[ n_{\Delta q}(k)]-\log2) \int \dk \chi_{A\overline{A}}^{(1)} (t,k) -\log \left[\sqrt{\frac{\pi^2}{2\tau}\left(\frac{4\tau^2}{\pi^2} -\Delta q^2\right)}\right].
\end{equation}
The arising of the $\log\tau$ term can be related to the $1/\sqrt{2\pi \mathcal{F}''(x)}$ appearing in the saddle point evaluation. In general, for the $\alpha$-th R\'enyi entropy this would be a term arising from 
$\sqrt{\frac{(2\pi  \mathcal F_1''(\lambda_1))^{\alpha}}{(2\pi \mathcal F_{\alpha}''(\lambda_{\alpha}))}}$. Indeed, since $\mathcal F_{\alpha}''(\lambda_{\alpha}) = \frac{4\tau}{\pi} \frac{e^{\alpha\lambda_{\alpha}}}{(e^{\alpha\lambda_{\alpha}} +1)^2}$, and recalling that $\lambda_1 = \alpha \lambda_{\alpha}$ in the Néel state, the total corrective term becomes
\beqa
    \sqrt{\frac{(2\pi \mathcal F_1''(\lambda_1))^{\alpha}}{(2\pi \mathcal F_{\alpha}''(\lambda_{\alpha}))}} &=& \sqrt{\left(\frac{4\tau}{\pi}\right)^{\alpha-1} (2\pi)^{\alpha-1} \left(\frac{e^{\lambda_1}}{(1+e^{\lambda_1})^2}\right)^{\alpha-1}} \\
    &=&\sqrt{(8\tau)^{\alpha-1}  \left(\frac{1}{4}-\frac{\pi^2 \Delta q^2}{16\tau}\right)^{\alpha-1}}
\eeqa
which gives precisely the corrective term appearing in \eqref{eq:logcorr}. 

We see that at small times our problem is related closely to the problem of evaluating the symmetry resolution of the entropies. However, allowing for larger times leads to differences, which arise especially because of the structures of the counting functions. For $t>\ell/2$, in particular, the integrand would not be simply be a function of $\sum_j\lambda_j$, which was crucial in this derivation. In fact, at arbitrary times, labeling
\begin{equation}
    J_t = \int \dk \min(2|v_k|t,\ell)
\end{equation}
the moments can be expressed in a way which does not allow for an exact solution,
\beqa
    \Tr[\{\rho^{\alpha}_{A}(t|{q_1,q_2})\}] &=& \frac{e^{ (1-\alpha) \log 2J_{t-\tau} - \alpha \log2 J_\tau}}{\mathcal{N}^{\alpha}} \int \frac{{\rm d} \vec{\lambda}}{(2\pi)^{\alpha}} \exp\left\{-i\sum_j \lambda_j \Delta q \right\} \left(e^{i\sum_j \lambda_j} +1\right)^{J_t-J_{t-\tau}} \nonumber \\
    &\cdot&  \exp\left\{-\frac{1}{2}J_\tau \sum_j \lambda_j\right\} \prod_j (e^{i\lambda_j} +1)^{J_\tau+J_{t-\tau}-J_t}
\eeqa
which for $\tau,t<\ell/2$ reduces to the above. The other limit which can be easily solved is $\tau$ fixed and $t\to \infty$, in which $J_t \approx J_{t-\tau}$. In this regime the integral reduces to $\alpha$ copies of a 1-dimensional integral,
\begin{equation}
     \Tr[\{\rho^{\alpha}_{A}(t|{q})\}] = \frac{e^{ (1-\alpha) \log 2J_{t-\tau} - \alpha \log2 J_\tau}}{\mathcal{N}^{\alpha}} \left(\int \frac{{\rm d}\lambda}{(2\pi)} \exp\left(-i\lambda \Delta q\right) (2\cos(\lambda/2))^{J_\tau}\right)^{\alpha}
\end{equation}
when evaluating the R\'enyi entropies, the full integral would be identical to the denominator arising from $\Tr[\rho_{A,q}]^{\alpha}$. Hence in this regime we find exactly as expected
\begin{equation}
    S^{(\alpha)}_{A}(t\gg \ell|\tau,q) =  S^{(\alpha)}_{A}(t\gg \ell) = J_{\infty}\log 2 = \ell \log 2 
\end{equation}
which implies that the effect of the measurement is simply washed away at long times, and the solution saturates to the same value as the unperturbed one.

The exact solution can be extended also to several measurements. Considering two measurements, \eqref{eq:momentsforexactsolution} gets upgraded to 
\begin{multline}
   \Tr[\{\rho^{\alpha}_{A}(t|{q_1,q_2})\}]= \frac{1}{\mathcal{N}^{\alpha}} \int \frac{{\rm d} \vec{\lambda}_1}{(2\pi)^{\alpha}}\frac{{\rm d} \vec{\lambda}_2}{(2\pi)^{\alpha}}\exp\left\{-i \left(q_1-\frac{\ell}{2}\right)\sum_j \lambda_{1,j} -i \left(q_2-\frac{\ell}{2}\right)\sum_j \lambda_{2,j} \right. \\
+ \frac{4\tau}{\pi} \left(\log\left(e^{i\sum_j(\lambda_{1,j}+\lambda_{2,j})}+1 \right)-\alpha\log 2 \right) + \frac{4\tau}{\pi} \left(\log\left(e^{i\sum_j\lambda_{2,j}}+1 \right)-\alpha\log 2 \right)  \\
+ \left. 4\hspace{0.1cm}\frac{t-2\tau}{\pi} (1-\alpha )\log2 -i\frac{2\tau}{\pi}\sum_j\lambda_{1,j}  -i\frac{4\tau}{\pi}\sum_j\lambda_{2,j} \right\}.
\end{multline}
Similar manipulations allow to reduce the number of integrals
\begin{multline}
    \Tr[\{\rho^{\alpha}_{A}(t|{q_1,q_2})\}]= \\ \frac{e^{\frac{4t}{\pi}(1-\alpha)\log2}}{\mathcal{N}^{\alpha}} \int \frac{{\rm d}\lambda_1}{2\pi}\frac{{\rm d}\lambda_2}{2\pi}e^{\left(-i (q_1-\frac{\ell}{2})\lambda_1 -i (q_2-\frac{\ell}{2})\lambda_2 \right) }
\left[\cos(\frac{\lambda_2}{2}) \cos(\frac{\lambda_1 + \lambda_2}{2})\right]^{4\tau/\pi}. \end{multline}
which by a simple change of variables simply reduces to a product of Beta functions,
\beqa
  \Tr[\{\rho^{\alpha}_{A}(t|{q_1,q_2})\}]&=& \frac{e^{\frac{4t}{\pi}(1-\alpha)\log2}}{\mathcal{N}^{\alpha}} \int \frac{{\rm d}\gamma}{2\pi}\frac{{\rm d}\beta}{2\pi}e^{\left(-i \Delta q_1\gamma -i \Delta q_2 \beta \right) }
\left[\cos(\frac{\beta}{2}) \cos(\frac{\gamma}{2})\right]^{4\tau/\pi}
\eeqa
where we have reintroduced the notation $\Delta q_i = q_{i}-q_{i-1}$. The R\'enyi entropies are then simply
\begin{equation}
    S_{A}^{(\alpha)}(t|q_1,q_2)(t) = S_A^{(\alpha)}(t) - \log\left( \prod_{i=1}^{2} 2^{\frac{4\tau}{\pi}} \left(\frac{4\tau}{\pi}+1\right) B\left(\frac{2\tau}{\pi} + \Delta q_i +1 ,\frac{2\tau}{\pi} - \Delta q_i +1\right)
 \right)
\end{equation}
which generalizes immediately to the case  of several measurements by extending the product inside the logarithm. Expanding the Beta function as shown for the single measurement case leads to 
\beqa
     S_{A}^{(\alpha)}(t|\tau,\{q_i\}) = S_A^{(\alpha)}(t)  + \sum_i \frac{4\tau}{\pi} (s[n_{\Delta q_i}(k)]-\log2) - \sum_i  \log \left[\sqrt{\frac{\pi^2}{2\tau}\left(\frac{4\tau^2}{\pi^2} -\Delta q_i^2\right)} \right].
\eeqa
Note that also in the exact solution the result can be expressed in terms of the successive variations of charge in the different measurements, $\Delta q_i$.

  \section{Connection with the full counting statistics}
  In the main text we commented on the relation of our discussion with previously known results regarding the full counting statistics of the $U(1)$ charge. In fact,  
 the Fourier representation of the projectors \eqref{eq:projector_fourier} implies a close relation between our measurement protocol, and the evaluation of the full counting statistics, which was performed for symmetric states in \cite{bertini2023nonequilibrium} and for the symmetry breaking states in \cite{bertini_asymmetric_2024}. In fact, evaluating the full counting statistics of the charge essentially amounts to the calculation of $\mathcal{F}_1(t,\lambda)$ with the condition $t=\tau$. In this regime, the number of terms appearing in the integral reduces significantly, giving for the symmetry breaking state,
 \beqa
      \mathcal{F}_\alpha(\tau,\{\lambda_j,\lambda_j'\}) &=& \exp \int\frac{{\rm d}k}{2\pi}\left(\chi^{(2)}_{AA} \sum_{j}  f_1(2\lambda_j+2\lambda'_j,k) + \chi_{A\overline{A}}^{(1)} f_\alpha(\sum_j [\lambda_j+\lambda_j'],k) \right) . \nonumber
  \eeqa
  Substituting the explicit values of the counting functions, for $\tau < \ell/2$ leads to 
  \begin{equation}
    \mathcal{F}_\alpha(\tau,\{\lambda_j,\lambda_j'\}) =  \exp \int\frac{{\rm d}k}{2\pi}\left\{
      \frac{\ell}{2}\sum_{j}  f_1(2\lambda_j+2\lambda'_j,k) +  |v_k|\tau \left(2f_\alpha(\sum_j[\lambda_j+\lambda_j'],k) - \sum_jf_1(2\lambda_j+2\lambda'_j,k)\right) \right\} .\nonumber
  \end{equation}
  Taking the limit $\alpha \to 1$ and setting $\lambda + \lambda' = \beta$, this leads to the full counting statistics,
  \begin{equation}
    \mathcal{F}_1(\tau,\beta) = \frac{\ell}{2} \int \dk \log(n e^{2i\beta} +1-n ) + \int \dk |v_k| \tau  \log\left(\frac{(ne^{i\beta} +(1-n))^2}{ne^{2i\beta} +(1-n)}\right)
  \end{equation}
  which coincides with the result of \cite{bertini_asymmetric_2024} specialized to the free case: in particular, the first term represents the initial state contribution and the second term the genuine contribution arising from the crossed channel in the spacetime duality treatment.

  A similar analysis can be performed also in the symmetric case, leading to the corresponding expression for $\tau<\ell/2$,
  \beqa
       \mathcal{F}_1(\tau,\beta) &=& i\frac{\beta}{2} \int\dk (\ell - 2|v_k|\tau) + 2\tau \int \dk |v_k| \log(n e^{i\beta} + (1-n)) \\
       &=& i\beta q_0 + 2\tau\int \dk |v_k| \log(n e^{i\beta/2} + (1-n)e^{-i\beta/2} )
   \eeqa
   where as above $q_0 = \frac{\ell}{2} $. While very different from the symmetry breaking case, this is in agreement with the result of \cite{bertini2023nonequilibrium,parez2021quasiparticle} specialized to free theories.

\section{Measurements on different intervals}
\label{appC}
As it should be clear from the discussion in the main text, the simple structure of the presented formulation of the quasiparticle picture allows to deal quite easily with several types of generalizations of the problem that can be considered. 
The main modification that has to be applied is on the form of the counting functions, but the structure of the solution is left unvaried. It is interesting for example to consider the situation in which the measurement is not performed on the interval $A$, but on its complement $\overline{A}$. Considering the squeezed state and $\tau = 0$, the solution is again
\begin{multline}
\label{eq:appbtrace}
    \Tr[\{\rho^{\alpha}_{A}(t|{q})\}] =  \frac{1}{\mathcal{N}^p} \int \frac{{\rm d} \vec{\lambda} {\rm d} \vec{\lambda}'}{(2\pi)^{2^\alpha}} e^{-i \sum_j(\lambda_j + \lambda_j') q}\times\\ \times\exp \int \frac{dk}{2\pi}\left(\tilde \chi_{AA}^{(2)} \sum_{j}  f_1(\lambda_j' + \lambda_{j+1},k) + \tilde \chi_{A\overline{A}}^{(2)} f_\alpha(\sum_j[\lambda_j+\lambda_j'],k) + \tilde \chi_{A\overline{A}}^{(0)} f_\alpha(0,k)\right)        
\end{multline}
where the role of the counting function is different: since $\tilde \chi_{A\overline{A}}^{(1)} $ counts the pairs which originated in the complement and are shared at time $t$, while $\tilde \chi_{A\overline{A}}^{(2)} $ counts the pairs which originated in $A$ and are shared at time $t$, they are just exchanged with respect to the case of section \ref{sec:squeezed}:
\begin{equation}
    \begin{cases}
        \tilde \chi_{A\overline{A}}^{(2)} =\min(|v_k|t,\ell)\vspace{0.2cm}\\
        \tilde \chi_{A\overline{A}}^{(0)}=\max(|v_k|t,\ell) - \max(\ell-|v_k|t, |v_k|t) 
    \end{cases}
\end{equation}
hence their sum is still $\min(2|v_k|t,\ell)$. The other term instead counts all pairs which were created in the complement at t=0 and are not shared at time t, it is simply
\begin{equation}
    \tilde \chi_{AA} = \frac{L-\ell}{2}  -\tilde \chi_{A\overline{A}}^{(1)}
\end{equation}
where $L$ is the length of the total system.
The solution of the saddle point should now be familiar,
\beqa
     S_{A}(t|0,q) &=& \int \dk \tilde \chi_{A\overline{A}}^{(2)}s[n_{2\Delta q}(k)] +  \int \dk \tilde \chi_{A\overline{A}}^{(0)}s[n(k)] \\ &=& S_A(t) + \int \dk \tilde \chi_{A\overline{A}}^{(2)}(s[n_{2\Delta q}(k)]-s[n(k)])
\eeqa
Contrarily to the situations considered in the main text, here $\tilde \chi_{A\overline{A}}^{(2)}$ does not vanish at long times, and instead leads precisely to a saturation to a configuration in which the measurement dominates, as would be expected,
\begin{equation}
      S_{A}(t\gg \ell|0,q) = \ell \int \dk s[n_{2\Delta q}(k)]
\end{equation}
regarding the evaluation of the saddle point, the situation is exactly analogous to the case above, with the sole difference that $\tilde \chi_{A\overline{A}}^{(2)} +\tilde \chi_{AA}^{(2)} =\frac{L-\ell}{2}$, which implies that the solution will be the same with $\overline{q} \to L/2 - \overline{q} $.

Another interesting situation which can be considered is that in which the measurement is performed on a different interval $B$, which can be placed at some distance $d$ from the interval $A$ of which we evaluate the entropy. In order to apply the hydrodynamic framework, it is clear that all quantities involved has to be either large with respect to lattice spacing, or zero. Hence the two regimes which can be studied are that of adjacent intervals and that of large intervals placed at a distance which is comparable with their length; here we focus on the regime of separate intervals. In this situation, since the excitations have a maximum velocity $v_{max}=1$, there will be a light cone effect, and the effect of the measurement will be visible on the entropy only after a time $t=d$. Since \eqref{eq:appbtrace} is still valid, this effect is simply encoded in the counting functions: $\tilde \chi_{A\overline{A}}^{(1)} $ now counts pairs which originated in $B$ and at time t are shared between $A$ and its complement, and therefore is clearly zero for $t<d$. On the other hand, $\tilde \chi_{A\overline{A}}^{(2)} $
 counts pairs which originated outside of $B$ and are shared  between $A$ and its complement at time t, and therefore for $t<d$ it is simply given by $\min(2|v_k|t,\ell)$.
Hence, the solution for $t<d$ is unperturbed,
 \begin{equation} 
    S_{A}(t<d|0,q)= \int \dk \tilde \chi_{A\overline{A}}^{(1)}s[n_{2\Delta q}(k)]  + \int \dk \tilde \chi_{A\overline{A}}^{(2)}s[n(k)] = S_A(t)  .
\end{equation}
For $t\sim d$, the effect of the measurement then becomes relevant for a time spike and goes back to zero as $t \gg d,\ell,\ell_B$.

We see therefore that the extension of the discussion is extremely natural and allows us to consider several different situations, all leading to interesting physical phenomena. For example, even the extension to multipartite geometries could be straightforwardly considered.

\bibliographystyle{ytphys}
\bibliography{bibliography}
\end{document}

%% file: setup.tex
\usepackage[a4paper, total={5.8in, 9in}]{geometry}
\usepackage[indent,skip=4pt]{parskip}
\usepackage{mathtools}
\usepackage{amsmath,amssymb,amsfonts,amsthm,mathrsfs,mathtools,cases}
\usepackage{dsfont}
\usepackage{authblk}
\usepackage[colorlinks=true,linkcolor=blue, citecolor=blue, bookmarks]{hyperref}
\usepackage{cleveref}
\usepackage{microtype}
\usepackage{comment}
\usepackage{graphicx,adjustbox} 
\usepackage{subfig}
\usepackage{wrapfig}
\usepackage[font=small,labelfont=bf]{caption}
\usepackage[usenames,dvipsnames]{xcolor}  
\usepackage{booktabs}
\usepackage{braket}
\usepackage[nottoc]{tocbibind}
\usepackage{nicematrix}
\usepackage{tikz}
\usepackage{pgfplots}
\usepackage{cite}
\pgfplotsset{compat=1.18}
\usepackage{enumitem}

\usetikzlibrary{matrix}
\usetikzlibrary{decorations.markings}
\usetikzlibrary{patterns}
\usetikzlibrary{arrows.meta}

\numberwithin{equation}{section}

\newcommand{\dk}{\frac{{\rm d}k}{2\pi}}

\DeclareMathOperator{\tr}{tr}

\crefname{figure}{Fig.}{Figs.}
\crefname{equation}{Eq.}{Eqs.}
\crefname{section}{Sec.}{Secs.}
\crefname{appendix}{Appendix}{Appendices}
  
\colorlet{darkerblue}{MidnightBlue!20!black}
\colorlet{lightblue}{blue!70!white}

\hypersetup{
    colorlinks,
    linkcolor={Blue},
    citecolor={Blue},
    urlcolor={Blue},
    backref=true
}

\newcommand{\beq}{\begin{equation}}
\newcommand{\eeq}{\end{equation}}
\newcommand{\beqa}{\begin{eqnarray}}
\newcommand{\eeqa}{\end{eqnarray}}

\newcommand{\Tr}{\operatorname{Tr}}

\usepackage[framemethod=TikZ]{mdframed}


%% file: main.bbl
\providecommand{\href}[2]{#2}\begingroup\begin{thebibliography}{100}

\bibitem{measurement1}
T.~Basch{\'e}, S.~Kummer, and C.~Br{\"a}uchle, {\slshape Direct spectroscopic observation of quantum jumps of a single molecule,} \href{https://doi.org/10.1038/373132a0}{{\em Nature} {\bfseries 373} (1995) 132}.

\bibitem{Measurement2}
S.~Gleyzes, S.~Kuhr, C.~Guerlin, J.~Bernu, S.~Del{\'e}glise, U.~Busk~Hoff, M.~Brune, J.-M. Raimond, and S.~Haroche, {\slshape Quantum jumps of light recording the birth and death of a photon in a cavity,} \href{https://doi.org/10.1038/nature05589}{{\em Nature} {\bfseries 446} (2007) 297}.

\bibitem{measurement3}
R.~Vijay, D.~H. Slichter, and I.~Siddiqi, {\slshape Observation of quantum jumps in a superconducting artificial atom,} \href{https://link.aps.org/doi/10.1103/PhysRevLett.106.110502}{{\em Phys. Rev. Lett.} {\bfseries 106} (2011) 110502}.

\bibitem{measurement4}
C.~Noel, P.~Niroula, D.~Zhu, A.~Risinger, L.~Egan, D.~Biswas, M.~Cetina, A.~V. Gorshkov, M.~J. Gullans, D.~A. Huse, and C.~Monroe, {\slshape Measurement-induced quantum phases realized in a trapped-ion quantum computer,} \href{https://doi.org/10.1038/s41567-022-01619-7}{{\em Nature Physics} {\bfseries 18} (2022) 760}.

\bibitem{Li_2018}
Y.~Li, X.~Chen, and M.~P.~A. Fisher, {\slshape Quantum {Z}eno effect and the many-body entanglement transition,} \href{https://link.aps.org/doi/10.1103/PhysRevB.98.205136}{{\em Phys. Rev. B} {\bfseries 98} (2018) 205136}.

\bibitem{Biella2021manybodyquantumzeno}
A.~Biella and M.~Schir{\'{o}}, {\slshape Many-body quantum {Z}eno effect and measurement-induced subradiance transition,} \href{https://doi.org/10.22331/q-2021-08-19-528}{{\em {Quantum}} {\bfseries 5} (2021) 528}.

\bibitem{Li_2019}
Y.~Li, X.~Chen, and M.~P.~A. Fisher, {\slshape Measurement-driven entanglement transition in hybrid quantum circuits,} \href{http://dx.doi.org/10.1103/PhysRevB.100.134306}{{\em Physical Review B} {\bfseries 100} (2019) }.

\bibitem{Skinner_2019}
B.~Skinner, J.~Ruhman, and A.~Nahum, {\slshape Measurement-induced phase transitions in the dynamics of entanglement,} \href{https://link.aps.org/doi/10.1103/PhysRevX.9.031009}{{\em Phys. Rev. X} {\bfseries 9} (2019) 031009}.

\bibitem{Alberton_2021}
O.~Alberton, M.~Buchhold, and S.~Diehl, {\slshape Entanglement transition in a monitored free-fermion chain: From extended criticality to area law,} \href{https://link.aps.org/doi/10.1103/PhysRevLett.126.170602}{{\em Phys. Rev. Lett.} {\bfseries 126} (2021) 170602}.

\bibitem{plenio1998}
M.~B. Plenio and P.~L. Knight, {\slshape The quantum-jump approach to dissipative dynamics in quantum optics,} \href{https://link.aps.org/doi/10.1103/RevModPhys.70.101}{{\em Rev. Mod. Phys.} {\bfseries 70} (1998) 101}.

\bibitem{holographic_tensor_networks}
R.~Vasseur, A.~C. Potter, Y.-Z. You, and A.~W.~W. Ludwig, {\slshape Entanglement transitions from holographic random tensor networks,} \href{https://link.aps.org/doi/10.1103/PhysRevB.100.134203}{{\em Phys. Rev. B} {\bfseries 100} (2019) 134203}.

\bibitem{zabalo2020criticalproperties}
A.~Zabalo, M.~J. Gullans, J.~H. Wilson, S.~Gopalakrishnan, D.~A. Huse, and J.~H. Pixley, {\slshape Critical properties of the measurement-induced transition in random quantum circuits,} \href{https://link.aps.org/doi/10.1103/PhysRevB.101.060301}{{\em Phys. Rev. B} {\bfseries 101} (2020) 060301}.

\bibitem{gullans2020scalableprobes}
M.~J. Gullans and D.~A. Huse, {\slshape Scalable probes of measurement-induced criticality,} \href{https://link.aps.org/doi/10.1103/PhysRevLett.125.070606}{{\em Phys. Rev. Lett.} {\bfseries 125} (2020) 070606}.

\bibitem{rossinivicari2020}
D.~Rossini and E.~Vicari, {\slshape Measurement-induced dynamics of many-body systems at quantum criticality,} \href{https://link.aps.org/doi/10.1103/PhysRevB.102.035119}{{\em Phys. Rev. B} {\bfseries 102} (2020) 035119}.

\bibitem{Fan2021selforganized}
R.~Fan, S.~Vijay, A.~Vishwanath, and Y.-Z. You, {\slshape Self-organized error correction in random unitary circuits with measurement,} \href{https://link.aps.org/doi/10.1103/PhysRevB.103.174309}{{\em Phys. Rev. B} {\bfseries 103} (2021) 174309}.

\bibitem{nahum2021}
A.~Nahum, S.~Roy, B.~Skinner, and J.~Ruhman, {\slshape Measurement and entanglement phase transitions in all-to-all quantum circuits, on quantum trees, and in {L}andau-{G}insburg theory,} \href{https://link.aps.org/doi/10.1103/PRXQuantum.2.010352}{{\em PRX Quantum} {\bfseries 2} (2021) 010352}.

\bibitem{Ware2021}
B.~Ware and R.~Vasseur, {\slshape Measurements make the phase,} \href{https://doi.org/10.1038/s41567-020-01131-w}{{\em Nature Physics} {\bfseries 17} (2021) 298}.

\bibitem{li2021conformal}
Y.~Li, X.~Chen, A.~W.~W. Ludwig, and M.~P.~A. Fisher, {\slshape Conformal invariance and quantum nonlocality in critical hybrid circuits,} \href{https://link.aps.org/doi/10.1103/PhysRevB.104.104305}{{\em Phys. Rev. B} {\bfseries 104} (2021) 104305}.

\bibitem{Sierant2022dissipativefloquet}
P.~Sierant, G.~Chiriac{\`{o}}, F.~M. Surace, S.~Sharma, X.~Turkeshi, M.~Dalmonte, R.~Fazio, and G.~Pagano, {\slshape Dissipative {F}loquet dynamics: from steady state to measurement induced criticality in trapped-ion chains,} \href{https://doi.org/10.22331/q-2022-02-02-638}{{\em {Quantum}} {\bfseries 6} (2022) 638}.

\bibitem{Zabalo2022operatorscaling}
A.~Zabalo, M.~J. Gullans, J.~H. Wilson, R.~Vasseur, A.~W.~W. Ludwig, S.~Gopalakrishnan, D.~A. Huse, and J.~H. Pixley, {\slshape Operator scaling dimensions and multifractality at measurement-induced transitions,} \href{https://link.aps.org/doi/10.1103/PhysRevLett.128.050602}{{\em Phys. Rev. Lett.} {\bfseries 128} (2022) 050602}.

\bibitem{vidal_2003}
G.~Vidal, J.~I. Latorre, E.~Rico, and A.~Kitaev, {\slshape Entanglement in quantum critical phenomena,} \href{https://link.aps.org/doi/10.1103/PhysRevLett.90.227902}{{\em Phys. Rev. Lett.} {\bfseries 90} (2003) 227902}.

\bibitem{Plenio2005AnIT}
M.~B. Plenio and S.~Virmani, {\slshape An introduction to entanglement measures,} \href{https://api.semanticscholar.org/CorpusID:7131013}{{\em Quantum Inf. Comput.} {\bfseries 7} (2005) 1}.

\bibitem{damico2008}
L.~Amico, R.~Fazio, A.~Osterloh, and V.~Vedral, {\slshape Entanglement in many-body systems,} \href{https://link.aps.org/doi/10.1103/RevModPhys.80.517}{{\em Rev. Mod. Phys.} {\bfseries 80} (2008) 517}.

\bibitem{horodecki2009}
R.~Horodecki, P.~Horodecki, M.~Horodecki, and K.~Horodecki, {\slshape Quantum entanglement,} \href{https://link.aps.org/doi/10.1103/RevModPhys.81.865}{{\em Rev. Mod. Phys.} {\bfseries 81} (2009) 865}.

\bibitem{Pasquale_Calabrese_2004}
P.~Calabrese and J.~Cardy, {\slshape Entanglement entropy and quantum field theory,} \href{https://dx.doi.org/10.1088/1742-5468/2004/06/P06002}{{\em J. Stat. Mech.} {\bfseries 2004} (2004) P06002}.

\bibitem{Calabrese_2009}
P.~Calabrese and J.~Cardy, {\slshape Entanglement entropy and conformal field theory,} \href{https://dx.doi.org/10.1088/1751-8113/42/50/504005}{{\em J. Phys. A} {\bfseries 42} (2009) 504005}.

\bibitem{Calabrese_2009_1}
P.~Calabrese, J.~Cardy, and B.~Doyon, {\slshape Entanglement entropy in extended quantum systems,} \href{https://dx.doi.org/10.1088/1751-8121/42/50/500301}{{\em J. Phys. A} {\bfseries 42} (2009) 500301}.

\bibitem{quench1}
P.~Calabrese and J.~Cardy, {\slshape Time dependence of correlation functions following a quantum quench,} \href{https://link.aps.org/doi/10.1103/PhysRevLett.96.136801}{{\em Phys. Rev. Lett.} {\bfseries 96} (2006) 136801}.

\bibitem{quench2}
P.~Calabrese and J.~Cardy, {\slshape Evolution of entanglement entropy in one-dimensional systems,} \href{https://dx.doi.org/10.1088/1742-5468/2005/04/P04010}{{\em J. Stat. Mech.} (2005) P04010}.

\bibitem{Deutsch1991}
J.~M. Deutsch, {\slshape Quantum statistical mechanics in a closed system,} \href{https://link.aps.org/doi/10.1103/PhysRevA.43.2046}{{\em Phys. Rev. A} {\bfseries 43} (1991) 2046}.

\bibitem{srednicki1}
M.~Srednicki, {\slshape Chaos and quantum thermalization,} \href{https://link.aps.org/doi/10.1103/PhysRevE.50.888}{{\em Phys. Rev. E} {\bfseries 50} (1994) 888}.

\bibitem{Rigol:2007juv}
M.~Rigol, V.~Dunjko, and M.~Olshanii, {\slshape {Thermalization and its mechanism for generic isolated quantum systems},} \href{http://dx.doi.org/10.1038/nature06838}{{\em Nature} {\bfseries 452} (2008) 854}.

\bibitem{polkovnikov}
A.~Polkovnikov, K.~Sengupta, A.~Silva, and M.~Vengalattore, {\slshape Colloquium: Nonequilibrium dynamics of closed interacting quantum systems,} \href{https://link.aps.org/doi/10.1103/RevModPhys.83.863}{{\em Rev. Mod. Phys.} {\bfseries 83} (2011) 863}.

\bibitem{DAlessio:2015qtq}
L.~D'Alessio, Y.~Kafri, A.~Polkovnikov, and M.~Rigol, {\slshape {From quantum chaos and eigenstate thermalization to statistical mechanics and thermodynamics},} \href{http://dx.doi.org/10.1080/00018732.2016.1198134}{{\em Adv. Phys.} {\bfseries 65} (2016) 239}.

\bibitem{Gogolin_2016}
C.~Gogolin and J.~Eisert, {\slshape Equilibration, thermalisation, and the emergence of statistical mechanics in closed quantum systems,} \href{https://dx.doi.org/10.1088/0034-4885/79/5/056001}{{\em Rep. Prog. Phys.} {\bfseries 79} (2016) 056001}.

\bibitem{ares2025simpler}
F.~Ares, C.~Rylands, and P.~Calabrese, {\slshape A simpler probe of the quantum {M}pemba effect in closed systems,} \href{http://arxiv.org/abs/2507.05946}{{ arXiv:2507.05946}}.

\bibitem{Collura_2014}
M.~Collura, M.~Kormos, and P.~Calabrese, {\slshape Stationary entanglement entropies following an interaction quench in 1d {Bose} gas,} \href{https://dx.doi.org/10.1088/1742-5468/2014/01/P01009}{{\em J. Stat. Mech.} {\bfseries 2014} (2014) P01009}.

\bibitem{CalabreseLN}
P.~Calabrese, {\slshape {Entanglement spreading in non-equilibrium integrable systems},} \href{https://scipost.org/10.21468/SciPostPhysLectNotes.20}{{\em SciPost Phys. Lect. Notes} (2020) 20}.

\bibitem{chan2019}
A.~Chan, R.~M. Nandkishore, M.~Pretko, and G.~Smith, {\slshape Unitary-projective entanglement dynamics,} \href{https://link.aps.org/doi/10.1103/PhysRevB.99.224307}{{\em Phys. Rev. B} {\bfseries 99} (2019) 224307}.

\bibitem{measurement_induced_criticality}
C.-M. Jian, Y.-Z. You, R.~Vasseur, and A.~W.~W. Ludwig, {\slshape Measurement-induced criticality in random quantum circuits,} \href{https://link.aps.org/doi/10.1103/PhysRevB.101.104302}{{\em Phys. Rev. B} {\bfseries 101} (2020) 104302}.

\bibitem{choi2020}
S.~Choi, Y.~Bao, X.-L. Qi, and E.~Altman, {\slshape Quantum error correction in scrambling dynamics and measurement-induced phase transition,} \href{https://link.aps.org/doi/10.1103/PhysRevLett.125.030505}{{\em Phys. Rev. Lett.} {\bfseries 125} (2020) 030505}.

\bibitem{Purifications}
M.~J. Gullans and D.~A. Huse, {\slshape Dynamical purification phase transition induced by quantum measurements,} \href{https://link.aps.org/doi/10.1103/PhysRevX.10.041020}{{\em Phys. Rev. X} {\bfseries 10} (2020) 041020}.

\bibitem{ippoliti_2020}
M.~Ippoliti, M.~J. Gullans, S.~Gopalakrishnan, D.~A. Huse, and V.~Khemani, {\slshape Entanglement phase transitions in measurement-only dynamics,} \href{https://link.aps.org/doi/10.1103/PhysRevX.11.011030}{{\em Phys. Rev. X} {\bfseries 11} (2021) 011030}.

\bibitem{bao2020}
Y.~Bao, S.~Choi, and E.~Altman, {\slshape Theory of the phase transition in random unitary circuits with measurements,} \href{https://link.aps.org/doi/10.1103/PhysRevB.101.104301}{{\em Phys. Rev. B} {\bfseries 101} (2020) 104301}.

\bibitem{turkeshi2021infzeroclicks}
X.~Turkeshi, A.~Biella, R.~Fazio, M.~Dalmonte, and M.~Schir\'o, {\slshape Measurement-induced entanglement transitions in the quantum {Ising} chain: From infinite to zero clicks,} \href{https://link.aps.org/doi/10.1103/PhysRevB.103.224210}{{\em Phys. Rev. B} {\bfseries 103} (2021) 224210}.

\bibitem{turkeshi20203d}
X.~Turkeshi, R.~Fazio, and M.~Dalmonte, {\slshape Measurement-induced criticality in $(2+1)$-dimensional hybrid quantum circuits,} \href{https://link.aps.org/doi/10.1103/PhysRevB.102.014315}{{\em Phys. Rev. B} {\bfseries 102} (Jul, 2020) 014315}.

\bibitem{Altland_2022}
A.~Altland, M.~Buchhold, S.~Diehl, and T.~Micklitz, {\slshape Dynamics of measured many-body quantum chaotic systems,} \href{https://link.aps.org/doi/10.1103/PhysRevResearch.4.L022066}{{\em Phys. Rev. Res.} {\bfseries 4} (2022) L022066}.

\bibitem{murciano_measurement_criticality}
S.~Murciano, P.~Sala, Y.~Liu, R.~S.~K. Mong, and J.~Alicea, {\slshape Measurement-altered ising quantum criticality,} \href{https://link.aps.org/doi/10.1103/PhysRevX.13.041042}{{\em Phys. Rev. X} {\bfseries 13} (2023) 041042}.

\bibitem{cao_2019}
X.~Cao, A.~Tilloy, and A.~D. Luca, {\slshape {Entanglement in a fermion chain under continuous monitoring},} \href{https://scipost.org/10.21468/SciPostPhys.7.2.024}{{\em SciPost Phys.} {\bfseries 7} (2019) 024}.

\bibitem{Turkeshi2022}
X.~Turkeshi, L.~Piroli, and M.~Schir\'o, {\slshape Enhanced entanglement negativity in boundary-driven monitored fermionic chains,} \href{https://link.aps.org/doi/10.1103/PhysRevB.106.024304}{{\em Phys. Rev. B} {\bfseries 106} (2022) 024304}.

\bibitem{coppola2022}
M.~Coppola, E.~Tirrito, D.~Karevski, and M.~Collura, {\slshape Growth of entanglement entropy under local projective measurements,} \href{https://link.aps.org/doi/10.1103/PhysRevB.105.094303}{{\em Phys. Rev. B} {\bfseries 105} (2022) 094303}.

\bibitem{entanglement_transition_quasiparticles}
X.~Turkeshi, M.~Dalmonte, R.~Fazio, and M.~Schir\`o, {\slshape Entanglement transitions from stochastic resetting of non-hermitian quasiparticles,} \href{https://link.aps.org/doi/10.1103/PhysRevB.105.L241114}{{\em Phys. Rev. B} {\bfseries 105} (2022) L241114}.

\bibitem{tirrito2022measindising}
E.~Tirrito, A.~Santini, R.~Fazio, and M.~Collura, {\slshape Full counting statistics as probe of measurement-induced transitions in the quantum {Ising} chain,} \href{https://scipost.org/10.21468/SciPostPhys.15.3.096}{{\em SciPost Phys.} {\bfseries 15} (2023) 096}.

\bibitem{turkeshi2023}
X.~Turkeshi and M.~Schir\'o, {\slshape Entanglement and correlation spreading in non-hermitian spin chains,} \href{https://link.aps.org/doi/10.1103/PhysRevB.107.L020403}{{\em Phys. Rev. B} {\bfseries 107} (2023) L020403}.

\bibitem{Romito1}
C.~Y. Leung, D.~Meidan, and A.~Romito, {\slshape Theory of free fermions dynamics under partial postselected monitoring,} \href{https://link.aps.org/doi/10.1103/PhysRevX.15.021020}{{\em Phys. Rev. X} {\bfseries 15} (2025) 021020}.

\bibitem{Romito2}
G.~Kells, D.~Meidan, and A.~Romito, {\slshape {Topological transitions in weakly monitored free fermions},} \href{https://scipost.org/10.21468/SciPostPhys.14.3.031}{{\em SciPost Phys.} {\bfseries 14} (2023) 031}.

\bibitem{rajabpour2015}
M.~A. Rajabpour, {\slshape Post-measurement bipartite entanglement entropy in conformal field theories,} \href{https://link.aps.org/doi/10.1103/PhysRevB.92.075108}{{\em Phys. Rev. B} {\bfseries 92} (2015) 075108}.

\bibitem{Rajabpour_2016}
M.~A. Rajabpour, {\slshape Entanglement entropy after a partial projective measurement in 1+1 dimensional conformal field theories: exact results,} \href{https://dx.doi.org/10.1088/1742-5468/2016/06/063109}{{\em J. Stat. Mech} {\bfseries 2016} (2016) 063109}.

\bibitem{Lin2023probingsign}
C.-J. Lin, W.~Ye, Y.~Zou, S.~Sang, and T.~H. Hsieh, {\slshape Probing sign structure using measurement-induced entanglement,} \href{https://doi.org/10.22331/q-2023-02-02-910}{{\em {Quantum}} {\bfseries 7} (2023) 910}.

\bibitem{khanna2025}
K.~Khanna and R.~Vasseur, {\slshape Measurement-induced entanglement in conformal field theory,} \href{http://arxiv.org/abs/2508.02788}{{ arXiv:2508.02788}}.

\bibitem{rottoli2024entanglementHamiltoniansquasiparticlepicture}
F.~Rottoli, C.~Rylands, and P.~Calabrese, {\slshape Entanglement {Hamiltonians} and the quasiparticle picture,} \href{https://link.aps.org/doi/10.1103/PhysRevB.111.L140302}{{\em Phys. Rev. B} {\bfseries 111} (Apr, 2025) L140302}.

\bibitem{fagotti2008evolution}
M.~Fagotti and P.~Calabrese, {\slshape {Evolution of entanglement entropy following a quantum quench: Analytic results for the $XY$ chain in a transverse magnetic field},} \href{https://link.aps.org/doi/10.1103/PhysRevA.78.010306}{{\em Phys. Rev. A} {\bfseries 78} (2008) 010306}.

\bibitem{Calabrese_2016_quench}
P.~Calabrese and J.~Cardy, {\slshape Quantum quenches in 1+1 dimensional conformal field theories,} \href{https://dx.doi.org/10.1088/1742-5468/2016/06/064003}{{\em J. Stat. Mech.} {\bfseries 2016} (2016) 064003}.

\bibitem{alba1}
V.~Alba and P.~Calabrese, {\slshape Entanglement and thermodynamics after a quantum quench in integrable systems,} \href{http://dx.doi.org/10.1073/pnas.1703516114}{{\em PNAS} {\bfseries 114} (2017) 7947}.

\bibitem{alba2}
V.~Alba and P.~Calabrese, {\slshape {Entanglement dynamics after quantum quenches in generic integrable systems},} \href{https://scipost.org/10.21468/SciPostPhys.4.3.017}{{\em SciPost Phys.} {\bfseries 4} (2018) 017}.

\bibitem{klobas2021}
K.~Klobas and B.~Bertini, {\slshape {Entanglement dynamics in Rule 54: Exact results and quasiparticle picture},} \href{https://scipost.org/10.21468/SciPostPhys.11.6.107}{{\em SciPost Phys.} {\bfseries 11} (2021) 107}.

\bibitem{PRX}
B.~Bertini, K.~Klobas, V.~Alba, G.~Lagnese, and P.~Calabrese, {\slshape Growth of {R\'enyi} entropies in interacting integrable models and the breakdown of the quasiparticle picture,} \href{https://link.aps.org/doi/10.1103/PhysRevX.12.031016}{{\em Phys. Rev. X} {\bfseries 12} (2022) 031016}.

\bibitem{bertini2023nonequilibrium}
B.~Bertini, P.~Calabrese, M.~Collura, K.~Klobas, and C.~Rylands, {\slshape Nonequilibrium full counting statistics and symmetry-resolved entanglement from space-time duality,} \href{https://link.aps.org/doi/10.1103/PhysRevLett.131.140401}{{\em Phys. Rev. Lett.} {\bfseries 131} (2023) 140401}.

\bibitem{bertini_asymmetric_2024}
B.~Bertini, K.~Klobas, M.~Collura, P.~Calabrese, and C.~Rylands, {\slshape Dynamics of charge fluctuations from asymmetric initial states,} \href{https://link.aps.org/doi/10.1103/PhysRevB.109.184312}{{\em Phys. Rev. B} {\bfseries 109} (2024) 184312}.

\bibitem{Bertini_2018}
B.~Bertini, M.~Fagotti, L.~Piroli, and P.~Calabrese, {\slshape Entanglement evolution and generalised hydrodynamics: {Noninteracting} systems,} \href{https://dx.doi.org/10.1088/1751-8121/aad82e}{{\em J. Phys. A} {\bfseries 51} (2018) 39LT01}.

\bibitem{Entanglement_GHD_interacting}
V.~Alba, B.~Bertini, and M.~Fagotti, {\slshape {Entanglement evolution and generalised hydrodynamics: interacting integrable systems},} \href{https://scipost.org/10.21468/SciPostPhys.7.1.005}{{\em SciPost Phys.} {\bfseries 7} (2019) 005}.

\bibitem{alba2019quantum}
V.~Alba and P.~Calabrese, {\slshape Quantum information dynamics in multipartite integrable systems,} \href{http://dx.doi.org/10.1209/0295-5075/126/60001}{{\em EPL} {\bfseries 126} (2019) 60001}.

\bibitem{AlbaQA}
V.~Alba and P.~Calabrese, {\slshape Quench action and {R\'enyi }entropies in integrable systems,} \href{https://link.aps.org/doi/10.1103/PhysRevB.96.115421}{{\em Phys. Rev. B} {\bfseries 96} (Sep, 2017) 115421}.

\bibitem{parez2021exact}
G.~Parez, R.~Bonsignori, and P.~Calabrese, {\slshape Exact quench dynamics of symmetry resolved entanglement in a free fermion chain,} \href{http://dx.doi.org/10.1088/1742-5468/ac21d7}{{\em J. Stat. Mech.} {\bfseries 2021} (2021) 093102}.

\bibitem{parez2021quasiparticle}
G.~Parez, R.~Bonsignori, and P.~Calabrese, {\slshape Quasiparticle dynamics of symmetry-resolved entanglement after a quench: Examples of conformal field theories and free fermions,} \href{https://link.aps.org/doi/10.1103/PhysRevB.103.L041104}{{\em Phys. Rev. B} {\bfseries 103} (2021) L041104}.

\bibitem{Parez_2022}
G.~Parez, R.~Bonsignori, and P.~Calabrese, {\slshape Dynamics of charge-imbalance-resolved entanglement negativity after a quench in a free-fermion model,} \href{https://dx.doi.org/10.1088/1742-5468/ac666c}{{\em J. Stat. Mech.} {\bfseries 2022} (2022) 053103}.

\bibitem{Murciano2022}
S.~Murciano, V.~Alba, and P.~Calabrese, {\slshape Quench dynamics of rényi negativities and the quasiparticle picture,} \href{http://dx.doi.org/10.1007/978-3-031-03998-0_14}{{\em Entanglement in Spin Chains} (2022) 397}.

\bibitem{ares2023entanglement}
F.~Ares, S.~Murciano, and P.~Calabrese, {\slshape Entanglement asymmetry as a probe of symmetry breaking,} \href{http://dx.doi.org/10.1038/s41467-023-37747-8}{{\em Nat. Commun.} {\bfseries 14} (2023) 2036}.

\bibitem{murciano2024entanglement}
S.~Murciano, F.~Ares, I.~Klich, and P.~Calabrese, {\slshape Entanglement asymmetry and quantum {Mpemba} effect in the {XY} spin chain,} \href{http://dx.doi.org/10.1088/1742-5468/ad17b4}{{\em J. Stat. Mech.} {\bfseries 2024} (2024) 013103}.

\bibitem{travaglino2024}
R.~Travaglino, C.~Rylands, and P.~Calabrese, {\slshape Quasiparticle picture for entanglement {H}amiltonians in higher dimensions,} \href{https://dx.doi.org/10.1088/1742-5468/adb7d3}{{\em J. Stat. Mech} {\bfseries 2025} (2025) 033102}.

\bibitem{travaglino2025}
R.~Travaglino, C.~Rylands, and P.~Calabrese, {\slshape Quench dynamics of negativity {H}amiltonians,} \href{http://arxiv.org/abs/2506.09561}{{ arXiv:2506.09561}}.

\bibitem{alba_carollo2021}
V.~Alba and F.~Carollo, {\slshape Spreading of correlations in markovian open quantum systems,} \href{https://link.aps.org/doi/10.1103/PhysRevB.103.L020302}{{\em Phys. Rev. B} {\bfseries 103} (2021) L020302}.

\bibitem{Alba_2022}
V.~Alba and F.~Carollo, {\slshape Hydrodynamics of quantum entropies in ising chains with linear dissipation,} \href{https://dx.doi.org/10.1088/1751-8121/ac48ec}{{\em J. Phys. A} {\bfseries 55} (2022) 074002}.

\bibitem{carollo_alba2022}
F.~Carollo and V.~Alba, {\slshape Dissipative quasiparticle picture for quadratic markovian open quantum systems,} \href{https://link.aps.org/doi/10.1103/PhysRevB.105.144305}{{\em Phys. Rev. B} {\bfseries 105} (2022) 144305}.

\bibitem{alba_carollo2023}
V.~Alba and F.~Carollo, {\slshape {Logarithmic negativity in out-of-equilibrium open free-fermion chains: An exactly solvable case},} \href{https://scipost.org/10.21468/SciPostPhys.15.3.124}{{\em SciPost Phys.} {\bfseries 15} (2023) 124}.

\bibitem{carollo2022}
F.~Carollo and V.~Alba, {\slshape Entangled multiplets and spreading of quantum correlations in a continuously monitored tight-binding chain,} \href{https://link.aps.org/doi/10.1103/PhysRevB.106.L220304}{{\em Phys. Rev. B} {\bfseries 106} (2022) L220304}.

\bibitem{Goldstein-sela}
M.~Goldstein and E.~Sela, {\slshape Symmetry-resolved entanglement in many-body systems,} \href{https://link.aps.org/doi/10.1103/PhysRevLett.120.200602}{{\em Phys. Rev. Lett.} {\bfseries 120} (2018) 200602}.

\bibitem{xavier_alcaraz_sierra}
J.~C. Xavier, F.~C. Alcaraz, and G.~Sierra, {\slshape Equipartition of the entanglement entropy,} \href{https://link.aps.org/doi/10.1103/PhysRevB.98.041106}{{\em Phys. Rev. B} {\bfseries 98} 041106}.

\bibitem{castro2024symmetry}
O.~A. Castro-Alvaredo and L.~Santamaría-Sanz, {\slshape Symmetry-resolved measures in quantum field theory: A short review,} \href{http://dx.doi.org/10.1142/S0217984924300023}{{\em Mod. Phys. Lett. B} {\bfseries 39} (2024) }.

\bibitem{doyon2020lecture}
B.~Doyon, {\slshape {Lecture Notes On Generalised Hydrodynamics},} \href{https://scipost.org/10.21468/SciPostPhysLectNotes.18}{{\em SciPost Phys. Lect. Notes} (2020) 18}.

\bibitem{horvath2024full}
D.~X. Horv\'ath and C.~Rylands, {\slshape Full counting statistics of charge in quenched quantum gases,} \href{https://link.aps.org/doi/10.1103/PhysRevA.109.043302}{{\em Phys. Rev. A} {\bfseries 109} (2024) 043302}.

\bibitem{Gour:2009abc}
G.~Gour, I.~Marvian, and R.~W. Spekkens, {\slshape {Measuring the quality of a quantum reference frame: The relative entropy of frameness},} \href{http://dx.doi.org/10.1103/physreva.80.012307}{{\em Phys. Rev. A} {\bfseries 80} (2009) 012307}.

\bibitem{Casini:2019kex}
H.~Casini, M.~Huerta, J.~M. Mag\'an, and D.~Pontello, {\slshape {Entanglement entropy and superselection sectors. Part I. Global symmetries},} \href{http://dx.doi.org/10.1007/JHEP02(2020)014}{{\em JHEP} {\bfseries 02} (2020) 014}.

\bibitem{Casini:2020rgj}
H.~Casini, M.~Huerta, J.~M. Magan, and D.~Pontello, {\slshape {Entropic order parameters for the phases of QFT},} \href{http://dx.doi.org/10.1007/JHEP04(2021)277}{{\em JHEP} {\bfseries 04} (2021) 277}.

\bibitem{Marvian:2014awa}
I.~Marvian and R.~W. Spekkens, {\slshape {Extending Noether's theorem by quantifying the asymmetry of quantum states},} \href{http://dx.doi.org/10.1038/ncomms4821}{{\em Nat. Commun.} {\bfseries 5} (2014) 3821}.

\bibitem{Capizzi:2023xaf}
L.~Capizzi and V.~Vitale, {\slshape {A universal formula for the entanglement asymmetry of matrix product states},} \href{http://dx.doi.org/10.1088/1751-8121/ad8796}{{\em J. Phys. A} {\bfseries 57} (2024) 45LT01}.

\bibitem{Ferro:2023sbn}
F.~Ferro, F.~Ares, and P.~Calabrese, {\slshape {Non-equilibrium entanglement asymmetry for discrete groups: the example of the XY spin chain},} \href{http://dx.doi.org/10.1088/1742-5468/ad138f}{{\em J. Stat. Mech.} {\bfseries 2402} (2024) 023101}.

\bibitem{Chen:2023gql}
M.~Chen and H.-H. Chen, {\slshape {R{\'e}nyi entanglement asymmetry in (1+1)-dimensional conformal field theories},} \href{http://dx.doi.org/10.1103/PhysRevD.109.065009}{{\em Phys. Rev. D} {\bfseries 109} (2024) 065009}.

\bibitem{Fossati:2024xtn}
M.~Fossati, F.~Ares, J.~Dubail, and P.~Calabrese, {\slshape {Entanglement asymmetry in CFT and its relation to non-topological defects},} \href{http://dx.doi.org/10.1007/JHEP05(2024)059}{{\em JHEP} {\bfseries 05} (2024) 059}.

\bibitem{Benini:2024xjv}
F.~Benini, V.~Godet, and A.~H. Singh, {\slshape Entanglement asymmetry in conformal field theory and holography,} \href{http://dx.doi.org/10.1093/ptep/ptaf080}{{\em Prog. Theor. Exp. Phys.} {\bfseries 2025} (2025) }.

\bibitem{Fossati:2024ekt}
M.~Fossati, C.~Rylands, and P.~Calabrese, {\slshape {Entanglement asymmetry in CFT with boundary symmetry breaking},} \href{http://dx.doi.org/10.1007/JHEP06(2025)089}{{\em JHEP} {\bfseries 06} (2025) 089}.

\bibitem{Kusuki:2024gss}
Y.~Kusuki, S.~Murciano, H.~Ooguri, and S.~Pal, {\slshape {Entanglement asymmetry and symmetry defects in boundary conformal field theory},} \href{http://dx.doi.org/10.1007/JHEP01(2025)057}{{\em JHEP} {\bfseries 01} (2025) 057}.

\bibitem{Ares:2023ggj}
F.~Ares, S.~Murciano, L.~Piroli, and P.~Calabrese, {\slshape {Entanglement asymmetry study of black hole radiation},} \href{http://dx.doi.org/10.1103/PhysRevD.110.L061901}{{\em Phys. Rev. D} {\bfseries 110} (2024) L061901}.

\bibitem{Russotto:2024pqg}
A.~Russotto, F.~Ares, and P.~Calabrese, {\slshape {Non-Abelian entanglement asymmetry in random states},} \href{http://dx.doi.org/10.1007/JHEP06(2025)149}{{\em JHEP} {\bfseries 06} (2025) 149}.

\bibitem{ares2023lack}
F.~Ares, S.~Murciano, E.~Vernier, and P.~Calabrese, {\slshape {Lack of symmetry restoration after a quantum quench: An entanglement asymmetry study},} \href{http://dx.doi.org/10.21468/SciPostPhys.15.3.089}{{\em SciPost Phys.} {\bfseries 15} (2023) 089}.

\bibitem{rylands2024microscopic}
C.~Rylands, K.~Klobas, F.~Ares, P.~Calabrese, S.~Murciano, and B.~Bertini, {\slshape Microscopic origin of the quantum {M}pemba effect in integrable systems,} \href{https://link.aps.org/doi/10.1103/PhysRevLett.133.010401}{{\em Phys. Rev. Lett.} {\bfseries 133} (2024) 010401}.

\bibitem{shion2}
S.~Yamashika, F.~Ares, and P.~Calabrese, {\slshape {Entanglement asymmetry and quantum {M}pemba effect in two-dimensional free-fermion systems},} \href{https://link.aps.org/doi/10.1103/PhysRevB.110.085126}{{\em Phys. Rev. B} {\bfseries 110} (2024) 085126}.

\bibitem{Summer:2025wsa}
A.~Summer, M.~Moroder, L.~P. Bettmann, X.~Turkeshi, I.~Marvian, and J.~Goold, {\slshape {A resource theoretical unification of Mpemba effects: classical and quantum},} \href{http://arxiv.org/abs/2507.16976}{{ arXiv:2507.16976}}.

\bibitem{tiltedferro1}
G.~M\"uller and R.~E. Shrock, {\slshape Implications of direct-product ground states in the one-dimensional quantum {XYZ} and {XY} spin chains,} \href{https://link.aps.org/doi/10.1103/PhysRevB.32.5845}{{\em Phys. Rev. B} {\bfseries 32} (1985) 5845}.

\bibitem{tiltedferro2}
J.~Kurmann, H.~Thomas, and G.~Müller, {\slshape Antiferromagnetic long-range order in the anisotropic quantum spin chain,} \href{https://www.sciencedirect.com/science/article/pii/0378437182902175}{{\em Physica A} {\bfseries 112} (1982) 235}.

\bibitem{drude_weights}
B.~Doyon and H.~Spohn, {\slshape Drude weight for the {Lieb-Liniger Bose} gas,} \href{https://scipost.org/10.21468/SciPostPhys.3.6.039}{{\em SciPost Phys.} {\bfseries 3} (2017) 039}.

\bibitem{abramowitz1948handbook}
M.~Abramowitz and I.~A. Stegun, {\em Handbook of mathematical functions with formulas, graphs, and mathematical tables}, vol.~55.
\newblock US Government printing office, 1948.

\bibitem{Andrews_Askey_Roy_1999}
G.~E. Andrews, R.~Askey, and R.~Roy, {\em The Gamma and Beta Functions}, p.~1–60.
\newblock Encyclopedia of Mathematics and its Applications.
\newblock Cambridge University Press, 1999.

\end{thebibliography}\endgroup
